\documentclass[a4paper,fleqn,usenatbib]{aa}

\usepackage[T1]{fontenc}
\usepackage{ae,aecompl}

\usepackage{graphicx}	
\usepackage{amsmath}	
\usepackage{amssymb}	
\usepackage{mathtools}
\usepackage{comment}
\usepackage{subfigure}
\usepackage{hyperref}
\hypersetup{
    colorlinks = true,
    citecolor = blue,
}

\graphicspath{ {./images/} }

\DeclareMathOperator{\Vect}{Vect}

\begin{document} 

\title{Axisymmetric equilibrium models for magnetised neutron stars in scalar-tensor theories}

   \author{J. Soldateschi
          \inst{1,2,3}
          \and
          N. Bucciantini\inst{2,1,3}
          \and
          L. Del Zanna\inst{1,2,3}
          }

	\offprints{J. Soldateschi, \email{soldateschi@arcetri.astro.it} or N. Bucciantini, \email{niccolo@arcetri.astro.it}}

   \institute{Dipartimento di Fisica e Astronomia, Universit\`a degli Studi di Firenze, Via G. Sansone 1, I-50019 Sesto F. no (Firenze), Italy
         \and
             INAF - Osservatorio Astrofisico di Arcetri, Largo E. Fermi 5, I-50125 Firenze, Italy
		\and
			INFN - Sezione di Firenze, Via G. Sansone 1, I-50019 Sesto F. no (Firenze), Italy
             }
    
\date{Received XXX; accepted YYY}

\abstract{
Among the possible extensions of general relativity that have been put forward to address some long-standing issues in our understanding of the Universe, scalar-tensor theories have received a lot of attention for their simplicity. Interestingly, some of these predict a potentially observable non-linear phenomenon, known as spontaneous scalarisation, in the presence of highly compact matter distributions, as in the case of neutron stars. Neutron stars are ideal laboratories for investigating the properties of matter under extreme conditions and, in particular, they are known to harbour the strongest magnetic fields in the Universe. Here, for the first time, we present a detailed study of magnetised neutron stars in scalar-tensor theories. First, we showed that the formalism developed for the study of magnetised neutron stars in general relativity, based on the `extended conformally flat condition', can easily be extended in the presence of a non-minimally coupled scalar field, retaining many of its numerical advantages. We then carried out a study of the parameter space considering the two extreme geometries of purely toroidal and purely poloidal magnetic fields, varying both the strength of the magnetic field and the intensity of scalarisation. We compared our results with magnetised general-relativistic solutions and un-magnetised scalarised solutions, showing how the mutual interplay between magnetic and scalar fields affect the magnetic and the scalarisation properties of neutron stars. In particular, we focus our discussion on magnetic deformability, maximum mass, and range of scalarisation.}

          
\titlerunning{Axisymmetric magnetised neutron stars in scalar-tensor theories}
\authorrunning{J. Soldateschi et al.}
\maketitle


\section{Introduction}
\label{sec:intro}

The recent observation of gravitational and electromagnetic radiation coming from the merger of a binary neutron star system \citep{abbott_gw170817:_2017} has given us a new opportunity to test the theory of general relativity (GR) in the strong-field regime \citep{will_confrontation_2014}, beyond the vacuum case of binary black hole (BH) mergers \citep{abbott_bbh_2016}, and to probe the physics of compact objects in unprecedented detail \citep{abbott_multi-messenger_2017,abbott_gravitational_2017}, fostering a renewed interest in neutron stars (NSs) as possible probes of new gravitational physics.

Indeed, it is well known that our understanding of the Universe lacks an explanation for what is called the `dark sector'. While a possible solution is to assume the existence of dark matter and dark energy \citep{trimble_existence_1987,peebles_cosmological_2003}, a different approach is to consider the possibility that GR is not the definitive theory of gravity. Moreover, on a more theoretical basis, a consistent quantum theory of GR does not yet exist \citep{bars_is_1989,deser_infinities_2000}. This led to the development of theoretical frameworks [e.g. the hypothesis of Strings \citep{green_superstring_1988}] which try to give an explanation of the fundamental interactions which is different from the mainstream one, leading to a modification, in the low energy limit, of GR itself.
\\\\
Many attempts have been made to extend GR to account for such issues, giving rise to many alternative theories of gravity \citep{capozziello_extended_2011}. Among the most studied are: $f(R)$ theories \citep{buchdahl_non-linear_1970,de_felice_f_2010}, where deviations from GR are introduced by modifying the functional dependence of the gravitational Lagrangian on the Ricci scalar $R$; Gauss-Bonnet gravity \citep{lovelock_einstein_1971}, which increases the dimensionality of the spacetime; scalar-tensor theories (STTs) - to some extent equivalent to $f(R)$ theories \citep{sotiriou_fr_2006} - which modify gravity with respect to GR replacing the gravitational constant $G$ with a dynamical scalar field. STTs have been widely studied in the past \citep{brans_machs_1961,nord_1970,wagoner_scalar-tensor_1970,matsuda_hydrodynamic_1973,damour_nonperturbative_1993,damour_tensor-scalar_1996,novak_spherical_1998,fujii_scalar-tensor_2003,faraoni_2004,shibata_coalescence_2014,langlois_2018,gong_2018,quiros_2019,zhang_2019} and are among the most promising alternatives to GR. This is due to a number of reasons: they are the most simple extensions of GR \citep[Sec.~1.2]{papantonopoulos_modifications_2015}; they are predicted to be the low-energy limit of some possible theories of Quantum Gravity \citep{damour_runaway_2002}; most of them respect the weak equivalence principle (WEP) - that is they are metric theories of gravity \citep{will_confrontation_2014} - which has been extremely well tested \citep{touboul_microscope_2017}. They also seem to be free of some of the pathologies affecting other extensions of GR \citep{defelice_2006,defelice_2010,bertolami_2016}. On the other hand, STTs violate the strong equivalence principle (SEP), which means that tests using self-gravitating bodies are ideal to constrain them \citep{barausse_testing_2017}.

The foundations of STTs were laid by \citet{brans_machs_1961} in a seminal paper, in which the authors modified the Einstein-Hilbert action of GR attempting to bring it in conformity with Mach's principle by replacing the gravitational constant $G$ by a scalar field non-minimally coupled to the spacetime metric, giving birth to the Jordan-Fierz-Brans-Dicke theory (BD). Unfortunately, observational tests in the Solar System seem to have proved BD wrong, unless its only parameter is precisely fine tuned, in contrast with the principle of naturalness \citep{scharer_testing_2014}. In this scenario, the study of NSs is especially important, because since the first work on massless mono-scalar STTs \citep{damour_nonperturbative_1993}, a non-perturbative strong field effect has been predicted, allowing the scalar field to exponentially grow in magnitude inside compact material objects. Even generalisations of STTs to massive scalar fields and other gravitational theories have been shown to be subject to a similar phenomenon \citep{salgado_1998,ramazanoglu_spontaneous_2016,rama_2017,silva_2018,andreou_2019}. Scalarisation can happen in various contexts: binary systems of merging NSs can undergo a `dynamical scalarisation' process \citep{barausse_neutron-star_2013}, in which the initially non-scalarised NSs become scalarised once they get closer to each other; again, in a binary NS system, one scalarised star can prompt an `induced scalarisation' on its non-scalarised companion \citep{barausse_neutron-star_2013}; or even in an isolated NS system, where `spontaneous scalarisation' can develop (this was the first discovered non-perturbative strong field effect in STTs, \citealt{damour_nonperturbative_1993}). 
The importance of scalarisation is that STTs which include such effects predict strong deviations from GR only inside compact objects, while  allowing the tight observational constraints in the weak-gravity regime to be fulfilled \citep{shao_constraining_2017}. 
As of today, the strongest limit on the strength of spontaneous scalarisation for massless STTs comes from observations of pulsars in binary systems, in particular in systems characterised by a large mass difference between the two stars, where STTs predict the emission of dipole scalar waves, potentially observable in the dynamics of the inspiral \citep{freire_relativistic_2012,will_confrontation_2014,shao_constraining_2017,anderson_2019}. These, however, are binaries with large separations and the constraints do not apply in the case of screening \citep{yaza_2016,doneva_rapidly_2016}. 

Scalarisation modifies the relation between the mass and radius of the NS and its central density. In general, scalarised NSs have larger radii and higher maximum masses than the corresponding GR solutions computed with the same equation of state (EoS). Moreover, scalarisation is more effective at higher compactness. The presence of a strong scalar charge could, in principle, have important consequences on the phenomenology of NSs, even if many of these effects might be degenerate with the EoS. A different dependence of the mass and radius from the central density could lead to appreciable changes in the thermal evolution of NSs \citep{dohi_2020}, given the dependence of many cooling processes on the density itself \citep{yako_2005}. Changes in radii could potentially be observable in the distribution function of millisecond pulsars \citep{papitto_2014}. The same holds for the distribution of NS masses, and the expected maximum mass (the recent measure of a 13km radius for a 1.44M$_\odot$ NS by NICER, \citealt{miller_2019}, suggests larger NS radii than previously thought, \citealt{ozel_2016}). Spontaneous scalarisation might impact the dynamics and evolution of the post merger remnant of binary NS coalescence \citep{raithel_2018,abbott_2017post}. Indeed, there is some observational evidence suggesting the presence of long lived NSs powering the X-ray afterglow of Short-GRBs \citep{rowlinson_2013}, suggesting  values of the  maximum NS mass  $\gtrsim 2.2 M_\odot$ \citep{gao_2015,margalit_2017}. Scalar fields can affect the deformability of NSs \citep{doneva_rapidly_2014,doneva_differentially_2018}, leaving an imprint in the pre-merger inspiral, and in the spin-down history of millisecond  proto-magnetar as possible engines of GRBs \citep{Dall'Osso_Shore+09a}. Scalarised NSs differ in the frequency of their normal modes \citep{sotani_2005}. On top of this STTs predicts also a new scalar wave emission, potentially detectable with future gravitational waves (GWs) observatories \citep{gerosa_numerical_2016,hagihara_2019}.
\\\\
So far, only non-magnetised models of NSs have been studied in STTs in the full non-linear regime (see e.g. \citealt{suvorov_monopolar_2018} for a perturbative approach to the magnetised scenario). Most of them focus on static \citep{damour_nonperturbative_1993,harada_neutron_1998,novak_neutron_1998,taniguchi_quasiequilibrium_2015,anderson_scalar_2019,doneva_topological_2019} or slowly rotating \citep{damour_tensor-scalar_1996,sotani_slowly_2012,pani_i-love-q_2014,silva_slowly_2015} stars, while recently some work has been done for rapidly \citep{doneva_rapidly_2014,doneva_rapidly_2016,pappas_multipole_2019} and differentially \citep{doneva_differentially_2018} rotating models. NSs have also been studied beyond the massless limit, and in the presence of a screening potential \citep{doneva_rapidly_2016,yaza_2016,brax_2017,staykov_2018,doneva_topological_2019,staykov_2019}. However, NSs are known to contain extremely powerful magnetic fields, inferred to be in the range $10^{8-12}$G for normal pulsars and up to $10^{16}$G at the surface of magnetars, while newly formed proto-NSs are hypothesised to store magnetic fields as high as $10^{17-18}$G in their core (\citealt{bonanno_mean-field_2003,rheinhardt_proto-neutron-star_2005,burrows_simulations_2007,spruit_source_2009,ferrario_magnetic_2015,popov_origins_2016}; see also \citealt{price_2006,kawamura_2016,ciolfi_2019} for simulations showing remnants of binary NSs merger harboring such intense magnetic fields). These magnetic fields substantially affect the electromagnetic phenomenology of NSs, can act as a potentially detectable source of deformation, can modify the torsional oscillations of NSs and can also alter the cooling properties of the crust. This shows that an accurate modelling of the magnetised structure of NSs is fundamental for a correct understanding of their properties. 

In GR, the first magnetised model of NS dates back to \citet{chandrasekhar_problems_1953}. Throughout the years, many magnetised models were proposed \citep{ferraro_equilibrium_1954,roberts_equilibrium_1955,prendergast_equilibrium_1956,woltjer_magnetostatic_1960,monaghan_magnetic_1965,monaghan_magnetic_1966,roxburgh_magnetostatic_1966,ostriker_rapidly_1968,miketinac_structure_1975}, up to more recent works \citep{tomimura_new_2005,yoshida_twisted-torus_2006,fujisawa_appearance_2015}.
 Due to the non-linearity of the general-relativistic magnetohydrodynamic (GRMHD) equations, an accurate study of the structure of NSs must be done in a numerical way, and only recently numerical results in the full GR regime have appeared. Many of these models focus on either purely toroidal \citep{kiuchi_relativistic_2008,kiuchi_equilibrium_2009,frieben_equilibrium_2012} or purely poloidal \citep{bocquet_rotating_1995,konno_moments_2001,yazadjiev_relativistic_2012} magnetic field configurations [see also \citet{pili_axisymmetric_2014,pili_general_2017}]. However, such models are shown to develop an instability which causes the magnetic field to rearrange in a mixed configuration, called Twisted Torus, which is roughly axisymmetric \citep{prendergast_equilibrium_1956,tayler_adiabatic_1973,wright_pinch_1973,braithwaite_stable_2006,braithwaite_evolution_2006,braithwaite_axisymmetric_2009,lasky_2011}. Twisted Torus configurations have been studied only very recently \citep{ciolfi_2013,pili_axisymmetric_2014,uryu_equilibrium_2014,bucciantini_role_2015,uryu_new_2019}, because they require to solve a large set of coupled non-linear elliptic PDEs, which can be numerically unstable.
\\\\
In this paper, we present the first numerical computations of a magnetised NS in a STT of gravity in the full non-linear regime. We wish to investigate how the mutual interplay of a strong magnetic field and a scalar field modifies both the magnetic properties of NSs, with respect to GR, and their scalarisation properties with respect to the un-magnetised case. For this reason we are going to provide a characterisation as complete as possible of our equilibrium configurations, including a parametrisation of their deformation, and to carry a comparison with GR, not just in terms of global quantities but also in the specific internal distribution of density and magnetic field.
The purpose is to quantify for example how much the presence of a scalar field affects the magnetic deformability of NSs, which is a key parameter to evaluate the relative importance of GW vs electromagnetic  dipole emission in the early spin-down of proto-NSs \citep{Dall'Osso_Shore+09a}, and to assess the validity of the millisecond-magnetar model for Long GRBs \citep{Metzger_Giannios+11a}. On the other hand, we also want to evaluate if the presence of a magnetic field favours or disfavours the scalarisation of NSs, and how it changes the scalarisation range, or the maximum NS mass. For this reason we limit our analysis only to the two extreme cases of purely poloidal or purely toroidal magnetic fields, neglecting rotation. In this sense our work is both an extension of the existing literature on magnetised models of NSs in GR, and of un-magnetised models in STTs.

We also take the opportunity to introduce a computational strategy, which, for the sake of simplicity, we discuss here just in the case of non-rotating NSs, but that can easily be generalised to rotating and even dynamical regimes and that allows a straightforward extension of well established algorithms for GRMHD to handle MHD in STTs. 
Our algorithm is an extension of the well-tested \texttt{XNS} solver \citep{pili_axisymmetric_2014,pili_general_2017} to the case of a generic STT. It is based on the eXtended Conformally Flat Condition (XCFC) for the metric \citep{wilson_relativistic_1996,wilson_mathews_2003,cordero-carrion_improved_2009,bucciantini_general_2011}, which, even if not formally exact, has proved to be highly accurate for rotating NSs \citep{Camelio_Dietrich+19a}. We wish to point here that the accuracy of the solution with respect to full GR depends on which parameter, that is the central rotation rate or the surface ellipticity, is held fixed in the comparison (larger deviations have been found for differentially rotating models having the same surface ellipticity \citealt{iosif_2014}). The XCFC system has several advantages from a numerical point of view. These, as we are going to show, are retained also in STTs, and that can easily be adapted to the more complex case of time dependent dynamical evolution.
\\\\
This paper is structured as follows. In Sect.~\ref{sec:stt} we introduce MHD within STTs, both from a Lagrangian point of view, and within the 3+1 formalism. In Sect.~\ref{sec:magnetised} we show how the formalism developed to model magnetised NSs in GR can be extended to STTs. In Sect.~\ref{sec:xns} we present the new version of \texttt{XNS} for STTs.  In Sect.~\ref{sec:results} we illustrate and discuss our results for various magnetic configurations and choice of STT and, finally, we conclude in  Sect.~\ref{sec:conclusions}.

\section{Scalar-tensor theories and 3+1}
\label{sec:stt}

In the following we assume a signature $\{-, +, +, +\}$ for the
spacetime metric and use Greek letters $\mu$, $\nu$, $\lambda$, ... (running from
0 to 3) for 4D spacetime tensor components, while Latin letters
$i$, $j$, $k$, ... (running from 1 to 3) are employed for 3D spatial tensor components. Moreover, we use the dimensionless units where $c = G = \mathrm{M}_\odot = 1$, and we
absorb the $\sqrt{4\pi}$ factors in the definition of the electromagnetic quantities. Variables denoted with a tilde, $\tilde{\cdot}$, are calculated in the Jordan frame, while quantities denoted with a bar, $\bar{\cdot}$, are expressed in the Einstein frame.

\subsection{STT frames and ideal MHD}
\label{sec:sttmhd}

The most general action $S_{\mathrm J} $ that describes the mutual interplay of an ideal magnetised fluid at thermodynamic equilibrium with a gravitational space-time containing one scalar field $\varphi $ non-minimally coupled to the metric $\tilde{g}_{\mu \nu}$, is invariant under space-time diffeomorphisms, is at most quadratic in the derivatives of the fields, and which satisfies the WEP, can be written as the sum of two terms. The first term, encoding the information about the gravitational fields, $\tilde{S}_{\mathrm g}[\tilde{g}_{\mu \nu},\varphi]$,  according to  the `Bergmann-Wagoner formulation' \citep{bergmann_comments_1968,wagoner_scalar-tensor_1970,berti_testing_2015} is
\begin{equation} \label{eq:joract}
	\tilde{S}_{\mathrm g}= \frac{1}{16\pi}\int d^4x \sqrt{-\tilde{g}}\left[ \varphi \tilde{R} - \frac{\omega (\varphi)}{\varphi} \tilde{\nabla} _\mu \varphi \tilde{\nabla} ^\mu \varphi - U(\varphi) \right] ,
\end{equation}
where $\tilde{g}$ is the determinant of the spacetime metric $\tilde{g}_{\mu \nu}$,  $\tilde{\nabla} _\mu$ its associated covariant derivative, $\tilde{R}$ its Ricci scalar, while $\omega (\varphi)$ and $U(\varphi)$ are, respectively, the coupling function and the potential of the scalar field $\varphi$. The second term $\tilde{S}_{\mathrm p}[\tilde{g}_{\mu \nu},\tilde{N}^\mu,\tilde{A}^\mu,\tilde{\varepsilon},\tilde{s}]$ contains information on the other physical fields and it is a function of the mass current density $\tilde{N}^\mu = \tilde{\rho}\tilde{u}^\mu$, expressed as a function of the rest mass density $\tilde{\rho}$ and four-velocity $\tilde{u}^\mu$, the specific entropy $\tilde{s}$, the internal energy density $\tilde{\varepsilon}(\tilde{\rho},\tilde{s})$, and the electromagnetic four-potential $\tilde{A}^\mu$. For an ideal fluid neglecting polarisation, magnetisation \citep{chatterjee_consistent_2015,franzon_internal_2016}, dynamo or resistivity \citep{bucciantini_fully_2013,del_zanna_fast_2016,del_zanna_chiral_2018,tomei2020}, it is 
\begin{equation} \label{eq:joract}
 \begin{split}
  \tilde{S}_{\mathrm p}&= \int d^4x \sqrt{-\tilde{g}}\bigg[ \tilde{\varepsilon}(\tilde{N}^\mu \tilde{N}_\mu,\tilde{s}) + \zeta \tilde{\nabla} _\mu \tilde{N}^\mu  + \eta \tilde{N}^\mu  \tilde{\nabla} _\mu \tilde{s} \\
    & \tilde{F}_{\mu\nu} \tilde{F}^{\mu\nu} + \tau_\nu \tilde{N}_\mu\tilde{F}^{\mu\nu}\bigg] \quad ,
  \end{split}
\end{equation}
where $\tilde{F}_{\mu\nu} \coloneqq \tilde{\nabla} _\mu \tilde{A}_\nu-\tilde{\nabla} _\nu \tilde{A}_\mu$ is the Faraday tensor, and $\zeta,\eta,\tau_{\nu}$, are Lagrangian multipliers that enforce mass conservation, entropy conservation, and the ideal MHD condition $\tilde{u}_\mu\tilde{F}^{\mu\nu} = 0$ respectively \citep{hawking_ellis_1973,Brown93a,Bekenstein_Oron00a}.

The frame where the action reads $S_{\mathrm J}  =  \tilde{S}_{\mathrm g} + \tilde{S}_{\mathrm p} $ is called the `Jordan frame' (J-frame).  Variation of the action with respect to the various fields (and Lagrangian multipliers) leads to the Euler-Lagrange field equations (and to the constraints). In particular, variations with respect to the four potential $\tilde{A}_\mu$ lead to the Maxwell equation:
\begin{equation}
\delta S_{\mathrm J}/\delta \tilde{A}_\mu =0 \quad\Rightarrow\quad \tilde{\nabla}_\mu \tilde{F}^{\mu\nu}  =-\tilde{J}^\nu \quad ,
\end{equation}
where $\tilde{J}^\nu$ is the electromagnetic four-current. Variations with respect to the matter four-current $\tilde{N}^\mu$ lead, ultimately, to the fluid Euler equation and to the momentum-energy conservation law:
\begin{equation}
\delta S_{\mathrm J}/\delta \tilde{N}^\mu =0 \quad\Rightarrow\quad \tilde{\nabla}_\mu \tilde{T}_{\mathrm p}^{\mu\nu}  = 0 \quad ,
\end{equation}
where the energy momentum tensor is
\begin{equation}
	\tilde{T}_{\mathrm p}^{\mu \nu}= [\tilde{\rho} + \tilde{\varepsilon} +\tilde{p}] \tilde{u}^\mu \tilde{u}^\nu + \tilde{p} \tilde{g}^{\mu \nu} + \tilde{F}^\mu _{\ \ \lambda} \tilde{F}^{\nu \lambda}-\frac{1}{4}\tilde{F}^{\lambda \kappa}\tilde{F}_{\lambda \kappa}\tilde{g}^{\mu \nu}
\end{equation}
and $\tilde{p}$ is the pressure. Given that the scalar field does not enter $\tilde{S}_\mathrm{p}$, the equations describing the behaviour of the physical quantities are unaffected by the presence of the scalar field.  Introducing the Hodge dual of the Faraday tensor $\tilde{F}^{\star \mu \nu} = \frac{1}{2}\tilde{\epsilon} ^{\mu \nu \lambda \kappa}\tilde{F}_{\lambda \kappa}$, where $\tilde{\epsilon} _{\mu \nu \lambda \kappa} = -(-\tilde{g})^{1/2}[\mu \nu \lambda \kappa]$ is the Levi-Civita pseudo-tensor and $[\mu \nu \lambda \kappa]$ is the alternating Levi-Civita symbol, one can write the energy momentum tensor of ideal MHD in terms of the comoving magnetic field $\tilde{b}^\mu= \tilde{u}_\nu\tilde{F}^{\star \mu \nu} $ as
 \begin{equation}\label{eq:mattem}
 	\tilde{T}_{\mathrm p}^{\mu \nu}=\left( \tilde{\rho} \tilde{h} +\tilde{b}^2 \right) \tilde{u}^\mu \tilde{u}^\nu -\tilde{b}^\mu \tilde{b}^\nu + \left( \tilde{p}+\frac{1}{2} \tilde{b}^2 \right) \tilde{g}^{\mu \nu} \quad ,
 \end{equation}
 where $\tilde{b}^2=\tilde{b}_\mu \tilde{b}^\mu$ and $\tilde{h} = 1+(\tilde{\varepsilon}+\tilde{p})/\tilde{\rho}$ is the specific enthalpy.
On the other hand, variations of the action with respect to the metric lead to the generalisation of Einstein's field equations:
\begin{equation}
\delta S_{\mathrm J}/\delta \tilde{g}_{\mu\nu} =0 \quad\Rightarrow\quad \tilde{G}^{\mu\nu} +\tilde{G}_\mathrm{s}^{\mu\nu} =8\pi \tilde{T}_{\mathrm p}^{\mu\nu} \quad ,
\end{equation}
where $\tilde{G}^{\mu\nu} $ is the standard Einstein tensor, while $\tilde{G}_\mathrm{s}^{\mu\nu} $ contains the contribution from  the non-minimally coupled scalar field. Tensor $\tilde{G}_\mathrm{s}^{\mu\nu} $ contains higher-order derivatives of the scalar field, and its associated energy density is not positively defined \citep{Santiago_Silbergleit00a}. As a consequence, in the J-frame the generalisation of Einstein's field equations has a different mathematical structure than in GR, implying that standard solution techniques and algorithms developed for GR cannot be naively applied.
However, it is possible to show \citep{Santiago_Silbergleit00a} that, by performing a conformal transformation of the metric,
\begin{equation}
\bar{g}_{\mu \nu} \coloneqq \varphi	\tilde{g}_{\mu \nu} \quad ,
\end{equation}
and introducing a new scalar field $\chi$ related to $\varphi$ according to
\begin{equation}
\frac{{\mathrm d}\chi}{{\mathrm d}\ln \varphi} \coloneqq \sqrt{\frac{\omega{(\varphi)} +3}{4}} \quad ,
\end{equation}
the gravitational part of the action becomes
\begin{equation} \label{eq:einact}
	\bar{S}_{\mathrm g} = \frac{1}{16\pi}\int d^4x \sqrt{-\bar{g}}\left[ \bar{R} - 2 \bar{\nabla} _\mu \chi \bar{\nabla}^\mu\chi - V(\chi) \right] \quad ,
\end{equation}
where $\bar{g}$ is the determinant of the spacetime metric $\bar{g}_{\mu \nu}$,  $\bar{\nabla} _\mu$ its associated covariant derivative, $\bar{R}$ its scalar curvature, and $V(\chi) = U(\varphi)/\varphi^2$ the potential of the scalar field $\chi$. The frame where the gravitational part of the action reads as in Eq.~\ref{eq:einact} is known as the `Einstein frame'  (E-frame). In the E-frame the  generalisation of Einstein's field equations reads
\begin{equation}\label{eq:einstt}
\bar{G}^{\mu\nu} =8\pi \left(\bar{T}_{\mathrm s}^{\mu\nu} + \bar{T}_{\mathrm p}^{\mu\nu}\right) \quad ,
\end {equation}
where $\bar{T}_{\mathrm p}^{\mu\nu} = \tilde{T}_{\mathrm p}^{\mu\nu}/\varphi^3$ and the contribution from the scalar field now has the form
\begin{equation}\label{eq:tmunuscal}
	\bar{T}_{\mathrm{s}}^{\mu \nu}=\frac{1}{4\pi}\left[ \bar{\nabla} ^\mu \chi \bar{\nabla} ^\nu \chi - \frac{1}{2} \bar{g}^{\mu \nu} \bar{\nabla} _\lambda \chi \bar{\nabla} ^\lambda \chi \right]\quad .
\end{equation}
It is evident that in the E-frame the metric field equations are equivalent to those of GR, and the scalar field acts only as an extra energy-momentum source term. 
However, this conformal transformation affects also the physical part of the action $\bar{S}_{\mathrm p}$, redefining both the metric (and its covariant derivative) but also the physical fields (e.g. the E-frame energy density is now a function also of the scalar field $\chi$). As a result $\bar{\nabla}_\mu\bar{T}^{\mu\nu} \neq 0$, and  $\bar{\nabla}_\mu(\bar{\rho}\bar{u}^{\mu} )\neq 0$. Interestingly Maxwell equations retain their form, as expected from their pre-metric nature \citep{Cartan86a,van-Dantzig_Dirac34a,Delphenich05a}. As a consequence, in the E-frame standard methods, techniques, and algorithms developed in MHD, based on the conserved nature of the various physical quantities, and the locality of the EoS, cannot be naively applied.

In the E-frame one can also derive an equation for the scalar field by varying the action with respect to $\chi$,
\begin{equation}\label{eq:scal}
	\bar{\nabla} _\mu \bar{\nabla} ^\mu\chi = -4\pi \alpha _\mathrm{s} \bar{T}_{\mathrm{p}} \quad ,
\end{equation}
where $\bar{T}_{\mathrm{p}}\coloneqq \bar{g}_{\mu \nu}\bar{T}_{\mathrm{p}}^{\mu \nu}=3\bar{p}-\bar{\varepsilon}-\bar{\rho}$ and
\begin{equation}\label{eq:scalalpha}
  \alpha _\mathrm{s}(\chi) \coloneqq -\frac{{\mathrm d} \ln \varphi(\chi)}{2{\mathrm d}\chi}\quad .
\end{equation}
We note that the only direct sources of a massless scalar field are those physical fields with a non-vanishing trace of the energy-momentum tensor; as such, the EM field is not a direct source of the scalar field, and for the same reason purely metric black holes in STTs are undistinguishable from those in GR \citep{hawking_black_1972,berti_testing_2015}. Analogously, in the ultra-relativistic asymptotically free regime, $\varepsilon+\rho =3p$ and the same considerations apply.

This suggests that a simultaneous use of the E-frame, to compute the metric and scalar field, and of the J-frame, to compute the physical field, by performing the conformal transformations between the two whenever necessary, will enable us to easily extend the standard techniques of GRMHD to the case of STTs. 

\subsection{3+1 decomposition}
\label{sec:3+1}

According to the 3+1 formalism \citep{alcubierre_introduction_2008,gourgoulhon_3+1_2012}, any globally hyperbolic spacetime admits a foliation with a family of spacelike hypersurfaces $\Sigma _t$ with normal timelike vector $n^\mu$ (which is, by definition, the velocity of the so-called `Eulerian observer', $n_\mu n^\mu =-1$). The three-metric induced on $\Sigma _t$ is $\gamma _{\mu \nu} = g_{\mu \nu} + n_\mu n_\nu$ (and the induced rank-3 Levi-Civita pseudo tensor is $\epsilon^{ijk} = \epsilon^{ijk\mu} n_\mu$). Calling $x^\mu =[t,x^i]$ the coordinates adapted to the foliation, the generic line element takes the form
 \begin{equation}
 	ds^2=-\alpha ^2 dt^2 + \gamma _{ij}\left( dx^i + \beta ^i dt \right)\left( dx^j + \beta ^j dt \right) \quad ,
 \end{equation}
 where $\alpha$ is the lapse function and $\beta ^i$ is the shift vector. If $\beta ^i = 0$, the spacetimes is said to be static.
 $n_\mu$ and $\gamma _{\mu \nu}$ allow one to project any tensor according to the foliation. The relation between the E-frame Eulerian observer and J-frame one is: $\tilde{n}_\mu = \mathcal{A}\bar{n}_\mu $,  $\tilde{\gamma} _{\mu \nu} = \mathcal{A}^2\bar{\gamma}_{\mu \nu} $, where we have introduced the  conformal function $\mathcal{A}=1/\sqrt{\varphi(\chi)}$ coupling the two frames.

The standard 3+1 decomposition of any vector is
\begin{align}
  U^\mu =  U_\parallel n^\mu + U_\perp ^\mu \quad , 
\end{align}
where $U_\parallel = - n_\mu U^\mu $ and $n_\mu U_\perp ^\mu= 0 $, while any rank-2 
 symmetric $X^{\mu\nu}$ and antisymmetric $A^{\mu\nu}$ tensor can be written as
\begin{align}
 X^{\mu\nu} &= Y n^\nu n^\nu + Z^\mu n^\nu + Z^\nu n^\mu + W^{\mu\nu} \quad , \\
A^{\mu\nu} &=  C^\mu n^\nu + C^\nu n^\mu  +\epsilon^{\mu\nu\lambda\kappa}D_\lambda n_\kappa \quad ,  
\end{align}    
where $n_\mu Z^\mu = 0 = n_\mu W^{\mu\nu}$ and $n_\mu C^\mu = n_\mu D^\mu =0$.  Recalling that the relations between the J-frame and E-frame physical energy-momentum and Faraday tensors are $\mathcal{A}^6\tilde{T}^{\mu\nu} = \bar{T}^{\mu\nu}$ and $\mathcal{A}^4\tilde{F}^{\mu\nu} = \bar{F}^{\mu\nu}$, one can easily recover the following relations among the various projections:
\begin{align}
  &\tilde{\Gamma} = -\tilde{n}_\mu \tilde{u}^\mu  = -\bar{n}_\mu \bar{u}^\mu = \bar{\Gamma} \quad , \\
 &   \mathcal{A}\tilde{v}^j =\tilde{\gamma}^{j}_{\;\mu} \mathcal{A}\tilde{u}^\mu = \bar{\gamma}^{j}_{\;\mu} \bar{u}^\mu = \bar{v}^j \quad , \\
 & \mathcal{A}^4\tilde{E}_{\mathrm p} = \mathcal{A}^{-2}\tilde{n}_\mu \tilde{n}_\nu \mathcal{A}^6\tilde{T}_{\mathrm p}^{\mu\nu} = \bar{n}_\mu \bar{n}_\nu \bar{T}_{\mathrm p}^{\mu\nu}  =\bar{E}_{\mathrm p} \quad ,\\
 &\mathcal{A}^5 {\tilde{S}^j}_{\mathrm p} = -\mathcal{A}^{-1}\tilde{n}_\mu \tilde{\gamma}^{j}_{\;\nu} \mathcal{A}^6\tilde{T}_{\mathrm p}^{\mu\nu} = -\bar{n}_\mu \bar{\gamma}^{j}_{\;\nu} \bar{T}_{\mathrm p}^{\mu\nu}  ={\bar{S}^j}_{\mathrm p} \quad ,\\
 &\mathcal{A}^6 {\tilde{W}^{ij}} _{\mathrm p} = \tilde{\gamma}^{i}_{\;\mu} \tilde{\gamma}^{j}_{\;\nu} \mathcal{A}^6\tilde{T}_{\mathrm p}^{\mu\nu} = \bar{\gamma}^{i}_{\;\mu} \bar{\gamma}^{j}_{\;\nu} \bar{T}_{\mathrm p}^{\mu\nu}  ={\bar{W}^{ij}}_{\mathrm p} \quad ,\\
 &\mathcal{A}^3 \tilde{B}^\mu = \mathcal{A}^4\tilde{F}^{\star \mu \nu}\tilde{n}_\nu \mathcal{A}^{-1}=\bar{F}^{\star \mu \nu}\bar{n}_\nu  = \bar{B}^\mu \quad ,\\
 &\mathcal{A}^3 \tilde{E}^\mu = \mathcal{A}^4\tilde{F}^{\mu \nu}\tilde{n}_\nu  \mathcal{A}^{-1}=\bar{F}^{\mu \nu}\bar{n}_\nu  = \bar{E}^\mu \quad ,
\end{align}
showing, for example, that the Lorentz factor $\Gamma$ is the same in the two frames.
The energy conservation law in J-frame, $\tilde{\nabla}_\mu\tilde{T}^{\mu\nu}= 0$, together with the mass conservation $\tilde{\nabla}(\tilde{\rho}\tilde{u}^{\mu})= 0$ and Maxwell equations, can be cast into a system for the evolution of the projected quantities $\tilde{E}_{\mathrm p} ,  {\tilde{S}^j}_{\mathrm p},   \tilde{B}^\mu,  \tilde{E}^\mu $, once an EoS and a closure for the electromagnetic currents (e.g. the Ideal MHD conditions) are provided, according for example to \citet{del_zanna_echo:_2007} and \citet{bucciantini_general_2011}. Then, the above equations allow to rescale those quantities to the E-frame, where they are used to solve the 3+1 evolutionary equations for the metric and the scalar field. For this purpose one needs also the 3+1 projection of the latter. This is only done in the E-frame, given that it is not needed in the J-frame, according to:
\begin{align}
  &\bar{\nabla}^\mu\chi = P \bar{n}^\mu + Q^\mu \quad, \\
  &\bar{E}_{\mathrm s} = \bar{n}_\mu \bar{n}_\nu \bar{T}_{\mathrm s}^{\mu\nu} = Q^2 + P^2 \quad, \\
   &{\bar{S}^j}_{\mathrm s} = -\bar{n}_\mu \bar{\gamma}^{j}_{\;\nu} \bar{T}_{\mathrm s}^{\mu\nu} = PQ^j \quad, \\
&{\bar{W}^{ij}}_{\mathrm s}= \bar{\gamma}^{i}_{\;\mu}  \bar{\gamma}^{j}_{\;\nu} \bar{T}_{\mathrm s}^{\mu\nu} = Q^iQ^j +(Q^2+P^2) \bar{\gamma}^{ij} \quad, 
\end{align}
where $Q^\mu $ is purely spatial and Eq.~\ref{eq:scal} can also be cast into a set of evolutionary equations for $P$ and $Q^i$ \citep{Salgado06a,Salgado_Martinez-del-Rio+08a}.  
\\\\
From now on, for the sake of clarity, and for ease of reading, we will drop the  $\bar{\cdot}, \;\tilde{\cdot}$ notation. All quantities referring either to the metric or the scalar field are assumed to be taken in the E-frame, while the MHD and fluid ones are to be considered in the J-frame. Whenever necessary, in case of possible ambiguity, the bar and tilde notation will be restored to specify the frame of reference for the given quantity. 

\section{Static magnetised configurations}
\label{sec:magnetised}

For the problem we are interested in,  we chose spherical-like coordinates $x^\mu = [t,r,\theta,\phi]$ and considered only configurations that are stationary and axisymmetric. This means that there exist two commuting Killing vectors, the timelike $t^\mu = (\partial _t)^\mu$ and the spacelike $\phi^\mu = (\partial _\phi)^\mu$ \citep{carter_commutation_1970,carter_black_1973_a,carter_black_1973_b}. These two vectors span a timelike two-plane $\Pi = \Vect (t^\mu , \phi ^\mu)$. Any vector $V^\mu \in \Pi$ is said to be toroidal, and takes the form $V^\mu = c_t t^\mu + c_\phi \phi ^\mu$; instead, it is said to be poloidal if it lies in the spacelike two-plane orthogonal to $\Pi$. Given the generalised Einstein's equations for the metric, Eq.~\ref{eq:einstt}, if both the scalar and physical energy-momentum tensors obey the relations  
\begin{equation}
\begin{split}\label{eq:circ}
	t_\mu & \bar{T}^{\mu [\nu}t^\kappa \phi ^{\lambda ]} = 0 \quad , \\
	\phi _\mu & \bar{T}^{\mu [\nu}t^\kappa \phi ^{\lambda ]} = 0 \quad ,
\end{split}
\end{equation}
where the square brackets mean anti-symmetrisation with respect to the enclosed indices, then  the spacetime has the additional property of being `circular' \citep{kundt_orthogonal_1966,carter_killing_1969}.  In this case, $\beta^r=\beta^\theta=0$, $\gamma_{r\phi}=\gamma_{\theta \phi}=\gamma_{r\theta}=0$ and all the remaining metric components depend solely on $r$ and $\theta$.

In case of circular spacetimes and spherical-like coordinates, the line element simplifies to
\begin{equation}\label{eq:quasi}
	ds^2 = -\alpha ^2 dt^2 + \psi ^4 \left( dr^2 + r^2 d\theta ^2 \right) +R_{\mathrm{qi}}^2 \left( d\phi + \beta^\phi dt \right)^2 ,
\end{equation}
where $R_{\mathrm{qi}} \coloneqq \sqrt{\gamma _{\phi \phi}}$ is the quasi-isotropic radius and $\psi$ is the conformal factor. A metric in the form of Eq.~\ref{eq:quasi}  is said to be `quasi-isotropic'.
Stationarity and axisymmetry are enough to ensure that $\bar{T}^{\mu \nu}_{\mathrm s}$ satisfies Eq.~\ref{eq:circ}. However they are not enough to ensure the same for the physical part $\bar{T}^{\mu \nu}_{\mathrm p}$. Given that the energy-momentum tensor of the E and J-frame are related by a simple conformal transformation, and the same holds for the Killing vectors and the metric, the conditions that ensure circularity in one of them will also ensure it in the other. For an ideal plasma, having an energy-momentum tensor as in Eq.~\ref{eq:mattem}, on top of stationarity and axisymmetry, circularity requires the four-velocity to be toroidal, $u^r=u^\theta =0$, and the magnetic field $b^\mu$ to be either purely toroidal or purely poloidal (in this latter case, rotation must also be uniform). On the contrary, even if the configuration is static and axisymmetric, for a magnetic field with a mixed configuration,  Eq.~\ref{eq:circ} does not hold, and in principle the metric of Eq.~\ref{eq:quasi} is no longer correct. However, even in this case it has been shown in GR \citep{oron_relativistic_2002,shibata_magnetohydrodynamics_2005,dimmelmeier_non-linear_2006,ott_rotating_2007,bucciantini_general_2011,pili_axisymmetric_2014, pili_general_2017} that Eq.~\ref{eq:quasi} provides a good approximation of the correct metric, and leads to small errors in the structure of rotating stars, mostly in the outer layers close to the surface, even in the extreme cases of a rotation at the mass-shedding limit, and magnetic fields as strong as 10$^{19}$G. Moreover it can be also  shown that in GR the difference $R_{\mathrm{qi}}-\psi ^2 r \sin \theta$ is at most of order of 10$^{-3}$ \citep{pili_general_2017}. Thus, to a good level of accuracy, the metric can be further simplified to the conformally flat (CFC) approximation \citep{wilson_mathews_2003,isenberg_waveless_2008}, for which
 \begin{equation}\label{eq:cfcmetric}
 	ds^2 \!=\! -\alpha ^2 dt^2\! +\! \psi ^4 \!\left[ dr^2\! +\! r^2 d\theta ^2\! +\! r^2 \!\sin ^2 \theta \left( d\phi +\!\beta^\phi dt \right)^2 \right] ,
 \end{equation}
where we have a common factor multiplying all flat-space metric terms in spherical coordinates.

From now on we shall restrict our analysis to \emph{static} configurations alone, that is to the case of non-rotating stars, for which $v^i=0$ and $\beta^i=0$ (see App.~\ref{app:rotating} for a discussion on rotators). As a consequence, the ideal-MHD electric field  $E_i= -\tilde{\epsilon}_{ijk} v^j B^k = -\mathcal{A}^{-3}\bar{\epsilon}_{ijk} v^j B^k = 0$ and $S^i=0$. Then, it can be shown that the extrinsic curvature  $K_{ij}=0$, which means that maximal slicing, $K=0$, holds [see \cite{gourgoulhon_3+1_2012} for a discussion of the interesting properties of this kind of slicing].  Under these assumptions, Einstein's equations reduce to a system of two Poisson-like elliptic equations for $\psi$ and $\alpha$:
 \begin{align}
 	\Delta & \psi = \left[ -2\pi \hat{E} \right] \psi ^{-1} \quad, \label{eq:psi} \\
 	\Delta &\left( \alpha \psi \right) = \left[ 2\pi \left( \hat{E}+2\hat{S} \right) \psi ^{-2} \right]\left( \alpha \psi \right) \quad , \label{eq:alpsi}
 \end{align}
 where $\Delta = f^{ij} \hat{\nabla} _i \hat{\nabla} _j$ and $\hat{\nabla} _i$ are, respectively, the 3D Laplacian and nabla operator of the flat space metric $f_{ij}$. We note that the two equations are decoupled, such that Eq.~\ref{eq:psi} can be solved before Eq.~\ref{eq:alpsi}.  The source terms take the form
 \begin{equation}
 \begin{split}
 	\hat E&= \psi ^6\left\{\mathcal{A}^4\left[e+\frac{1}{2}B^2\right]+ \frac{1}{8 \pi} Q^2 \right\}\quad , \\
 	\hat S &= \psi ^6\left\{\mathcal{A}^4 \left[3 p + \frac{1}{2}  B^2 \right]- \frac{1}{8\pi} Q^2 \right\}\quad .
 \end{split}
 \end{equation}
Under the same conditions, it can be shown that Eq.~\ref{eq:scal} reduces to
\begin{equation}\label{eq:scal1}
	\Delta \chi = -4\pi \psi ^4 \alpha _\mathrm{s}(\chi) \mathcal{A}^4T_\mathrm{p}-\partial \ln \left( \alpha \psi ^2 \right) \partial \chi \quad ,
\end{equation}
where $\partial f \partial g \coloneqq \partial _r f \partial _r g + (\partial _\theta f \partial _\theta g) /r^2$ and $T_\mathrm{p} = 3p- \varepsilon-\rho$ is the trace of the J-frame energy momentum tensor.

We note that the Poisson-like equations Eqs.~\ref{eq:psi},\ref{eq:alpsi} for $\psi$ and $\alpha \psi$ have the form $\Delta u = s u ^q$. In GR ($\mathcal{A} = 1$, $Q^i=0$) they satisfy the criterion for local uniqueness, $sq \geq 0$.  In STTs ($Q^i\neq 0$), this is no longer true; in fact, the source term in Eq.~\ref{eq:alpsi},  $s=2\pi \psi^{-2}\{\mathcal{A}^4[ \varepsilon + 6 p + 3 B^2 /2] - Q^2/8\pi\}$ includes an additional factor $- Q^2/8\pi$ such that it cannot be excluded that in particular conditions, when the scalar field is extremely strong one has $s<0$. However we verified that this does not happen in any of the many configurations we computed, not even the most compact ones. Still, it remains to be verified that this holds also in the case of collapse to BH. 
Concerning instead Eq.~\ref{eq:scal1} at first order in $\chi$, neglecting the higher order second term on the right, it has the form $\Delta u = s f(u) $. It can be shown that the condition for local uniqueness is $s( {\mathrm d} f/{\mathrm d} u) \geq 0$. 
Now $s =-4\pi \psi ^4T_\mathrm{p} > 0$. This implies that if  $\alpha _\mathrm{s}(\chi) \mathcal{A}^4$ is a decreasing function of $\chi$, as it happens to be for STTs with spontaneous scalarisation, Eq.~\ref{eq:scal1} will not satisfy local uniqueness, and multiple solutions are expected. This will be further investigated and discussed in Sect.~\ref{sec:multiple} 

\subsubsection{Purely poloidal configuration}
\label{sec:polo}

We begin by showing how the Grad-Shafranov formalism used in GR \citep{del_zanna_exact_1996,pili_general_2017}, for the case of equilibrium configurations with a purely poloidal magnetic field, can be extended to the case of STTs. The solenoidal condition of the magnetic field allows us to write it as a function of the $\phi$-component of the vector potential, $A_\phi$. In conformally-flat metric
\begin{equation}\label{eq:bpolo}
	B^r = \frac{\partial _\theta A_\phi}{\mathcal{A}^3 \psi^6 r^2 \sin \theta} \quad , \quad
	B^\theta = -\frac{\partial _r A_\phi}{\mathcal{A}^3 \psi ^6 r^2 \sin \theta} \quad , 
\end{equation}
and we recall that all metric terms are in the E-frame. Function $A_\phi$ is also called the magnetic flux function, and its iso-surfaces $A_\phi = \mathrm{const}$, called magnetic surfaces, contain the magnetic poloidal field lines.

The Euler equation describing the static MHD equilibrium is
\begin{equation}\label{eq:euler}
	\partial _i p + \left( \varepsilon+p \right) \partial _i \ln (\mathcal{A} \alpha) = \epsilon _{ijk} J^i B^k /\mathcal{A}^3 =  L_i \quad ,
\end{equation}
where $J^i =  \mathcal{A}^{2}\alpha ^{-1} \epsilon ^{ijk} \partial _j (\mathcal{A} \alpha B_k)$ and $L_i$ is the Lorentz force.

NSs are often assumed to be well described by a barotropic EoS, that is $\varepsilon=\varepsilon(\rho)$ and $p=p(\rho)$. Then, also $h=h(\rho)$ and Eq.~\ref{eq:euler} becomes \citep{pili_axisymmetric_2014} the `generalised Bernoulli integral'
\footnote{In analogy with the non-relativistic case, the relativistic Bernoulli integral can be defined, in hydrodynamics, from the conservation law of $h u_t$ along the trajectories of a stationary flow (see \citealt{friedman_2013}). This is a special case of the global first integral of Euler's equation for iso-entropic flows which, for stationary cases, reduces to Eq.~\ref{eq:bernoulli}. This is the reason why we refer to Eq.~\ref{eq:bernoulli} as the generalised Bernoulli integral.}
\begin{equation}\label{eq:bernoulli}
	\ln \left( \frac{h}{h_{\mathrm c}} \right) + \ln \left( \frac{\mathcal{A}  \alpha}{\mathcal{A}_{\mathrm c}  \alpha _{\mathrm c}} \right) - \mathcal{M}=0 \quad ,
\end{equation}
where the magnetisation function $\mathcal{M}(A_\phi)$ defines the Lorentz force through
\begin{equation}\label{eq:lorenzf}
	L_i = \rho h \frac{d \mathcal{M}}{d A_\phi}\partial _i A_\phi \quad ,
\end{equation}
and $h_{\mathrm c}$, $\alpha _{\mathrm c}$, and $\mathcal{A}_{\mathrm c}$ are the values of $h$, $\alpha$ and $\mathcal{A}$ at the center of the star, respectively (we have assumed $\mathcal{M}_{\mathrm c}=0$).
By working out the derivatives of the poloidal components of the magnetic field, one can find an equation for $J^\phi$:
\begin{equation}\label{eq:jphi2}
	J^\phi = -\frac{1}{\mathcal{A}^4 \psi ^8 r^2 \sin ^2 \theta} \left[ \Delta _* A_\phi + \partial A_\phi \partial \ln \left( \alpha \psi ^{-2} \right)  \right] \quad ,
\end{equation}
where $\Delta _* = \partial _r ^2 + r^{-2} \partial _\theta ^2 - r^{-2} (\tan \theta)^{-1} \partial _\theta$. Given that, from  Eq.~\ref{eq:lorenzf}, $J^\phi = \rho h ({\mathrm d} \mathcal{M}/{\mathrm d} A_\phi)  $, we can obtain the Grad-Shafranov equation
\begin{equation}\label{eq:gradsh}
	\check{\Delta}_3 \check{A}_\phi + \frac{\partial A_\phi \partial \ln \left( \alpha\psi ^{-2} \right)}{r \sin \theta} + \mathcal{A}^4 \psi ^8 r \sin \theta \left( \rho h \frac{d \mathcal{M}}{dA_\phi} \right) = 0,
\end{equation}
where $\check{A}_\phi \coloneqq A_\phi / (r \sin \theta)$ and $\check{\Delta}_3 \check{A}_\phi = \Delta _* A_\phi / (r\sin \theta)$. 
Eq.~\ref{eq:gradsh} allows one to find the magnetic field and current components once the metric ($\alpha$ and $\psi$) is known and the free function $\mathcal{M}$ has been chosen. The simplest choice, found for example in \citet{pili_axisymmetric_2014}, is
\begin{equation}\label{eq:eosmagp}
	\mathcal{M}=k_{\mathrm{pol}}A_\phi \quad ,
\end{equation}
where $k_{\mathrm{pol}}$ is the poloidal magnetisation constant. This leads to dipolar magnetic field configurations and guarantees that the currents are confined within the star.

\subsubsection{Purely toroidal configuration}
\label{sec:toro}

For a purely toroidal magnetic field,  $\mathcal{M}$ in Eq.~\ref{eq:bernoulli} is no longer a function of $A_\phi$ and $L_i=\rho h \partial _i \mathcal{M}$. Deriving the generalised Bernoulli integral and writing the Lorentz force in terms of the magnetic field components, we obtain
\begin{equation}
	\partial _i \ln h + \partial _i \ln (\mathcal{A} \alpha) + \frac{\mathcal{A} \alpha B_\phi \partial _i \left( \mathcal{A} \alpha B_\phi \right)}{\rho h \mathcal{A}^4 \mathcal{R}^2} =0 \quad ,
\end{equation}
where $\mathcal{R}^2 = \alpha ^2 \psi ^4 r^2 \sin ^2 \theta$. This equation becomes integrable if we assume that the last term can be written as the gradient of a scalar function. Defining
\begin{equation}
	\mathcal{G} = \rho h \mathcal{A}^4\mathcal{R}^2 \quad ,
\end{equation}
this becomes possible if
\begin{equation}\label{eq:btoro}
	B_\phi = \frac{\mathcal{I}(\mathcal{G})}{\mathcal{A}\alpha}, \quad {\mathrm{and}}\quad
	\mathcal{M}(\mathcal{G}) = - \int \frac{\mathcal{I}}{\mathcal{G}}\frac{d\mathcal{I}}{d\mathcal{G}}d\mathcal{G} \quad . 
\end{equation} 
It is customary to assume a barotropic expression for $\mathcal{I}$ \citep{kiuchi_relativistic_2008,frieben_equilibrium_2012}:
\begin{equation}\label{eq:eosmagt}
\mathcal{I} = k_\mathrm{tor}\mathcal{G}^\mathrm{m} \quad {\mathrm{and}}\quad\mathcal{M} = -\frac{m k_\mathrm{tor}^2}{2m-1}\mathcal{G}^{2m-1} \quad ,
\end{equation}
where $k_\mathrm{tor}$ is the toroidal magnetisation constant and $m \geq 1$ is the toroidal magnetisation index. This form of $\mathcal{I}$ ensures that the magnetic field is confined within the star and that its configuration is symmetric with respect to the equatorial plane. The generalised Bernoulli integral then becomes
\begin{equation}\label{eq:torobernoulli}
	\ln \left( \frac{h}{h_{\mathrm c}} \right) + \ln \left( \frac{\mathcal{A} \alpha}{\mathcal{A}_{\mathrm c}\alpha _{\mathrm c}} \right) + \frac{m k_\mathrm{tor}^2}{2m-1} \left(\rho h \mathcal{A}^4\mathcal{R}^2\right)^{2m-1}=0 \quad .
\end{equation}

\section{The XNS code}
\label{sec:xns}

The \texttt{XNS} code \citep{bucciantini_general_2011,pili_axisymmetric_2014,pili_general_2015,pili_general_2017}  solves the coupled equations for the metric, scalar field, and MHD structure of a NS under the assumptions of stationarity and  axisymmetry, adopting conformal flatness and maximal slicing. It is based on an iterative scheme, which computes the various quantities separately. It has been applied also to the case of white dwarves \citep{Das_Mukhopadhyay15a} and to non-barotropic NSs \citep{Camelio_Dietrich+19a}.

Given that the equations for the scalar quantities $\psi , \alpha \psi , \chi$ involve the $\Delta$ operator, and that the Grad-Shafranov equation can be reduced to a non-linear vector Poisson equation for $\check A_\phi$,  the solutions for $u(r,\theta)=\psi , \alpha \psi , \chi$ are found as a sum of spherical harmonics $Y_l (\theta)$ with coefficients $A_l (r)$ according to
\begin{equation}
	u(r,\theta) = \sum ^\infty _{l=0} \left[ A_l(r)Y_l(\theta) \right] \quad ,
      \end{equation}
      and similarly for the vector potential,
\begin{equation}
	\check A_\phi(r,\theta) = \sum ^\infty _{l=0} \left[ C_l(r)\partial _\theta Y_l(\theta) \right] \quad .
\end{equation}
This choice leads to a series of radial, second order boundary values ODEs for the coefficients $A_l (r)$ and $C_l(r)$, which are solved using a tridiagonal matrix inversion. The decomposition in terms of spherical harmonics ensures the correct behaviour of the solutions on the symmetry axis, and allows us to enforce the proper boundary conditions  at $r=0$, where $A_l (r)$ and $C_l (r)$  go to zero with parity $(-1)^l$, and at the outer radial boundary, where we assume that $A_l (r)$ and $C_l (r)$ go to zero as $r^{-(l+1)}$.

Given the non-linear nature of the various elliptic equations, these are solved iteratively. If the source terms do not satisfy local-uniqueness, iterative schemes might fail to converge.  This issue is particularly relevant for Eq.~\ref{eq:scal1}  for the scalar field. As we discussed, the very nature of spontaneous scalarisation is tied to the non-uniqueness of the solutions. In the iterative scheme used to solve  Eq.~\ref{eq:scal1} we opted to keep fixed the trace of the energy-momentum tensor in the J-frame, and not in the E-frame. Fixing the trace in the E-frame leads to a source term of the form  $-4\pi \psi ^4 \alpha _{\mathrm s}\bar T_{\mathrm p}$, which can be shown to violate local uniqueness for all values of $\chi$. Fixing it in the J-frame instead leads to a source term of the form  $-4\pi \psi ^4 \alpha _{\mathrm s}\mathcal{A}^4 T_{\mathrm p}$, and it can be shown that local uniqueness is  violated only in a finite range of values for $\chi$. This ensures at least the boundedness of the solution.
\\\\
In the following we briefly describe the flow structure of \texttt{XNS}. The code computes at the beginning  the solution for a spherically symmetric non-rotating and un-magnetised NS in isotropic coordinates, at the desired central density $\rho_{\mathrm c}$, solving the generalisation of the Tolman-Oppenheimer-Volkoff (TOV) equations \citep{Tolman39a,Oppenheimer_Volkoff39a} to STTs (the `S-TOV' system, see App.~\ref{app:stov}). This is achieved with a nested shooting technique requiring that in the final solution the ratio $Q_r/\partial_r\alpha$ is constant outside the NS, and that the conformal factor $\psi$ corresponds to the Just metric \citep{Just59a} in isotropic coordinates. Then, starting with an initial guess, the \texttt{XNS} code performs iteratively the following steps until a converged solution is found:
\begin{enumerate}
\item Given a distribution of the physical and scalar fields,  Eqs.~\ref{eq:psi},\ref{eq:alpsi} for a new space-time metric in the E-frame are solved in sequence;
\item Using the new metric in the E-frame and the old physical fields, scalar field Eq.~\ref{eq:scal1} is solved, allowing one to define a new metric in the J-frame;
 \item   If the magnetic field is purely toroidal, Eq.~\ref{eq:torobernoulli} is solved, and new values of the physical fields, including the magnetic field components through Eq.~\ref{eq:btoro},  are found in the J-frame.  If the magnetic field is purely poloidal, first the equation for the vector potential Eq.~\ref{eq:gradsh} and then Eq.~\ref{eq:bernoulli} are solved, determining the new physical fields in the J-frame. 
 \item Convergence is checked and, if not reached, the new physical metric and scalar fields are used to define a new starting model.
\end{enumerate}

\section{Results}
\label{sec:results}

In this section, we present various equilibrium configurations, analysing how the global quantities that parametrise the resulting models depend on the strength and geometry of the magnetic field. All our models, unless otherwise specified, have been computed on a 2D grid in spherical coordinates extending over the range $r=[0,100]$ {in dimensionless units, corresponding to a range of $\sim$150 km, and $\theta=[0,\pi]$.  The grid has 400 points in the $r$-direction, with the first 200 points equally spaced, and covering the range $r=[0,20]$, and the remaining 200 points logarithmically spaced ($\Delta r_i/\Delta r_{i-1} = \mathrm{const}$), and 200 equally spaced point in the angular direction.
For the reference models shown in Sects.~\ref{subs:toro},\ref{subs:polo}, the radial resolution was doubled.  We have verified that at these resolutions our results have an accuracy of the order of $10^{-3}$, that the radius of the outer edge is far enough not to affect the solution, and the same holds for the choice of a stretched grid. In all cases the elliptic solvers use 20 spherical harmonics.
We found that in order to avoid strongly oscillatory behaviours in the relaxation scheme of \texttt{XNS}, iterations over the various quantities $Q$ had to be under-relaxed according to: $Q_{\mathrm{new} }=[Q_{\mathrm{new} }+Q_{\mathrm{old}}]/2$.

For the ease of comparison, and in line with previous literature in GR \citep{bocquet_rotating_1995,kiuchi_relativistic_2008,frieben_equilibrium_2012,pili_axisymmetric_2014} we adopted a simple polytropic EoS $p=K_{\mathrm a} \rho^{\gamma_{\mathrm a}}$ , with an adiabatic index $\gamma_{\mathrm a}=2$  and a polytropic constant $K_{\mathrm a}=110$ (in dimensionless units).  Concerning the magnetic field structure, for purely toroidal magnetic fields we chose a magnetic barotropic law, Eq.~\ref{eq:eosmagt}, with toroidal magnetisation index $m=1$, while for purely poloidal magnetic fields we opted for the simplest choice Eq.~\ref{eq:eosmagp} (for more complex choices see \citealt{pili_axisymmetric_2014}).

The coupling function $\mathcal{A}(\chi)$ is the only free function of a STT with zero potential. As introduced in \citet{damour_nonperturbative_1993}, and used in many subsequent works \citep{novak_spherical_1998,mendes_highly_2016}, we adopt the choice of an exponential coupling function:
\begin{equation}\label{eq:coupl}
	\mathcal{A}\left( \chi \right) \coloneqq \exp \left[ \alpha _0 \chi + \frac{\beta _0}{2} \chi ^2 \right] \quad ,
\end{equation}
where $\alpha _0$ and $\beta _0$ are parameters whose values are constrained by observations.  It can be shown \citep{ramazanoglu_spontaneous_2016} that, if $\beta _0 \bar T _\mathrm{p} >0$, a tachyonic instability is triggered, and modes with wavelength smaller than the NS radius grow exponentially, leading to spontaneous scalarisation, which only depends on the value of the parameter $\beta _0$. Instead, $\alpha _0$ is constrained by weak-field observations \citep{will_confrontation_2014}, and has no role in this instability. It is customary (and we will follow this choice) to choose STTs with $\beta _0<0$, because for most EoSs of NSs in the literature $\bar T _\mathrm{p} <0$. We note that, in principle, spontaneous scalarisation can happen for positive values of $\beta _0$ if the EoS predicts a strongly interacting behaviour of matter in the NS core, such that $\bar T _\mathrm{p} >0$ \citep{mendes_highly_2016}.  The most stringent constraints to this day require that, for massless scalar fields, $|\alpha _0| \lesssim 3 \times 10^{-3}$ and $\beta _0 \gtrsim -4.5$. However, for a scalar field with mass, screening effects come into play and much more negative values of $\beta _0$ are in principle allowed \citep{yaza_2016}. We chose  $\alpha _0 =-2\times 10^{-4}$ and varied $\beta_0$ in the range $[-6,-4.5]$. In order to enhance and highlight the effect of spontaneous scalarisation, a particular focus will be devoted to the case $\beta_0=-6$.

The global quantities used in the following are defined in App.~\ref{app:glob}. It can be shown that in the E-frame the Komar and ADM masses have the same value, while in the J-frame they differ by an amount proportional to the scalar charge. For this reason, in the following, when referring generically to the mass of the NS, we always mean the Komar mass in the E-frame ($M \coloneqq \bar{M}_\mathrm{k}$). On the other hand, given that the circumferential radius is a potentially measurable quantity, when referring to it we always mean its value in the J-frame. Moreover, since the metric field equations in the E-frame have the same mathematical structure as in GR, it is most natural to provide the quadrupole deformations in the E-frame, as this is where GWs should be studied.

\subsection{Uniqueness of scalarised NSs}
\label{sec:multiple}

\begin{table*}
\caption{Values of various physical quantities describing the solutions $\mathcal{S}_{\mathrm w}, \mathcal{S}_{\mathrm s}^+$ and $ \mathcal{S}_{\mathrm s}^-$, for $\alpha _0 =-0.05$ and $\beta _0=-6$, and for selected values of the central density $\rho _{\mathrm c}$ (in the J-frame), corresponding from top to bottom to: $\rho_\mathrm{b} <\rho_{\mathrm c} <\rho_{\mathrm 1}$, $\rho_{\mathrm 3} <\rho_{\mathrm c} <\rho_\mathrm{t}$, $\rho _{\mathrm c} = \rho_{\mathrm 1}, \rho_{\mathrm 2}, \rho_{\mathrm 3}$.
 $M$ is the Komar mass in the E-frame, $Q_\mathrm{s}$ the scalar charge in the E-frame, $R_{\mathrm c}$ the circumferential radius in the J-frame, $W$ the gravitational binding energy in the E-frame. See App.~\ref{app:glob} for their definition: Eqs.~\ref{eq:mk},\ref{eq:qs},\ref{eq:rc},\ref{eq:w}.}
\label{tab:multseq}
\centering

\begin{tabular}{c c c c c} 
 \hline\hline
 \noalign{\smallskip}
  $\rho _{\mathrm c}[10^{15}$g cm$^{-3}]$ & $M[M_\odot]$ & $Q_{\mathrm s}[M_\odot]$ & $R_{\mathrm c}[$km$]$& $|W|[M_\odot]$ \\ [0.5ex]
{}& $\mathcal{S}_{\mathrm w}; \mathcal{S}_{\mathrm s}^+; \mathcal{S}_{\mathrm s}^-$ & $\mathcal{S}_{\mathrm w}; \mathcal{S}_{\mathrm s}^+; \mathcal{S}_{\mathrm s}^-$ & $\mathcal{S}_{\mathrm w}; \mathcal{S}_{\mathrm s}^+; \mathcal{S}_{\mathrm s}^-$& $\mathcal{S}_{\mathrm w}; \mathcal{S}_{\mathrm s}^+; \mathcal{S}_{\mathrm s}^-$ \\ [0.5ex] 
  \noalign{\smallskip}
 \hline
  \noalign{\smallskip}
1.000 & 1.601; 1.402; 1.307 & -0.154; -0.679; 0.815 & 13.46; 13.68; 13.83 & 0.2709; 0.1304; 0.0779 \\ 
2.500 & 1.696; 1.986; 2.166 & -0.149; -0.894; 1.190 & 10.60; 12.40; 13.68 & 0.4870; 0.3607; 0.3022 \\
1.648 & 1.714; 1.683; 1.683 & -0.113; -0.996; 1.150 & 11.90; 13.33; 13.89 & 0.4021; 0.1562; 0.1044 \\
1.695 & 1.715; 1.708; 1.715 & -0.112; -1.010; 1.170 & 11.81; 13.33; 13.93 & 0.4088; 0.1618; 0.1093 \\
1.710 & 1.716; 1.716; 1.726 & -0.112; -1.010; 1.180 & 11.78; 13.33; 13.93 & 0.4109; 0.1637; 0.1110 \\
 \noalign{\smallskip}
 \hline
\end{tabular}
\end{table*}

It can be shown that, given a central density $\rho _{\mathrm c}$, NSs in STTs admit multiple solutions. If $\mathcal{A}$ is an even function of $\chi$ then $\alpha_{\mathrm s}$ is an odd-function (e.g. if $\alpha _0 = 0$ in Eq.~\ref{eq:coupl}) and Eq.~\ref{eq:scal1} is invariant under the transformation $\chi \rightarrow -\chi$ (the same holds for Eqs.~\ref{eq:psi},\ref{eq:alpsi} and  Eqs.~\ref{eq:bernoulli},\ref{eq:torobernoulli}). This implies that there are three possible NS solutions: one corresponding to $\chi=0$, identical to GR, and two with $\chi\neq0$, that only differ by the sign of $\chi$.  If $\alpha_{\mathrm s}$ is an arbitrary function of $\chi$, this symmetry breaks.  If $\alpha _0 \neq  0$ in Eq.~\ref{eq:coupl}, then these three solutions, split into three branches: the GR solution becomes a `weakly scalarised' solution $\mathcal{S}_{\mathrm w}$, where the total scalar charge $Q_{\mathrm s}$ is such that  $\alpha_{\mathrm 0}Q_{\mathrm s}>0$, while the other two scalarised branches split into two `strongly scalarised' solutions: one,  $\mathcal{S}_{\mathrm s}^+$, with $\alpha_{\mathrm 0}Q_{\mathrm s}>0$, the other, $\mathcal{S}_{\mathrm s}^-$, with $\alpha_{\mathrm 0}Q_{\mathrm s}<0$.

In Fig.~\ref{fig:multseq1}, we illustrate qualitatively how these three branches behave  in terms of their mass $M$ as a function of the central density $\rho_{\mathrm c}$.  The range of spontaneous scalarisation, $\rho_\mathrm{b} <\rho_{\mathrm c} <\rho_\mathrm{t}$, can be divided into 4 subregions depending on the relative values of the masses of the branches:
\begin{itemize}
\item for $\rho_\mathrm{b} <\rho_{\mathrm c} <\rho_{\mathrm 1}$ we have $M[\mathcal{S}_{\mathrm s}^-]<M[\mathcal{S}_{\mathrm s}^+]<M[\mathcal{S}_{\mathrm w}]$;
  \item for $\rho_{\mathrm 1} <\rho_{\mathrm c} <\rho_{\mathrm 2}$ we have $M[\mathcal{S}_{\mathrm s}^+]<M[\mathcal{S}_{\mathrm s}^-]<M[\mathcal{S}_{\mathrm w}]$; 
  \item for $\rho_{\mathrm 2} <\rho_{\mathrm c}, <\rho_{\mathrm 3}$ we have $M[\mathcal{S}_{\mathrm s}^+]<M[\mathcal{S}_{\mathrm w}]<M[\mathcal{S}_{\mathrm s}^-]$; 
  \item for $\rho_{\mathrm 3} <\rho_{\mathrm c}, <\rho_\mathrm{t}$ we have $M[\mathcal{S}_{\mathrm w}]<M[\mathcal{S}_{\mathrm s}^+]<M[\mathcal{S}_{\mathrm s}^-]$.
\end{itemize}
The densities $\rho_{1,2,3}$ correspond to the points two branches have the same mass. Almost  always, the $\mathcal{S}_{\mathrm s}^-$ branch is the one where the mass shows the largest deviation from the GR (or from $\mathcal{S}_{\mathrm w}$) and is also the one with the maximum mass.  In Table~\ref{tab:multseq}, we report the values of global quantities characterising solutions of the three branches, for  few selected values of the central density, assuming $\alpha_0=-0.05$ and $\beta_0=-6$, for spherically symmetric un-magnetised and non-rotating NSs. Such a non-physical high value of $\alpha_0$ was chosen in order to enhance the differences between the $\mathcal{S}_{\mathrm s}^-$  and $\mathcal{S}_{\mathrm s}^+$ branches.  We found that, in terms of the net scalar charge, $Q_{\mathrm s}[\mathcal{S}_{\mathrm w}]<Q_{\mathrm s}[\mathcal{S}_{\mathrm s}^+]<Q_{\mathrm s}[\mathcal{S}_{\mathrm s}^-]$, and similarly in terms of the NS circumferential radius $R_{\mathrm c}[\mathcal{S}_{\mathrm w}]<R_{\mathrm c}[\mathcal{S}_{\mathrm s}^+]<R_{\mathrm c}[\mathcal{S}_{\mathrm s}^-]$. In this sense the $\mathcal{S}_{\mathrm s}^-$ solution is the one with the largest deviation from GR. One can compare the three branches also in terms of their compactness $\mathcal{C} \coloneqq M/R_{\mathrm c}$, or in terms of their gravitational binding energy, defined as the difference between the Komar and proper masses in the E-frame, $W\coloneqq M-M_{\mathrm p}$. We find that $\mathcal{S}_{\mathrm s}^-$ is the one with the smallest compactness and highest gravitational binding energy.

If we interpret spontaneous scalarisation as an effective phase-transition \citep{damour_tensor-scalar_1996}, then the difference in binding energy between the $\mathcal{S}_{\mathrm s}^{\pm}$ and $\mathcal{S}_{\mathrm w}$ branches can be though of as an effective latent heat that the appearance of a scalar field releases into the system, inflating the star and reducing $|W|$. Within this interpretation, it is reasonable to expect that NSs undergoing spontaneous scalarisation should settle in the $\mathcal{S}_{\mathrm s}^-$ branch, which is the one with the lowest $|W|$. Indeed we find that our code always selects the $\mathcal{S}_{\mathrm s}^-$ solution [we note that for $\alpha_0=0$, \texttt{XNS} always selects the GR solution, and that $\alpha_0\neq 0$ is required to get a scalarised one; see \citet{bucciantini_role_2015} for a discussion of this issue with relaxation schemes for elliptic equations]. It remains to be understood, in a dynamical evolving system, which branch is selected and under what physical conditions.  
 \begin{figure}
   	\centering
         \includegraphics[width=\columnwidth]{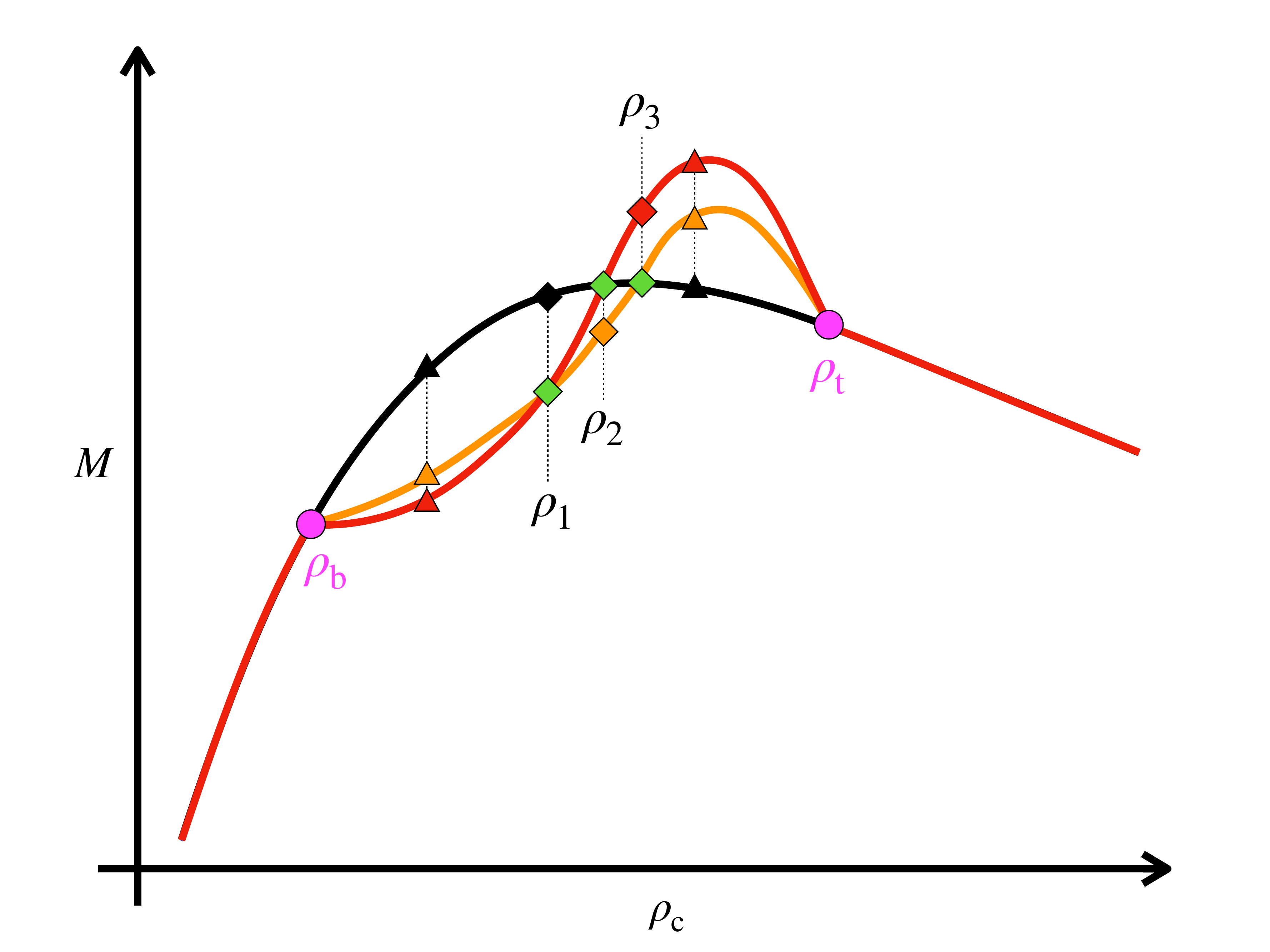}
         \caption{Qualitative behaviour of multiple solutions for NSs in STTs, in terms of the relation of their mass to the central density $\rho _{\mathrm c}$ . The black, orange and red sequences represent, respectively, the weakly scalarised solutions $\mathcal{S}_\mathrm{w}$ and the strongly scalarised solutions $\mathcal{S}_\mathrm{s}^+$ and $\mathcal{S}_\mathrm{s}^-$. Green diamonds mark the position with central densities $\rho_{\mathrm c}=\rho _1,\rho_ 2,\rho _3$ where two branches have the same mass; triangles select intermediate densities (see e.g. the values  in Table~\ref{tab:multseq}); $\rho _\mathrm{b}$ and $\rho _\mathrm{t}$ (magenta circles) represent the lower and upper limits of the central density for which spontaneous scalarisation happens.}
         \label{fig:multseq1}
 \end{figure}
\\\\
In the following, we will refer to strongly scalarised solutions, in the regime where spontaneous scalarisation leads to sizeable scalar charges, simply as `scalarised', while weakly scalarised solutions or in general solutions showing a negligible scalar charge, will be referred to as `de-scalarised' or `GR-like'.

\subsection{Toroidal field models with $\beta _0=-6$}
\label{subs:toro}

\begin{figure*}
  \centering
  \includegraphics[width=0.31\textwidth]{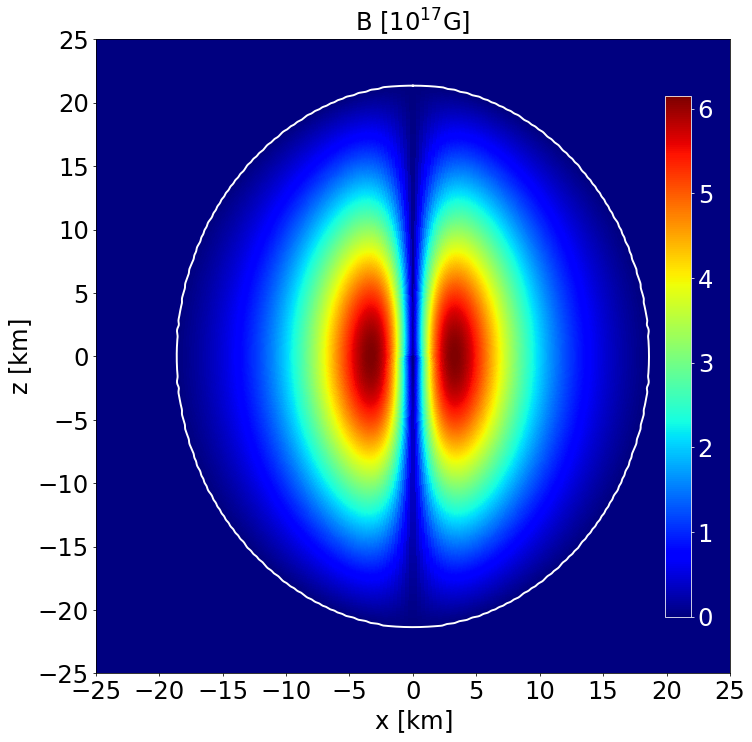}
  \includegraphics[width=0.31\textwidth]{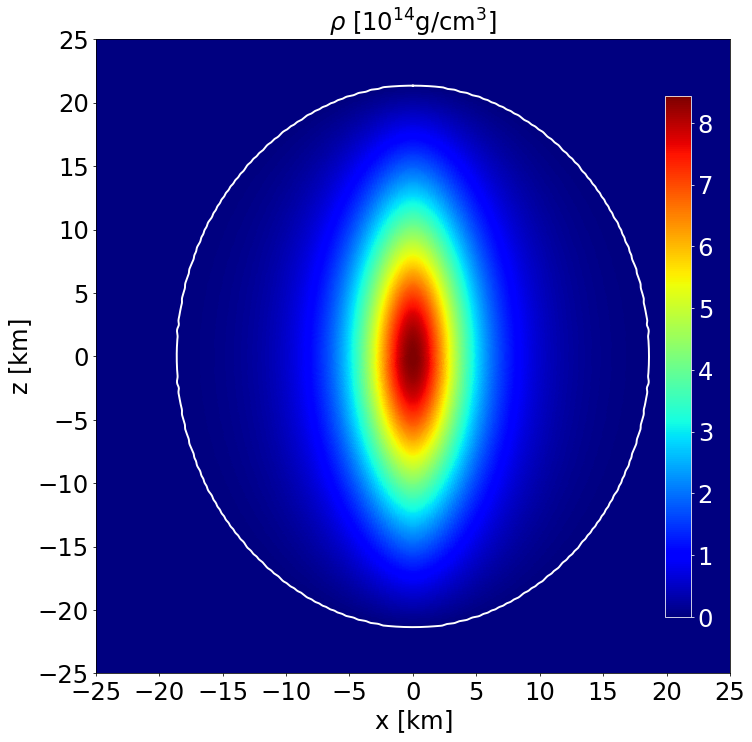}
  \includegraphics[width=0.31\textwidth]{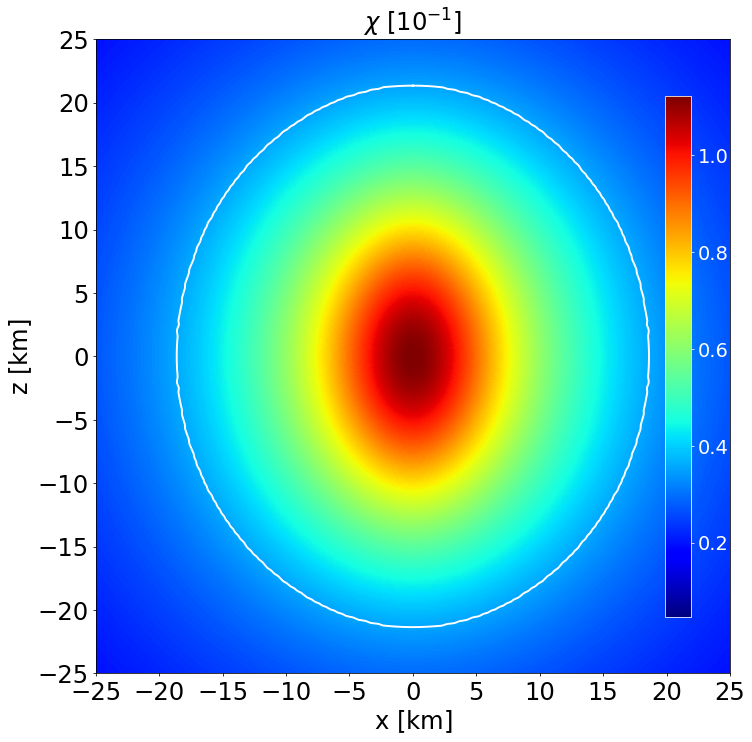}\\
  \caption{From left to right: Meridional distribution of the magnetic field strength $B=\sqrt{B^\phi B_\phi}$, of the density $\rho$ and of the scalar field $\chi$ for a model with a toroidal magnetic field of maximum strength $B_{\textrm{max}}=6.134\times 10^{17}$G and central density $\rho _{\mathrm c} = 8.440 \times 10^{14}$g cm$^{-3}$. The white curve represents the surface of the star. More quantitative details on this configuration can be found in Table~\ref{tab:ref}, where it is named `model T'.}
  \label{fig:toro1}
\end{figure*}
\begin{table*}

\caption{Global quantities (see App.~\ref{app:glob}) of the reference equilibrium models with a toroidal (T)  and poloidal (P) magnetic field, displayed in Figs.~\ref{fig:toro1},\ref{fig:polo1} respectively, together with  their un-magnetised counterparts,  T$_0$ and P$_0$.}
\label{tab:ref}
\centering
\begin{tabular}{c c c c c c c c c c c} 
 \hline\hline
  \noalign{\smallskip}
  Model & $\rho _{\mathrm c}$ & $M_\mathrm k$ & $M_0$ & $Q_{\mathrm s}$ & $R_{\mathrm c}$ & $r_\mathrm p / r_\mathrm e$ & $e$ & $e_\mathrm s$ & $\Phi$ & $\mu$ \\ [0.5ex] 
  {} & $[10^{14}$g cm$^{-3}]$ & $[M_\odot]$ & $[M_\odot]$ & $[M_\odot]$ & $[$km$]$ & {} & $[10^{-1}]$ & $[10^{-1}]$ & $[10^{30}$g cm$^{-2}]$ & $[10^{35}$erg G$^{-1}]$ \\ [0.5ex] 
  \noalign{\smallskip}
 \hline
  \noalign{\smallskip}
T$_0$ & 8.44 & 1.30 & 1.38 & 0.64 & 14.08 & 1.00 & 0.00 & 0.00 & 0.00 & 0.00 \\
T & 8.44 & 1.46 & 1.52 & 0.47 & 20.59 & 1.15 & -8.71 & 1.91 & 1.48 & 0.00 \\
P$_0$ & 5.15 & 1.25 & 1.33 & 0.17 & 15.73 & 1.00 & 0.00 & 0.00 & 0.00 & 0.00 \\
P & 5.15 & 1.36 & 1.42 & 0.56 & 16.71 & 0.67 & 2.90 & -1.52 & 0.00 & 2.20 \\  
 \noalign{\smallskip}
 \hline
\end{tabular}

\end{table*}


To illustrate how a purely toroidal magnetic field affects the properties of scalarised NSs, and to allow a comparison with GR, in Fig.~\ref{fig:toro1} we show the distribution of the magnetic field strength $B = \sqrt{B^\phi B_\phi}$, of the density $\rho$, and of the scalar field $\chi$, for a reference model chosen in order to have the same central density, $\rho _{\mathrm c} = 8.440 \times 10^{14}$g$~$cm$^{-3}$,  and the same maximum value of the magnetic field,  $B_\mathrm{max} =6.134\times 10^{17}$G,  as in \citet{pili_axisymmetric_2014}, for $\alpha _0 =-2\times 10^{-4}$ and $\beta _0 =-6$. Comparing Fig.~\ref{fig:toro1} to the GR solution (\citealt[Fig.~1]{pili_axisymmetric_2014}), we see that the overall distribution of the magnetic field and of the density are very similar, both in their shape and in their values: as expected for a toroidal field, the magnetic field vanishes on the symmetry axis  and reaches a maximum deep inside the star, close to its center. Again, as expected, the star displays a prolate shape in density, caused by the magnetic field stress, and the outer layers are inflated to large radii by the magnetic pressure. We note that this deformation is much more pronounced in the inner parts of the star compared to its outer layers, where the density isosurfaces show only a mild deviation from a spherical shape. On the other hand, we see that the effect of the magnetic stress on the shape of the scalar field is far less evident than on the density, and the scalar field isosurfaces show the same level of prolateness throughout the star.

In Table~\ref{tab:ref}, we give the values of various global quantities characterising this model (T). Its mass $M=1.460$M$_\odot$ is lower than that of its GR counterpart, $1.596$M$_\odot$, by roughly $10\%$. The same holds for the baryonic mass which now is $M_0=1.520$M$_\odot$, lower than in  the GR case where its value is 1.680M$_\odot$. With reference to the regimes shown in Fig.~\ref{fig:multseq1}, our reference model sits between $\rho _\mathrm{b}$ and $\rho _2$, on the $\mathcal{S}^{-}_\mathrm{s}$ sequence. Interestingly, the circumferential radius $R_\mathrm{c}=20.59$km is just $2\%$ higher than in GR. The `radius ratio' between the surface radial coordinate at the pole, $r_p$, and at the equator, $r_e$, is $r_p/r_e=1.15$, not much higher than 1, and only marginally higher than the corresponding GR value. The same holds for the quadrupole deformation $e$ (see App.~\ref{app:glob} for its definition). This might seem counterintuitive, because the scalar field is known to make NSs more spherical \citep{doneva_rapidly_2014}, in part because the contribution of the scalar field to the quadrupole deformation has the opposite sign with respect to the matter, in part because the scalar field pressure tends to counteract matter deformations. We also provide an estimate of the quadrupolar deformation of the scalar field through the quantity $e_\mathrm{s}$, that corresponds to the quadrupolar deformation of the trace of $\bar{T}_\mathrm{p}^{\mu\nu}$ (see App.~\ref{app:glob}). 

It is meaningful to compare our reference model also to an un-magnetised model in STT with the same central density, which is characterised in Table~\ref{tab:ref} as T$_0$. The main differences to note are the lower values of both the Komar and baryonic mass, and of the circumferential radius with respect to the magnetised case. This gives a quantitative estimate of how strong the effects of the magnetic field are and, as in GR, it shows that the magnetic field can provide extra pressure support to sustain a larger total mass.
On the other hand, the compactness is higher: $\mathcal{C}=0.09$ without a magnetic field versus $\mathcal{C}=0.07$ in the magnetised model. This reflects in the fact the the scalar charge $Q_\mathrm{s}$ is higher in the un-magnetised model, by about one third. 

To provide a more accurate comparison of model T with the corresponding GR one, in Fig.~\ref{fig:toro2} we plot for both of them the profiles of $B$ and $\rho$, normalised to their maximum value. In particular, we clearly see that the STT profiles are virtually coincident with the GR ones: only the polar radius gets slightly larger.  This agrees with the fact that apart from integrated quantities, that differ at most $\sim 10\%$, all other quantities characterising those models are very close, suggesting that it is not the dynamical action of the scalar field that gives rise to the differences in mass, but more likely changes in the volume element, associated to small changes in the metric.
In the same figure we also compare model T to the un-magnetised model T$_0$, clearly showing the magnetic induced deformation on the density profile, that affects mostly the low-density outer part of the NS, nearly doubling the star's polar radius.
We also compare the profiles of $\chi$, normalised to its maximum value $\chi _\mathrm{max}$. While in the central part of the star, $r \lesssim 7$km, the equatorial and polar profiles are respectively steeper and shallower than in the un-magnetised case, in the outer part of the star and outside it they are both shallower than in the un-magnetised case, as expected for a lower total scalar charge.
\begin{figure}
  \centering
  \includegraphics[width=0.45\textwidth]{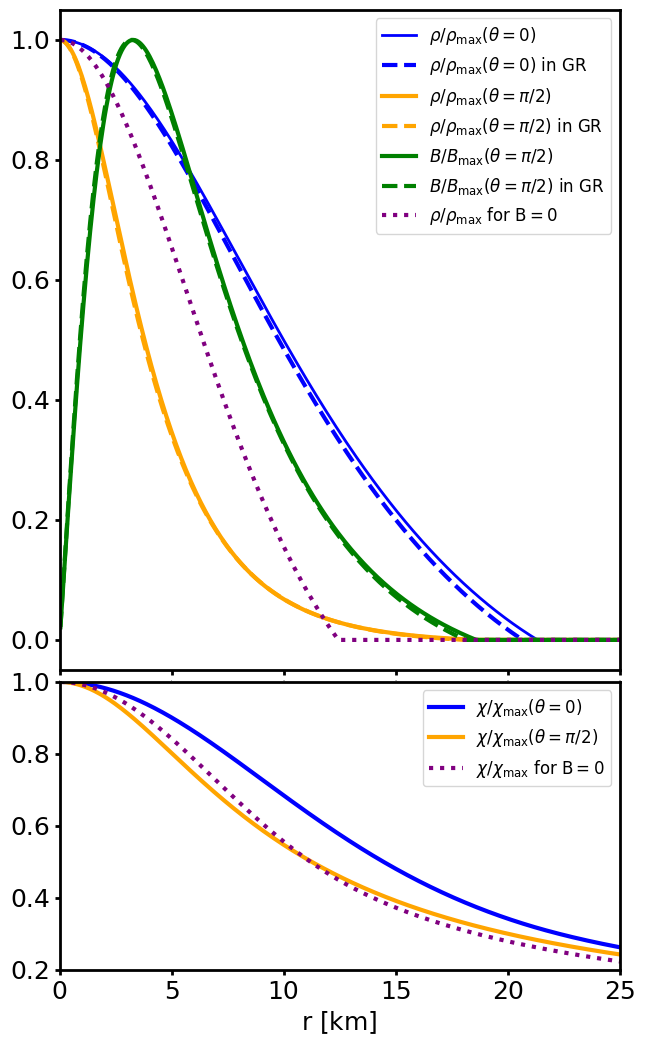}
  \caption{Upper panel: Profile of the polar (solid blue lines) and equatorial (solid orange lines) density, and of the magnetic field strength at the equator (solid green lines), normalised to their maximum values, for the equilibrium model T (with purely toroidal magnetic field) of Table~\ref{tab:ref}. These are to be compared to the corresponding GR model at the same $\rho _{\mathrm c}$ and $B_\mathrm{max}$ (dashed lines), and with the density of the scalarised and un-magnetised model at the same $\rho _{\mathrm c}$, T$_0$ (dotted purple line). Lower panel: Profile of the equatorial (orange line) and polar (blue line) scalar field, normalised to the maximum value, for the equilibrium model T (solid), compared to the un-magnetised model T$_0$ (dotted purple).}
  \label{fig:toro2}
\end{figure}
\begin{figure*}
  \centering
  \subfigure{\includegraphics[width=0.45\textwidth]{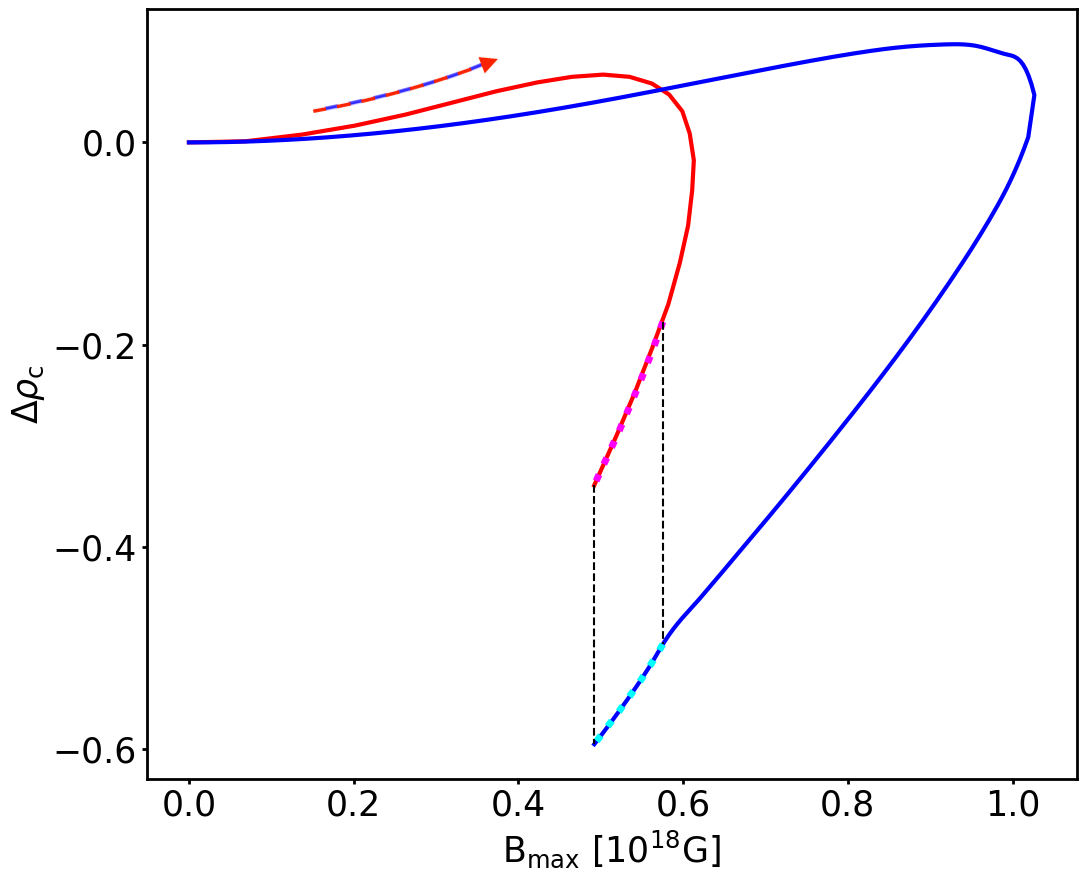}}
  \subfigure{\includegraphics[width=0.44\textwidth]{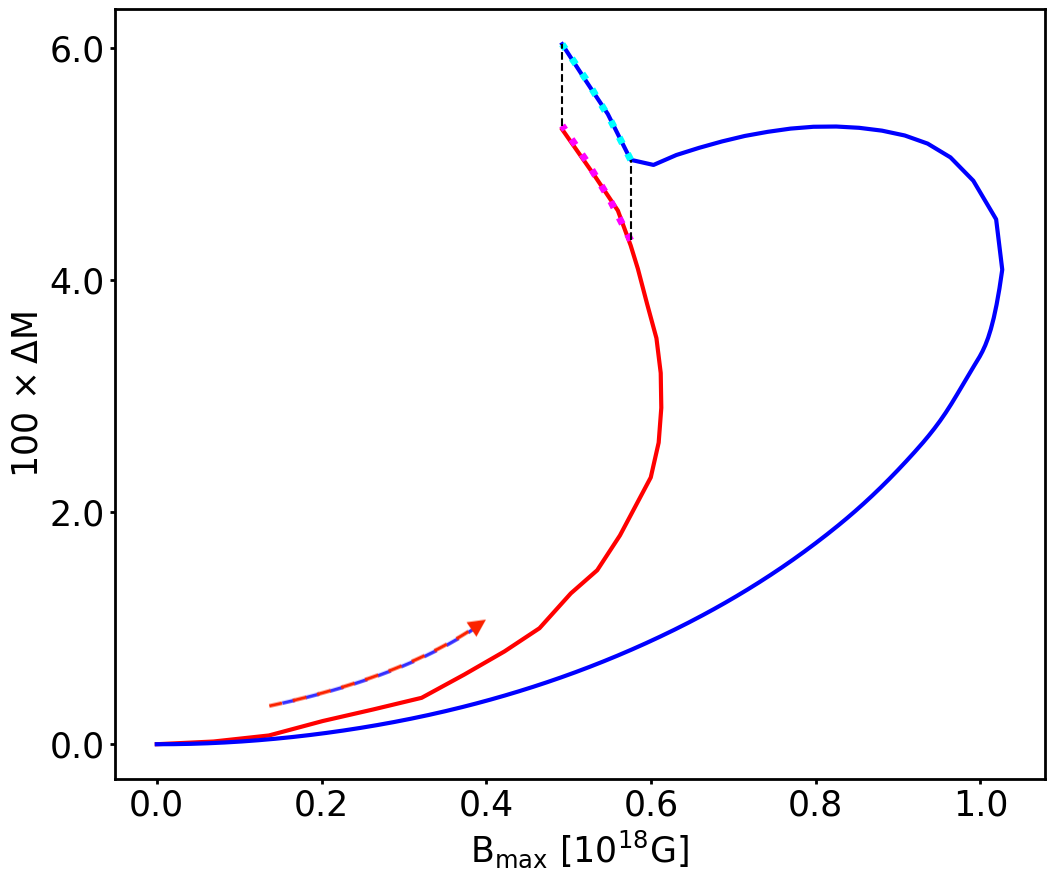}}
  \subfigure{\includegraphics[width=0.45\textwidth]{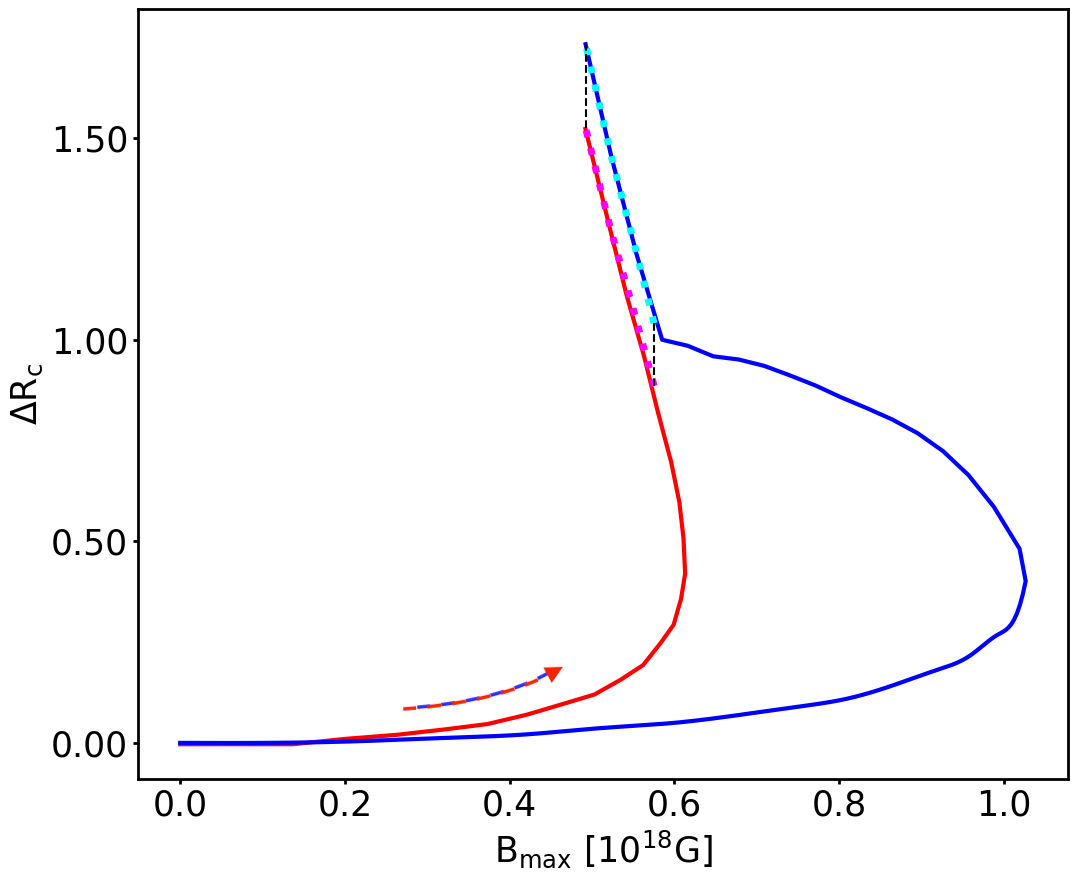}}
  \subfigure{\includegraphics[width=0.44\textwidth]{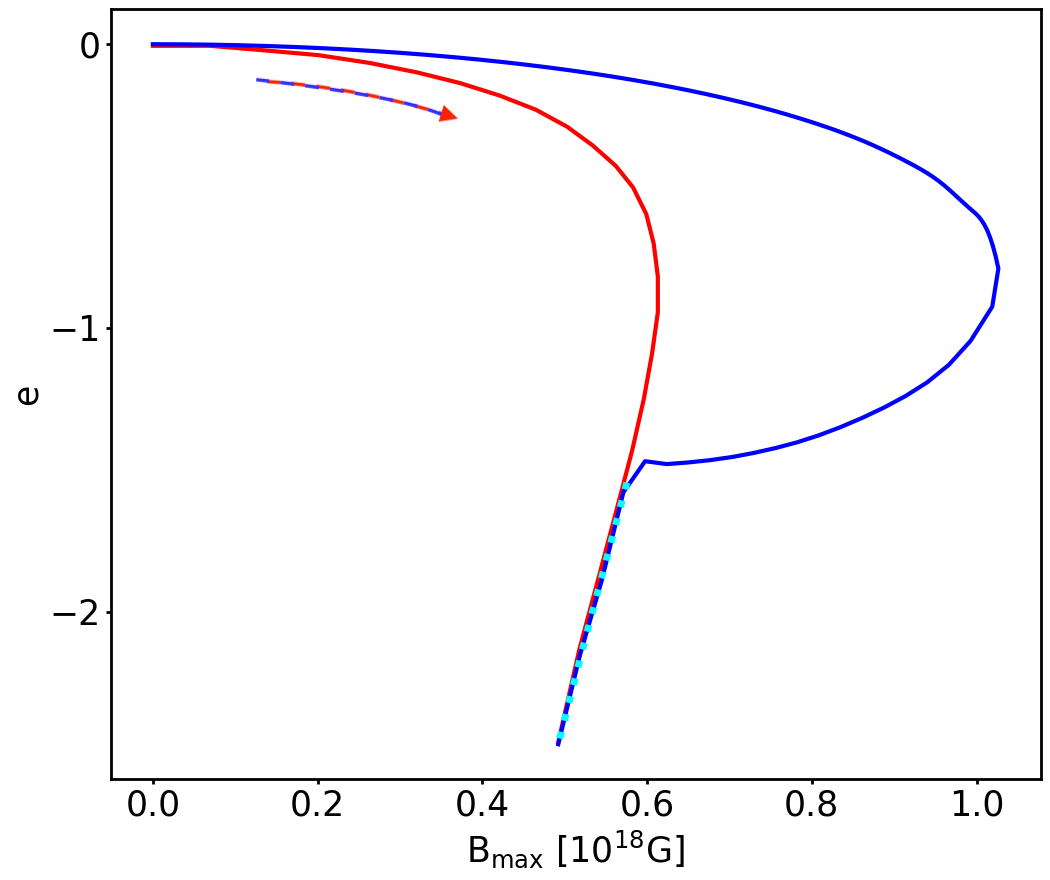}}
  \caption{Variation, with respect to the un-magnetised model, of various quantities along the equilibrium sequence with constant $M_0=1.68$M$_\odot$ for purely toroidal magnetic field. From left to right, top to bottom: Central density $\rho _\mathrm{c}$, Komar mass $M_\mathrm{k}$, circumferential radius $R_\mathrm{c}$ and quadrupole deformation $e$. The blue lines represent our STT results, to be compared to the red lines, describing the GR models in \citet[Fig.~2]{pili_axisymmetric_2014}. The cyan dotted lines highlight the de-scalarised configurations; it is connected by the black dashed segments to the magenta dotted lines, which represent the same STT deviations when calculated with respect to the un-magnetised model in GR. The arrows show the direction of increasing magnetisation.}
  \label{fig:toro3}
\end{figure*}
\\\\
In line with \citet{pili_axisymmetric_2014}, in order to characterize the interplay of the scalar and magnetic field, in Fig.~\ref{fig:toro3},  for equilibrium models having all the same baryonic mass $M_0=1.68$M$_\odot$, we plot the deviations $\Delta$ of $\rho _\mathrm{c}$, $M$, $R_\mathrm{c}$ and $e$ with respect to the un-magnetised case, as functions of the maximum value of the magnetic field strength inside the star $B_\mathrm{max}$. The deviation of a quantity $f$ is defined as
\begin{equation}
	\Delta f \coloneqq \frac{f\left( B_\mathrm{max},M_0 \right) - f\left( 0,M_0\right)}{f\left( 0,M_0\right)} \quad ,
\end{equation}
except for $e$, in which case we just plot its value, since $e(0,M_0)=0$. The results are compared with the GR sequence having the same baryonic mass.

It is immediately evident that the qualitative trends are unchanged. The sequence shows that at a fixed baryonic mass there is a limit to the strength of the magnetic field that a NS can host. We find that in out STT models this value is 1.05$\times 10^{18}$G, almost twice with respect to the one of the equivalant GR sequence, 6.13$\times 10^{17}$G. As the magnetisation parameter $k_{\mathrm m}$ increases, so does at the beginning also $B_\mathrm{max}$, until it reaches its limiting value. A further increase of $k_{\mathrm m}$ leads to a reduction of the magnetic field. The central density first rises with $k_{\mathrm m}$, reaching a value about 10$\%$ larger at $B_\mathrm{max} \simeq 9\times 10^{17}$G and then beginning to decrease. For weak magnetisations, we find that,  for the same $B_\mathrm{max}$, the deviation is about one fourth than in GR. However, once the magnetisation parameter $k_{\mathrm m}$ increases beyond the point where the limiting magnetic field is reached, the deviation of our STT models becomes about a factor two higher than GR. We also find that, as the magnetisation increases even farther, solutions de-scalarise (cyan dotted line), becoming equivalent to GR. 
When looking at  $\Delta M$ or $\Delta R_\mathrm{c}$, one recovers similar trends, with deviations that are smaller than in GR for weak magnetic fields. Interestingly, along the scalarised part of our sequence, there seems to be a maximum value of $\Delta M=0.05$ at $B_\mathrm{max}=8\times 10^{17}$G, a behaviour not present in GR. Similarly, the quadrupolar deformation $e$ is about one fourth than that of GR for weak magnetisations and, again, GR  is recovered at high magnetisations, when the NS de-scalarises. Just focusing on the weakly magnetised part of the sequence, before the limiting magnetic field is reached, we found that the same deviations are usually achieved at twice the value of $B_\mathrm{max}$ with respect to GR. This indicates that NSs in STTs are far less deformable than their GR counterparts of the same baryonic mass. The origin of this behaviour is to be looked for in the effective pressure support provided by the scalar field. A purely toroidal magnetic field exerts a stress on the star that leads to a prolate matter distribution. This, as a consequence, acting  as a source for the scalar field, leads to a prolate distribution of the scalar field itself. Given that the effective pressure of $\chi$ depends on its gradient, a prolate distributions leads, with respect to a spherically symmetric one, to an increased outward-pointing force along the equator and a decreased one along the polar axis (see e.g. the scalar field profiles on a prolate system shown in Fig.~\ref{fig:toro2}). This might seem to contradict what was found before, where we showed only marginal differences between STT and GR. But while previously the comparison was done at the same central density, here is instead done at the same baryonic mass.

In Fig.~\ref{fig:qh_toro}, we show how the magnetic energy $\mathcal{H}$ and the scalar charge $Q_\mathrm{s}$ change with $B_\mathrm{max}$. As the magnetisation parameter $k_{\mathrm m}$ rises, the magnetic energy scales with good approximation as $\mathcal{H}=1.25\times 10^{39}(B_\mathrm{max}/10^{18}\mathrm{G})^2$erg up to $B_\mathrm{max} \simeq 10^{18}$G. As the magnetisation rises beyond the  point where $B_\mathrm{max}=1.03 \times 10^{18}$G, the magnetic field energy, in the scalarised part, reaches a maximum of $\mathcal{H}=1.65\times 10^{39}$erg at $B_\mathrm{max}=9.7\times 10^{17}$G, finally relaxing to the GR profile when the sequence de-scalarises around $B_\mathrm{max}=3\times 10^{17}$G. The scalar charge, instead, drops with increasing magnetisation, being about 10$\%$ smaller at $B_\mathrm{max}=5.8 \times 10^{17}$G. Beyond this point, the scalar charge drops substantially until the NS completely de-scalarises.

In Fig.~\ref{fig:const-toro_seq}, we show how the Komar mass changes with central density holding fixed the magnetic flux $\Phi$ (top panel) or the baryonic mass $M_0$ (middle panel). The lower bound for scalarised models, $\rho _{\mathrm{b}}$, moves to higher densities from $\rho_\mathrm{b}=5\times 10^{14}$g cm$^{-3}$ for $\Phi =0$ to $\rho_\mathrm{b}=7.5\times 10^{14}$g cm$^{-3}$ for $\Phi =2.55\times 10^{30}$G cm$^2$, while the corresponding Komar (baryonic) mass changes from 1.25M$_\odot$ (1.33M$_\odot$) to 1.75M$_\odot$ (1.81M$_\odot$). We find no evidence suggesting the existence of an upper bound to the mass of the possible de-scalarised models. Analogously, the upper bound $\rho_\mathrm{t}$ for scalarised models increases from $\rho_\mathrm{t}=3.5\times 10^{15}$g cm$^{-3}$ for $\Phi =0$ to $\rho_\mathrm{t}=4\times 10^{15}$g cm$^{-3}$ for $\Phi =1.46\times 10^{30}$G cm$^2$, while the corresponding Komar (baryonic) mass changes from 1.60M$_\odot$ (1.73M$_\odot$) to 1.62M$_\odot$ (1.71M$_\odot$). 
Contrary to GR, where it is found that the maximum mass of sequences at fixed $\Phi$ increases with the magnetic flux while the central density of the related models first rises and then drops \citep[Fig.~4]{pili_axisymmetric_2014}, in our STT sequences we found that the behaviour is more complex. At densities just above $\rho_\mathrm{b}$, the mass of magnetised models is found to be always larger than the un-magnetised one. However, as the density increases, the trend is reversed and we find magnetised models having a lower mass than the un-magnetised configuration at the same central density. This is reversed again once the density exceeds 2.72$\times 10^{15}$g cm$^{-3}$ as a consequence of the shift of the position of the maximum mass. This trend is also evident by looking at configurations at fixed baryonic mass and when sequences are parametrised at fixed values of $B_\mathrm{max}$ or at fixed $e$, in Fig.~\ref{fig:const_seq-t}. It is interesting to notice that close to $\rho _{\mathrm c} \simeq 2.72\times 10^{15}$g cm$^{-3}$ the Komar mass is independent of the magnetisation. Quantitatively, the density at which the maximum is reached always increases from $\rho _{\mathrm c}=2.55\times 10^{15}$g cm$^{-3}$ for $\Phi =0$ to $\rho _{\mathrm c}=2.95\times 10^{15}$g cm$^{-3}$ for $\Phi =2.55 \times 10^{30}$G cm$^2$, while the value of the maximum mass drops initially from 2.08M$_\odot$ to 2.04M$_\odot$ for $\Phi=1.46\times 10^{30}$G cm$^2$ and then rises again to 2.08M$_\odot$ for $\Phi =2.55 \times 10^{30}$G cm$^2$. The full characterisation of the models at maximum mass is given in Table~\ref{tab:const_seq}.

In a similar way, in Fig.~\ref{fig:const_seq-t}, we have also analysed how the scalar charge $Q_\mathrm{s}$ changes with magnetisation. The maximum of the scalar charge goes from $Q_s=1.16$M$_\odot$ at  $\Phi = 0$,to $Q_s=1.14$M$_\odot$ when $\Phi = 2.55 \times 10^{30}$G cm$^2$, while the density at which this maximum is reached increases from $2.09\times 10^{15}$g cm$^{-3}$ to $2.46\times 10^{15}$g cm$^{-3}$. Globally, this appears as a shift to higher density of the sequences. The maximum of the scalar charge is always reached before the maximum of the mass. Analogously to the mass, we find that close to $\rho _{\mathrm c} \simeq 2.33\times 10^{15}$g cm$^{-3}$ the scalar charge is independent of the magnetisation.
\begin{figure*}
  \centering
  \includegraphics[width=0.45\textwidth]{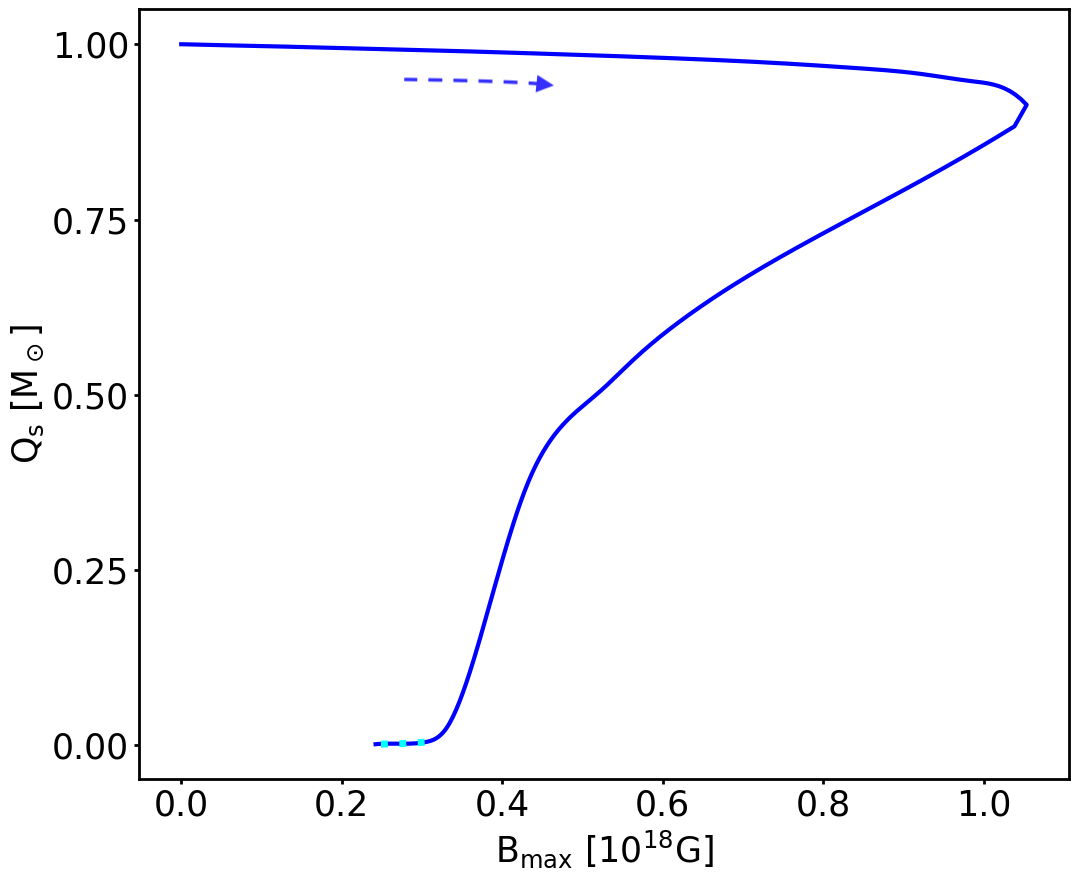}
  \includegraphics[width=0.45\textwidth]{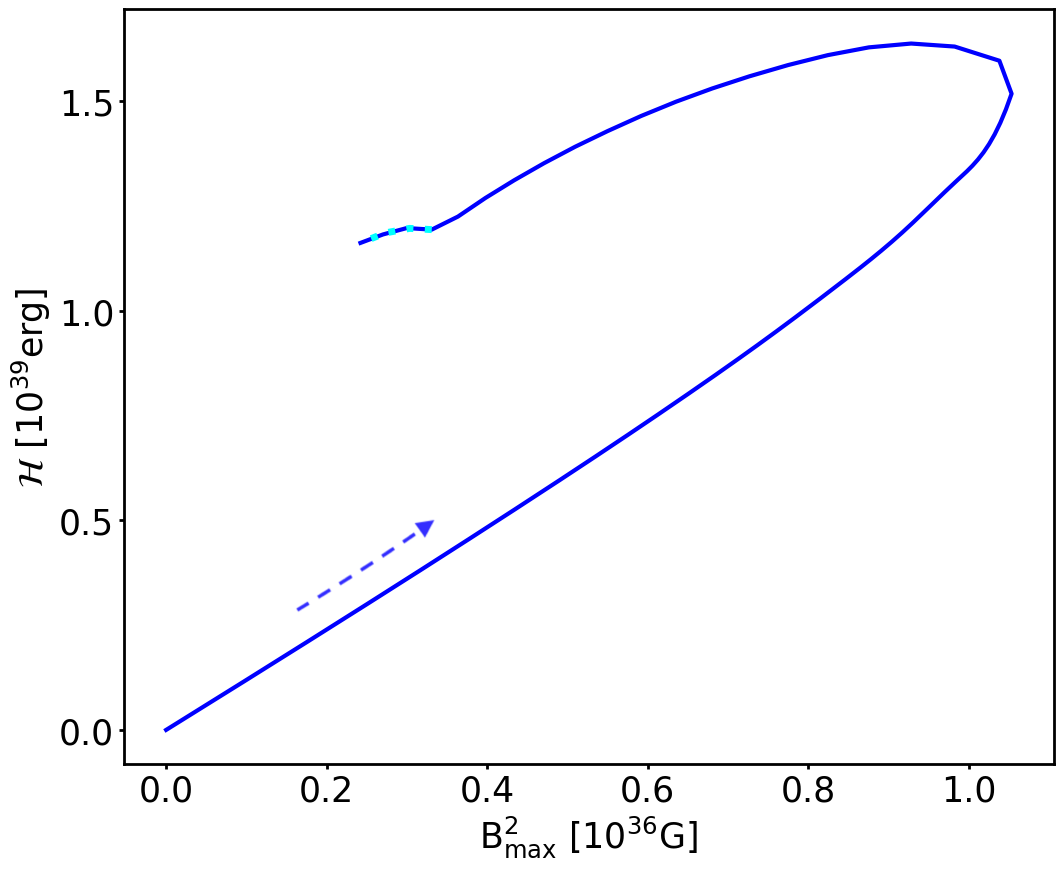}
  \caption{Scalar charge $Q_\mathrm{s}$, normalised to its value for the un-magnetised model (left panel), and magnetic field energy $\mathcal{H}$ (right panel) as functions of $B_\mathrm{max}$ along the equilibrium sequence with constant $M_0=1.68$M$_\odot$ and purely toroidal magnetic field. The cyan dotted line highlights the de-scalarised configurations. The arrows show the direction of increasing magnetisation.}
  \label{fig:qh_toro}
\end{figure*}
\begin{figure}
  \centering
  \includegraphics[width=0.5\textwidth]{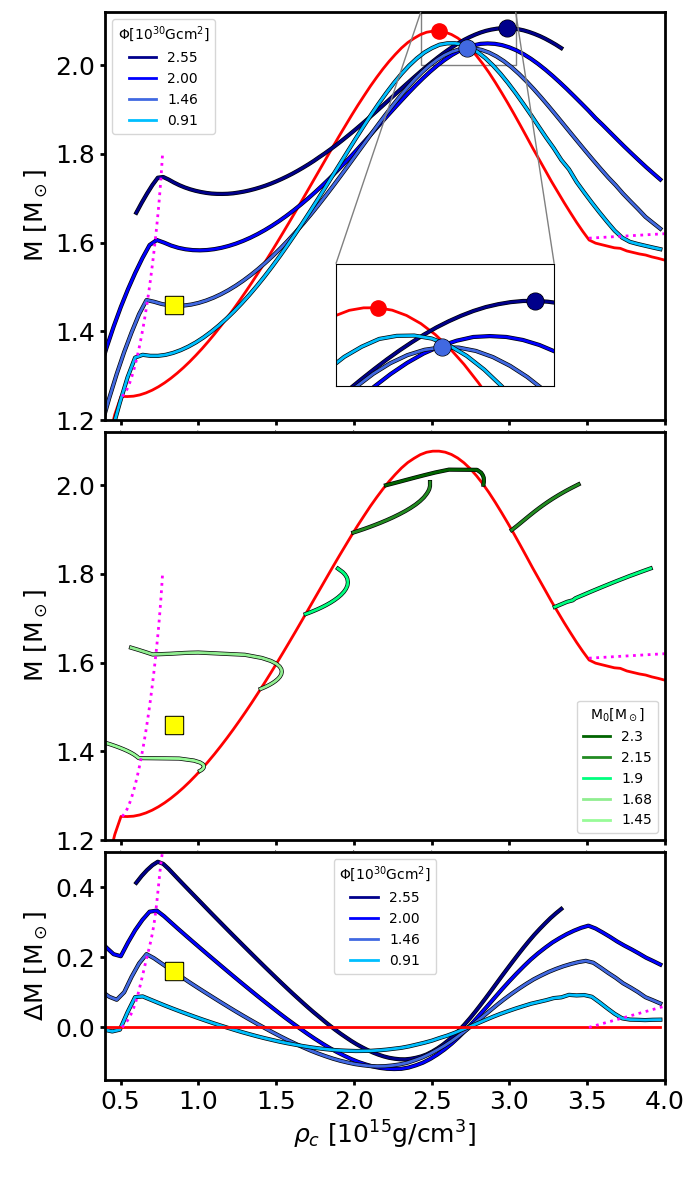}
  \caption{Mass-density sequences for models with purely toroidal magnetic field and $\beta _0=-6$. Upper panel: Sequences computed at fixed values of the magnetic flux $\Phi$ (blue lines), compared with the un-magnetised case (red line). The dotted magenta lines represent the limit for spontaneous scalarisation. Dots mark the position of the maximum mass models UM$_0$ (red), TM$_1$ (light blue) and TM$_2$ (dark blue) of Table~\ref{tab:const_seq}. The yellow square represents model T of Fig.~\ref{fig:toro1}. Middle panel: Sequences computed at fixed baryonic mass (green lines). Lower panel: Mass difference of sequences at fixed $\Phi$ with respect to the un-magnetised one.}
  \label{fig:const-toro_seq}
\end{figure}
\begin{table*}
\caption{Global quantities (see App.~\ref{app:glob}) of the maximum mass models with a purely toroidal (TM$_1$,TM$_2$)  and purely poloidal (PM$_1$,PM$_2$) magnetic field, displayed in Figs.~\ref{fig:const-toro_seq},\ref{fig:const-polo_seq} respectively, together with  their un-magnetised counterpart (UM$_0$).}
\label{tab:const_seq}
\centering
\begin{tabular}{c c c c c c c c c c c} 
 \hline\hline
 \noalign{\smallskip}
  Model & $\rho _{\mathrm c}$ & $M_\mathrm k$ & $M_0$ & $Q_{\mathrm s}$ & $R_{\mathrm c}$ & $B_\mathrm{max}$ & $\Phi$ & $\mu$ & $e$ & $e_\mathrm s$ \\ [0.5ex] 
  {} & $[10^{15}$g cm$^{-3}]$ & $[M_\odot]$ & $[M_\odot]$ & $[M_\odot]$ & $[$km$]$ & $[10^{18}]$G & $[10^{30}$g cm$^{-2}]$ & $[10^{35}$erg G$^{-1}]$ & $[10^{-1}]$ & $[10^{-1}]$ \\ [0.5ex] 
 \noalign{\smallskip}
 \hline
 \noalign{\smallskip}
UM$_0$ & 2.55 & 2.08 & 2.41 & 1.01 & 12.1 & 0.0 & 0.0 & 0.0 & 0.0 & 0.0 \\
TM$_1$ & 2.72 & 2.04 & 2.29 & 1.01 & 13.2 & 1.37 & 1.46 & 0.0 & -0.236 & 0.107 \\
TM$_2$ & 2.95 & 2.08 & 2.26 & 1.04 & 15.8 & 1.99 & 2.55 & 0.0 & -0.656 & 0.200 \\
PM$_1$ & 2.46 & 2.12 & 2.45 & 1.04 & 12.3 & 1.33 & 1.06 & 1.16 & 0.074 & -0.048 \\
PM$_2$ & 2.42 & 2.15 & 2.49 & 1.04 & 12.5 & 1.76 & 1.40 & 1.57 & 0.118 & -0.078 \\[1ex] 
\noalign{\smallskip}
 \hline
\end{tabular}
\end{table*}
\begin{figure*}
  \centering
  \includegraphics[width=0.31\textwidth]{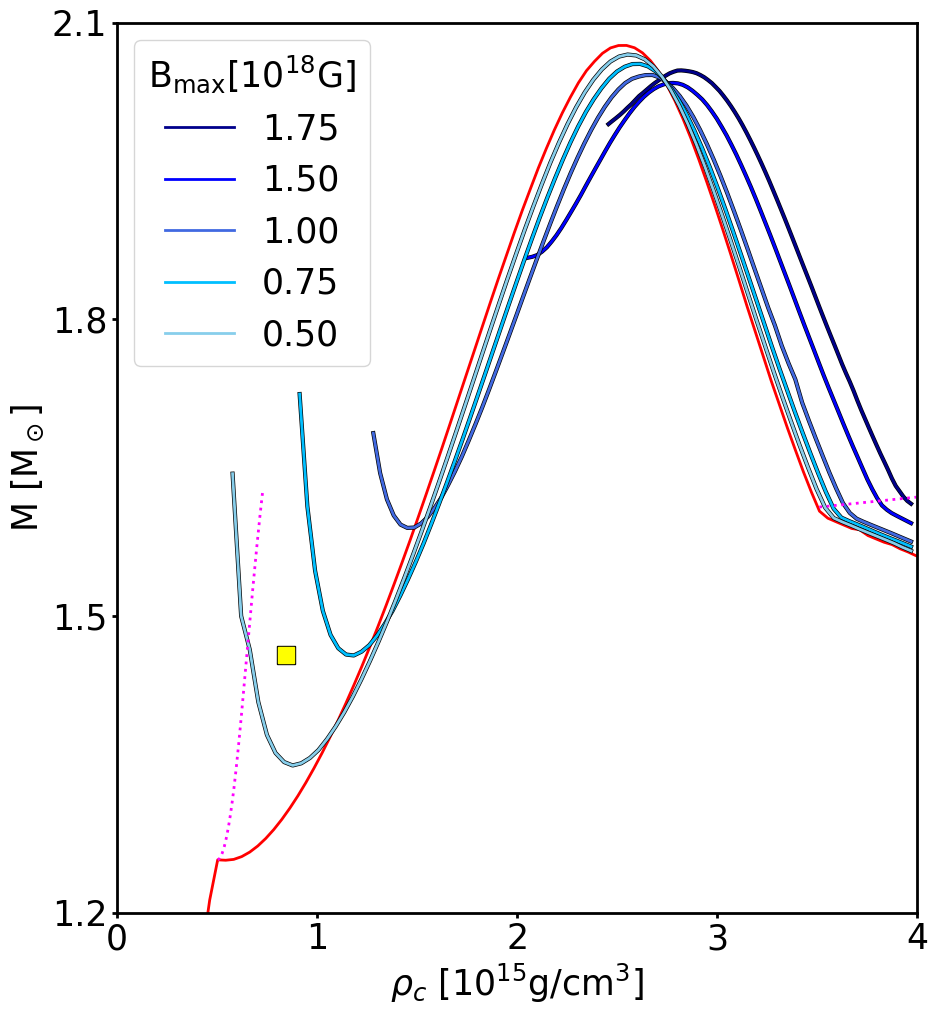}
  \includegraphics[width=0.31\textwidth]{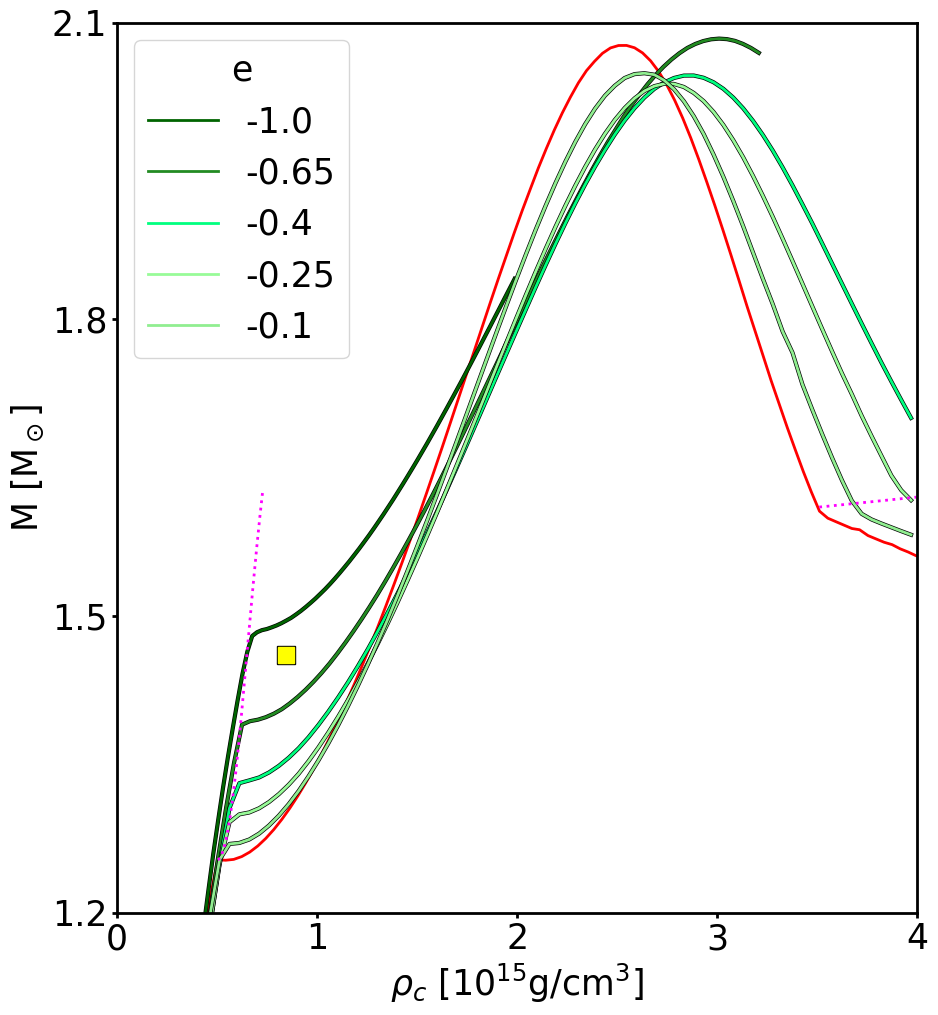}
  \includegraphics[width=0.31\textwidth]{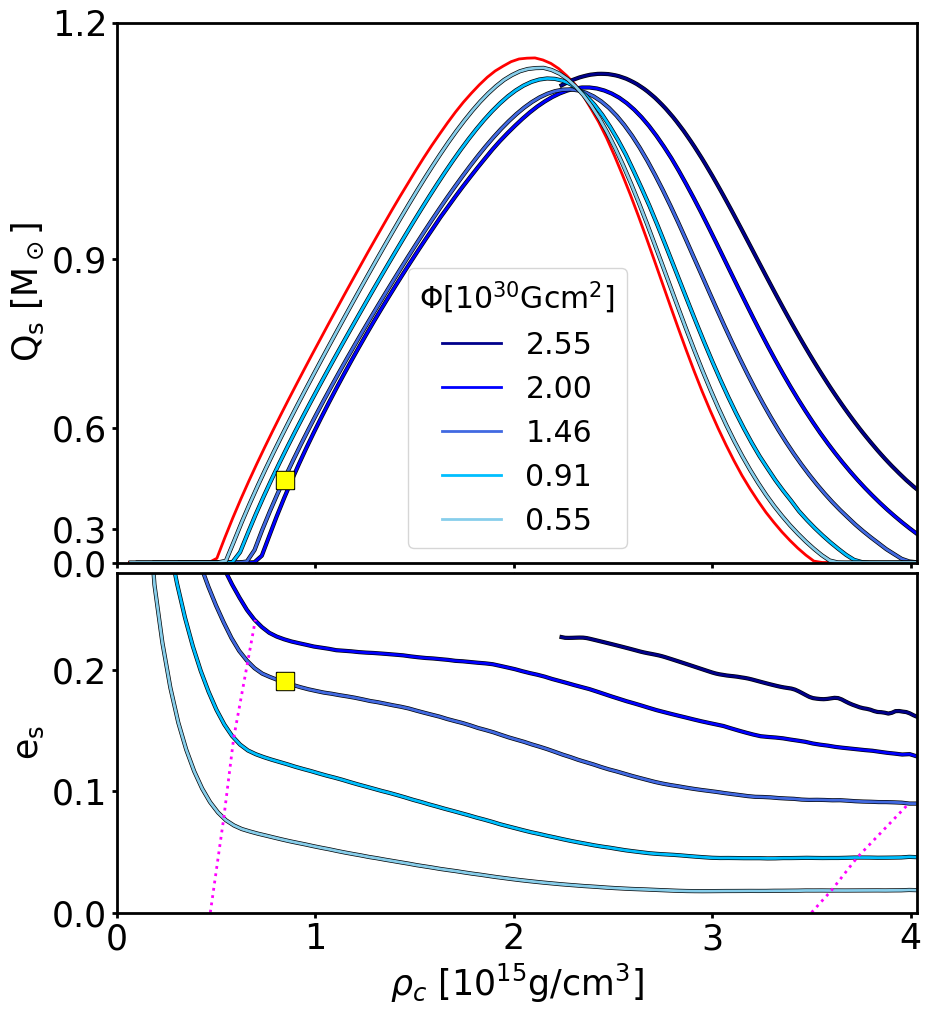}
  \caption{Sequences for the models with purely toroidal magnetic field and $\beta _0=-6$. Left panel: Mass-density relation computed at fixed $B_\mathrm{max}$ (blue lines) compared with the un-magnetised sequence (red line). Middle panel: Mass-density relation computed at fixed $e$ (green lines) compared with the un-magnetised sequence (red line). Right panel: On top, scalar charge computed at fixed $\Phi$ (blue lines) compared with the un-magnetised sequence (red line); on bottom, trace quadrupole deformation $e_\mathrm{s}$. In all panels, the dotted magenta lines represent the limit for spontaneous scalarisation and the yellow square represents model T of Fig.~\ref{fig:toro1}.}
  \label{fig:const_seq-t}
\end{figure*}
 
\subsection{Poloidal field models with $\beta _0=-6$}\label{subs:polo}

As it was done in the toroidal case, also for purely poloidal magnetic fields, our reference model was chosen in order to have the same central density $\rho _{\mathrm c}=5.15\times 10^{14}$g cm$^{-3}$ and the same maximum value of the magnetic field $B_\mathrm{max} = 6.256\times 10^{17}$G, as in \citet{pili_axisymmetric_2014}. Analogously to the previous toroidal case, this model sits in the part of Fig.~\ref{fig:multseq1} between $\rho _\mathrm{b}$ and $\rho _2$, on the sequence $\mathcal{S}^{-}_\mathrm{s}$. In Fig.~\ref{fig:polo1}, we show the distribution of the magnetic field strength $B = \sqrt{B^r B_r + B^\theta B_\theta}$, of the density $\rho$ and of the scalar field $\chi$ for this model. Comparing them to the GR ones in \citet[Fig.~5]{pili_axisymmetric_2014}, we see that, even for a purely poloidal magnetic field, the overall distributions of the various quantities are very similar to GR, both in their shape and in their values. As expected for a poloidal field, the magnetic field reaches a maximum at the center of the star, and vanishes in an equatorial ring located at $r \simeq 12$km. The star displays an oblate shape in density, caused by the magnetic field stress, with an equatorial density profile which is almost flat close to the center. As in GR, increasing farther the magnetic field strength produces configurations where the density maximum is no longer at the center (analogously to \citealt[Fig.~6]{pili_axisymmetric_2014}).  Again, we see that the effect of the magnetic stress on the shape of the scalar field is far less pronouced than on the density.

In Table~\ref{tab:ref}, we give the values of various global quantities characterizing this model (P).  The Komar mass $M=1.360$M$_\odot$, is lower than the GR mass,  $1.597$M$_\odot$ by roughly $15\%$, and the same holds for the baryonic mass which is $M_0=1.42$M$_\odot$, compared to the value of the GR counterpart,  1.680M$_\odot$. The radius ratio $r_p/r_e=0.67$ is instead marginally smaller than the GR value of $0.69$. On the other hand, its circumferential radius $R_\mathrm{c}=16.71$km is less than $1\%$ smaller than the GR one. The quadrupole deformation $e$ is the same as in GR. As before, it seems that the presence of a scalar field, at the same central density and for the same maximum magnetic field, does not affect the distribution of fluid quantities. Moreover, we provide an estimate of the quadrupolar deformation of the scalar field through the quantity $e_\mathrm{s}$, which is comparable in strength to the quadrupole deformation $e$.

We can also make a comparison to the un-magnetised model with the same central density, characterised in Table~\ref{tab:ref} under the name P$_0$. The main differences are the values of the masses and of the circumferential radius, that are smaller for $B=0$. Also the compactness is slightly lower: $\mathcal{C}=0.0795$ without a magnetic field versus $\mathcal{C}=0.0814$ in the magnetised model. Differently than in the toroidal case, the scalar charge $Q_\mathrm{s}$ is much higher in the magnetised model.

In Fig.~\ref{fig:polo2}, we show the profiles of the magnetic field $B$ and and density$\rho$, normalised to their maximum value, for the model P (solid lines) and for the corresponding GR model (dashed lines) with the same $B_{\textrm{max}}$ and $\rho _{\mathrm c}$ together with the un-magnetised model P$_0$. We also plot the profiles of $\chi$, normalised to its maximum value $\chi _\mathrm{max}$, for the models P and P$_0$. Again, the STT profiles are almost coincident with the GR ones: only the equatorial radius gets marginally increased. This is slightly different than the effect of the magnetic field, which changes the density profile and decreases the star's polar radius and increases the equatorial one. The profile of the scalar field reflects the oblateness of the matter distribution, showing deviations that are somewhat smaller than the toroidal case. The same conclusions drawn in the toroidal case apply here too.

In Fig.~\ref{fig:polo3}, we show the deviations $\Delta$ as it was done in Fig.~\ref{fig:toro3}. The qualitative trends are the same as in GR, and do not show the complexity of the toroidal case. In GR there was some evidence indicating that the maximum magnetic field for a NS of $1.68$M$_\odot$ could not exceed $\approx 6.2\times 10^{17}$G. In STT we found instead that up to values or order of $1\times 10^{18}$G there is no evidence of a saturation or limit of the maximum value of the magnetic field, which does not rule out the possibility that it might exist above $10^{18}$G.
The behaviour of all quantities appears to be monotonic in $B_\mathrm{max}$: the central density decreases, while the mass, the circumferential radius and the quadrupole deformation rise. As in the toroidal case, for a given value of $B_\mathrm{max}$ the deviation appears to be about one fourth than in GR, while the same deviation is reached for values of $B_\mathrm{max}$ about twice higher than in GR. There is no evidence that the sequence would de-scalarise. As in the poloidal case, this trend can again be understood based on the effective pressure support provided by the scalar field. A purely poloidal magnetic field exerts a stress on the star that leads to an oblate matter distribution. This leads to an oblate distribution of the scalar field itself which, in turn, increases the outward-pointing force along the pole and decreases the one along the equator with respect to a spherically symmetric model.
We found that, up to $B_\mathrm{max}\approx 10^{18}$G, the total magnetic field energy $\mathcal{H}$ scales with a good approximation as $\mathcal{H}=0.55\times 10^{39}(B_\mathrm{max}/10^{18}\mathrm{G})^2$erg, and the scalar charge increases by about $2\%$ with respect to the un-magnetised case. We also found that the magnetic dipole scales as $\mu = 1.5 \times 10^{35}(B_\mathrm{max}/10^{18}\mathrm{G})$erg G$^{-1}$, about $30\%$ less than in GR. Given that the dipole moment is ultimately a measure of the net toroidal current, this can be considered a kind of global measure of a quantity integrated throughout the NS; as such, even in this case strongly affected by variations in the value of the volume element, related to the metric itself.

In Fig.~\ref{fig:const-polo_seq}, we show how the Komar mass changes with central density holding fixed the magnetic dipole moment $\mu$ or the baryonic mass $M_0$ (top panel). The lower bound $\rho_\mathrm{b}$ for scalarised models now moves to lower densities - from $\rho _{\mathrm b}=5\times 10^{14}$g cm$^{-3}$ for $\mu =0$ to $\rho _{\mathrm b}=4.3\times 10^{14}$g cm$^{-3}$ for $\mu =1.57\times 10^{35}$erg/G - while the corresponding Komar (baryonic) mass rises, going to $1.31$M$_\odot$ (1.38M$_\odot$). Contrary to the toroidal case, we see from Fig.~\ref{fig:const_seq-p} (left panel) that, for purely poloidal magnetic fields, above a Komar mass of 1.34M$_\odot$ there are no de-scalarised models.
Analogously, the upper bound $\rho_\mathrm{t}$ for scalarised models decreases - from $\rho _{\mathrm t}=3.5\times 10^{15}$g cm$^{-3}$ for $\mu =0$ to $\rho _{\mathrm t}=3.42\times 10^{15}$g cm$^{-3}$ for $\mu =0.45\times 10^{35}$erg/G - while the Komar mass remains almost unchanged.
As in GR, it is found that the maximum mass of sequences at fixed $\mu$ increases with the magnetic dipole moment, and the central density at which the maximum is reached drops. The characterisation of the models at maximum mass is given in Table~\ref{tab:const_seq}.
Similarly to GR we found that, at a given central density, the mass of equilibrium configurations is always above the un-magnetised case (in the stable part of the sequence). 
This same trend is also evident when sequences are parametrised at fixed values of $B_\mathrm{max}$ or at fixed $e$, in Fig.~\ref{fig:const_seq-p}. Again, close to $\rho _{\mathrm c} \simeq 2.72\times 10^{15}$g cm$^{-3}$ the Komar mass is independent on the magnetisation.
We have also analysed in Fig.~\ref{fig:const_seq-p} how the scalar charge changes with magnetisation. The maximum of the scalar charge changes from $Q_s=1.16$M$_\odot$ to $Q_s=1.21$M$_\odot$ when $\mu = 0.54 \times 10^{35}$erg/G, while the density at which the maximum is reached drops to $1.96\times 10^{15}$g cm$^{-3}$. Globally, this appears as a shift to lower density of the sequences. Analogously to the mass, we find that close to $\rho _{\mathrm c} \simeq 2.33\times 10^{15}$g cm$^{-3}$ the scalar charge is independent on the magnetisation.
\begin{figure*}
   	\centering
         \includegraphics[width=0.31\textwidth]{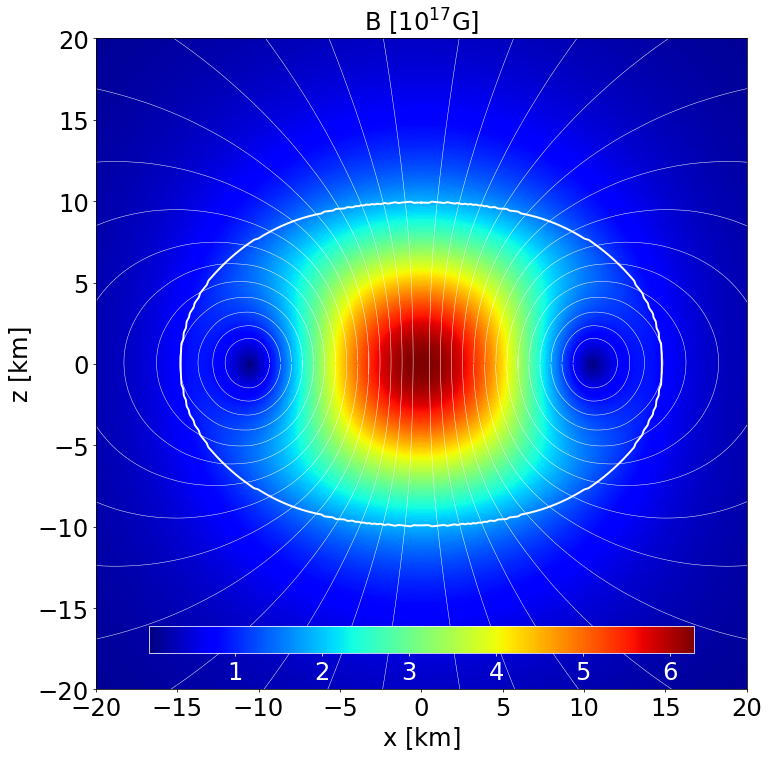}
         \includegraphics[width=0.31\textwidth]{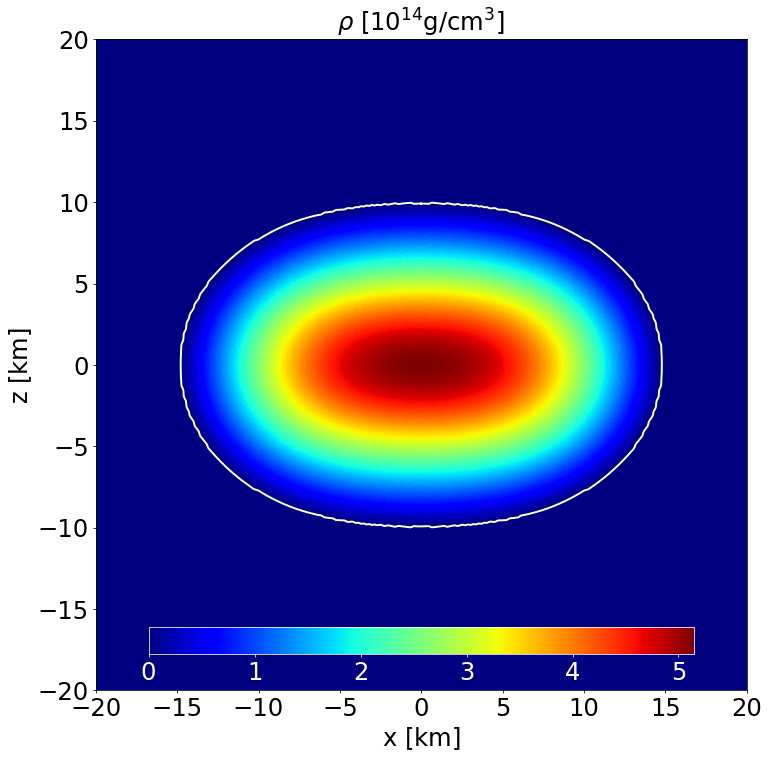}
         \includegraphics[width=0.31\textwidth]{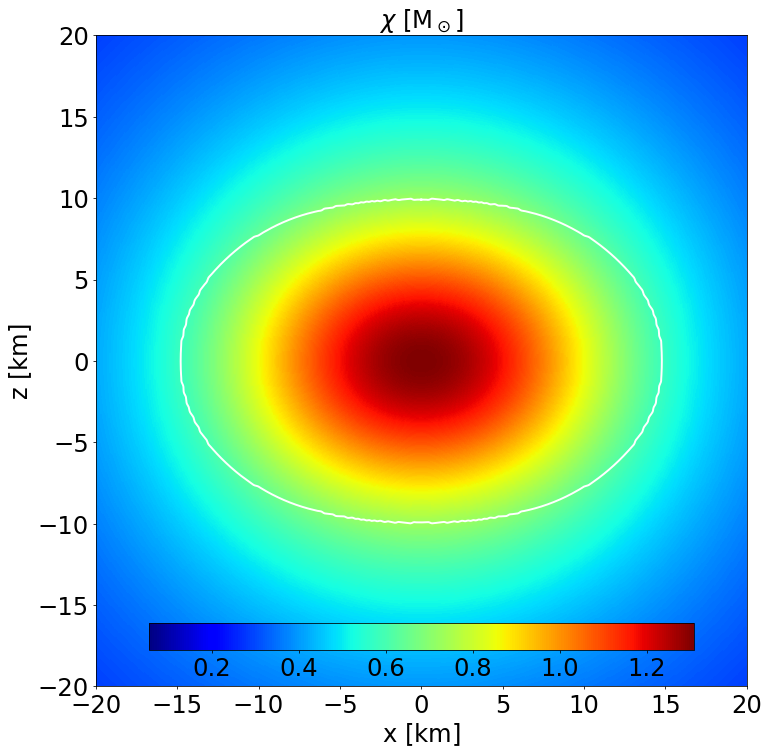}\\
         \caption{From left to right: Meridional distribution of the magnetic field strength $B=\sqrt{B^r B_r+B^\theta B_\theta}$, of the density $\rho$ and of the scalar field $\chi$ for a model with a poloidal magnetic field of maximum strength $B_{\textrm{max}}=6.256\times 10^{17}$G and central density $\rho _{\mathrm c} = 5.15 \times 10^{14}$g~cm$^{-3}$. The white curve represents the surface of the star. The light white lines on the left panel represent magnetic surfaces. More quantitative details on this configuration can be found in Table~\ref{tab:ref}, where it is named `model P'.}
         \label{fig:polo1}
\end{figure*}
\begin{figure}
   	\centering
         \includegraphics[width=0.4\textwidth]{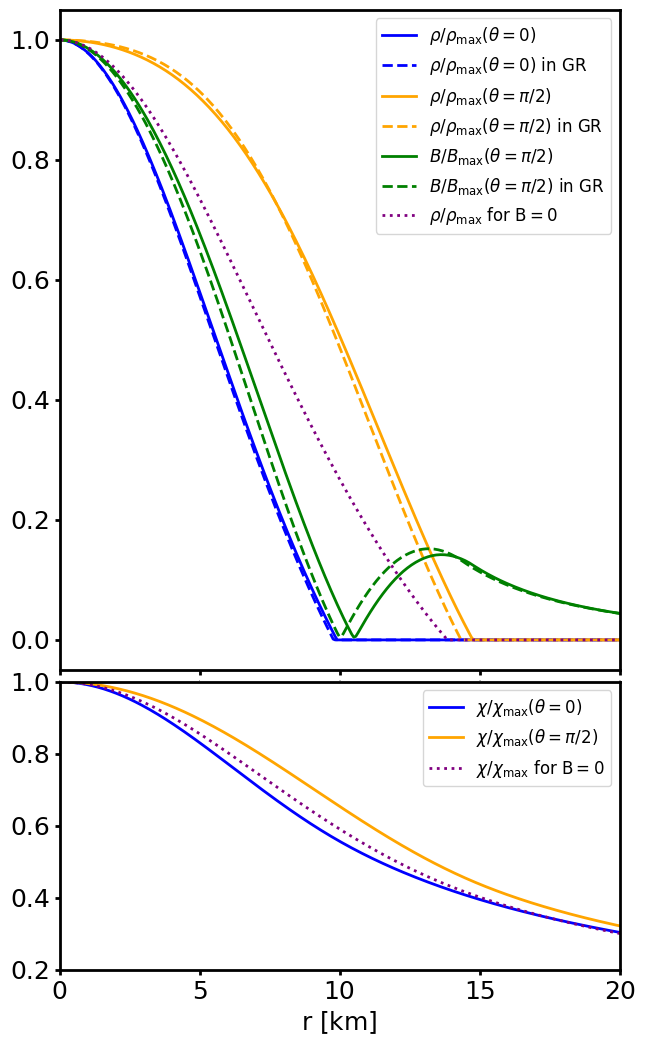}
         \caption{Top panel: Profile of the polar (solid blue lines) and equatorial (solid orange lines) 
         density, and of the magnetic field strength (solid green lines) at the equator, normalised to their maximum values, for the equilibrium model P (with purely poloidal magnetic field) of Table~\ref{tab:ref}. These are to be compared to the corresponding GR model at the same $\rho _{\mathrm c}$ and $B_\mathrm{max}$ (dashed), and with the density of the scalarised and un-magnetised model at the same $\rho _{\mathrm c}$, P0 (dotted purple line). Bottom panel: Profile of the equatorial (orange line) and polar (blue line) scalar field, normalised to their maximum value, for the equilibrium model P (solid), compared to the un-magnetised model P0 (dotted purple).}
         \label{fig:polo2}
\end{figure}
\begin{figure*}
   	\centering
         \subfigure{\includegraphics[width=0.455\textwidth]{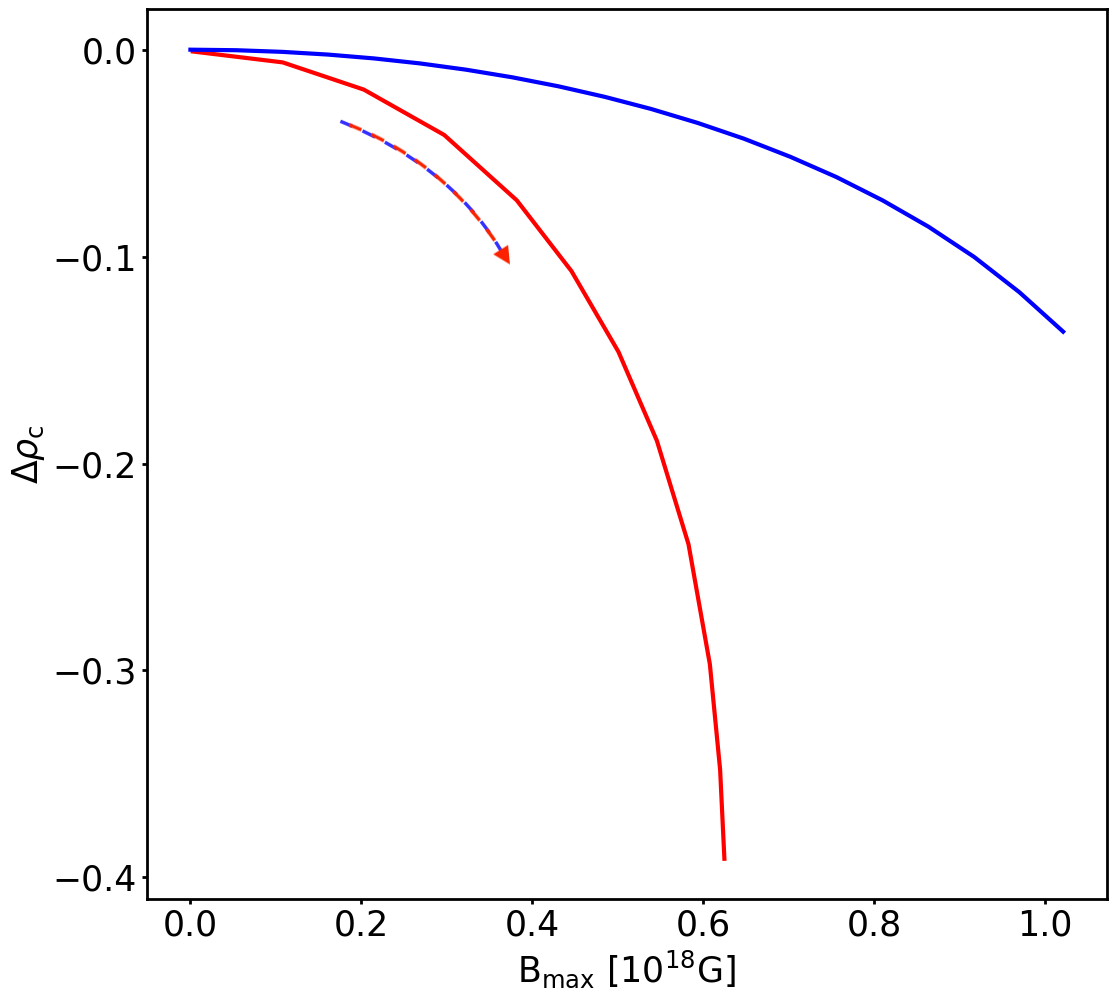}}
         \subfigure{\includegraphics[width=0.44\textwidth]{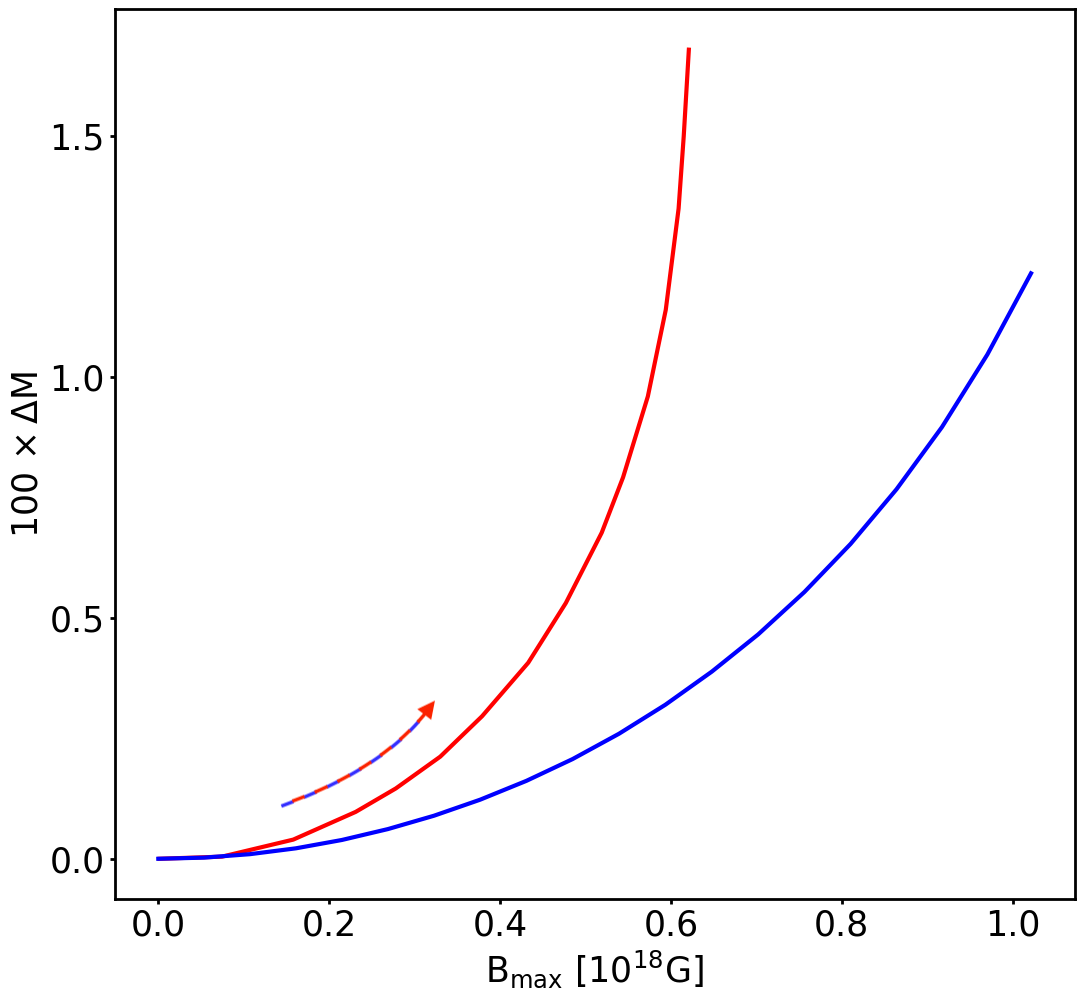}}
         \subfigure{\includegraphics[width=0.45\textwidth]{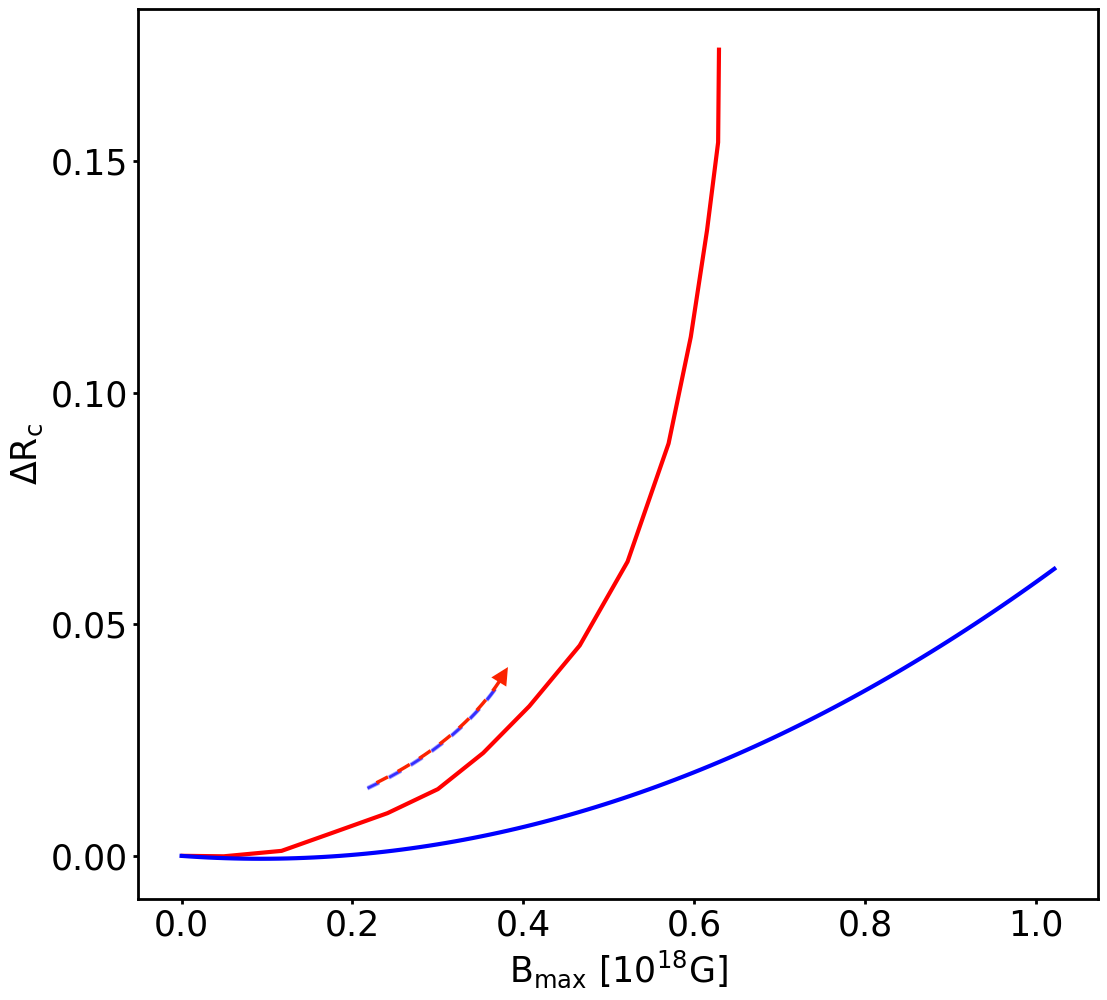}}
         \subfigure{\includegraphics[width=0.44\textwidth]{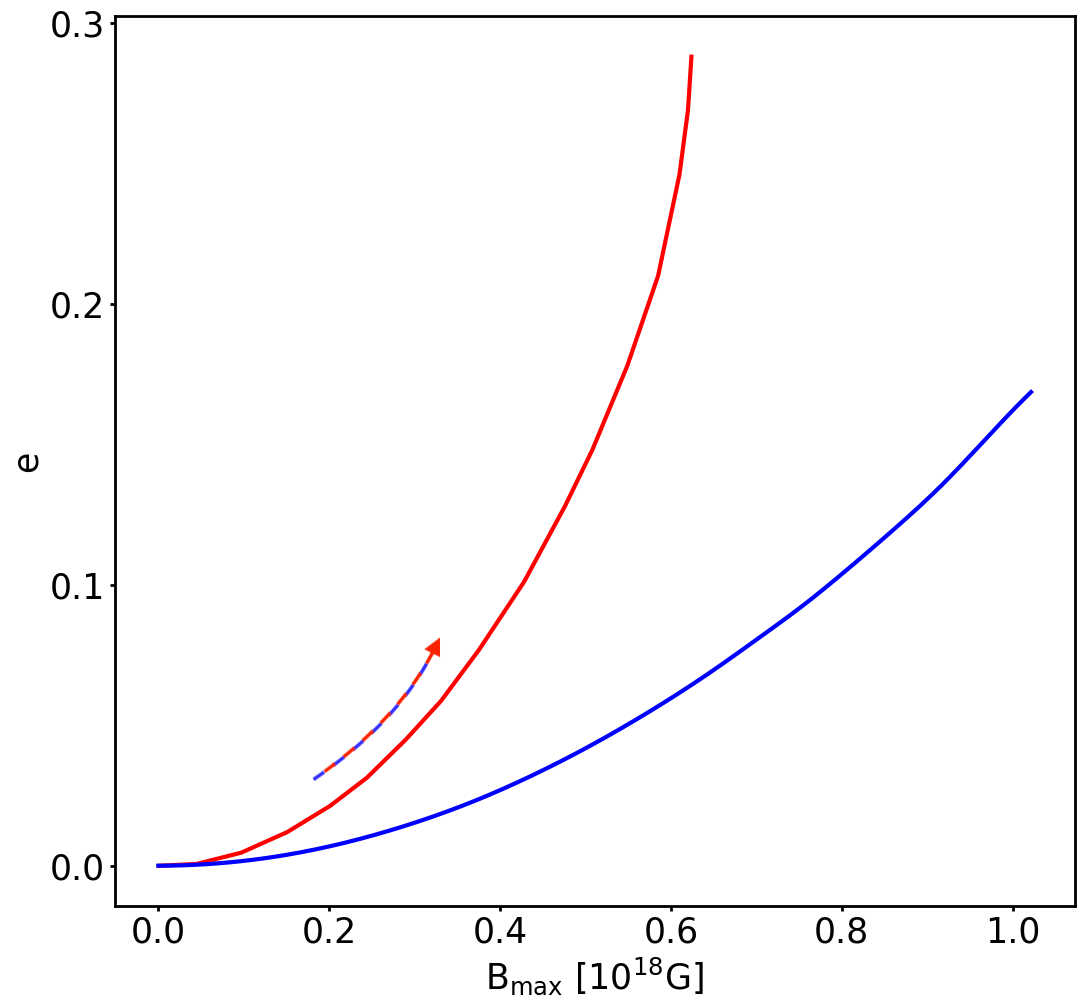}}
         \caption{Variation, with respect to the un-magnetised model, of various quantities along the equilibrium sequence with constant $M_0=1.68$M$_\odot$ for purely poloidal magnetic field. From left to right, top to bottom: Central density $\rho _\mathrm{c}$, Komar mass $M_\mathrm{k}$, circumferential radius $R_\mathrm{c}$ and quadrupole deformation $e$. The blue line represents our STT results, to be compared to the red line, describing the GR models of \citet[Fig.~7]{pili_axisymmetric_2014}. The arrows show the direction of increasing magnetisation.}
         \label{fig:polo3}
\end{figure*}
\begin{figure}
  \centering
  \includegraphics[width=0.45\textwidth]{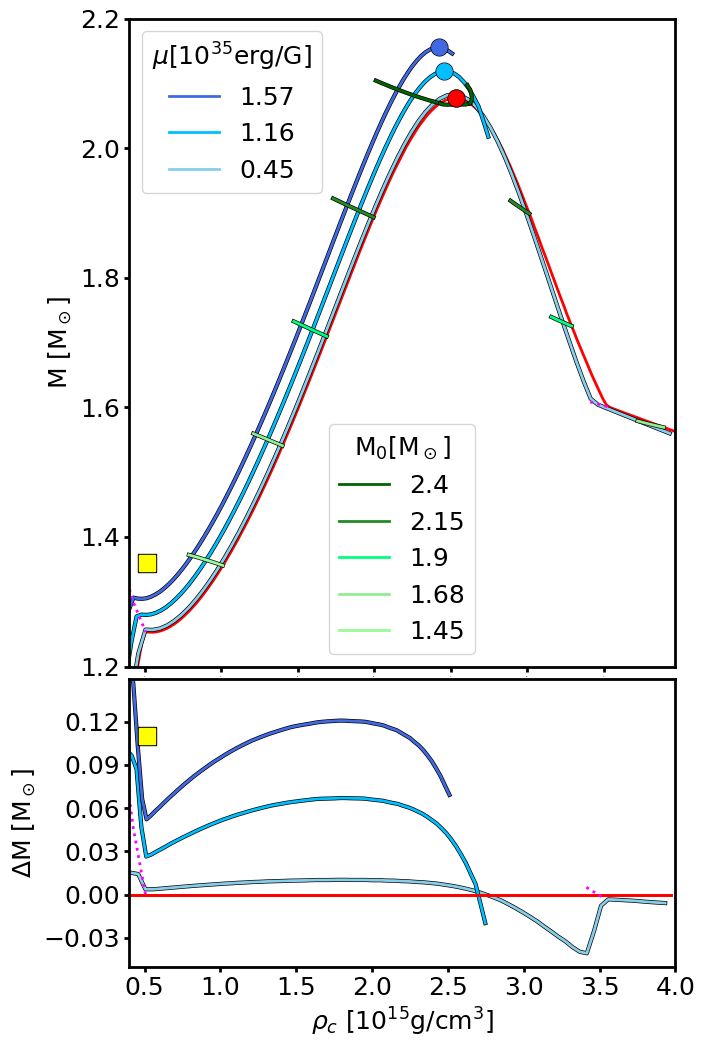}
  \caption{Mass-density sequences for models with purely poloidal magnetic field and $\beta _0=-6$. Upper panel: Sequences computed at fixed values of the magnetic dipole moment $\mu$ (blue lines) and at fixed baryonic mass (green lines), compared with the un-magnetised case (red line). The dotted magenta lines represent the limit for spontaneous scalarisation. Dots mark the position of the maximum mass models UM$_0$ (red), PM$_1$ (light blue) and PM$_2$ (dark blue) of Table~\ref{tab:const_seq}. The yellow square represents the model of Fig.~\ref{fig:polo1}. Lower panel: Mass difference of sequences at fixed $\mu$ with respect to the un-magnetised one.}
  \label{fig:const-polo_seq}
\end{figure}
\begin{figure*}
  \centering
  \includegraphics[width=0.31\textwidth]{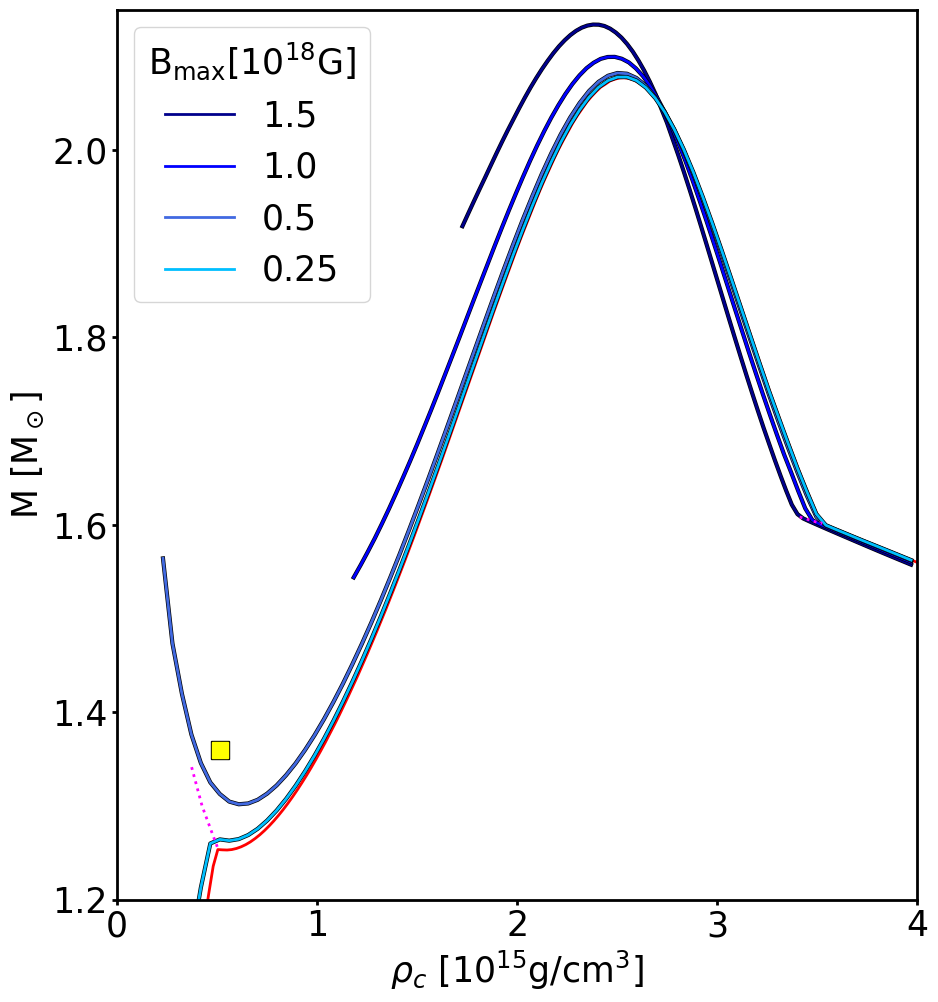}
  \includegraphics[width=0.31\textwidth]{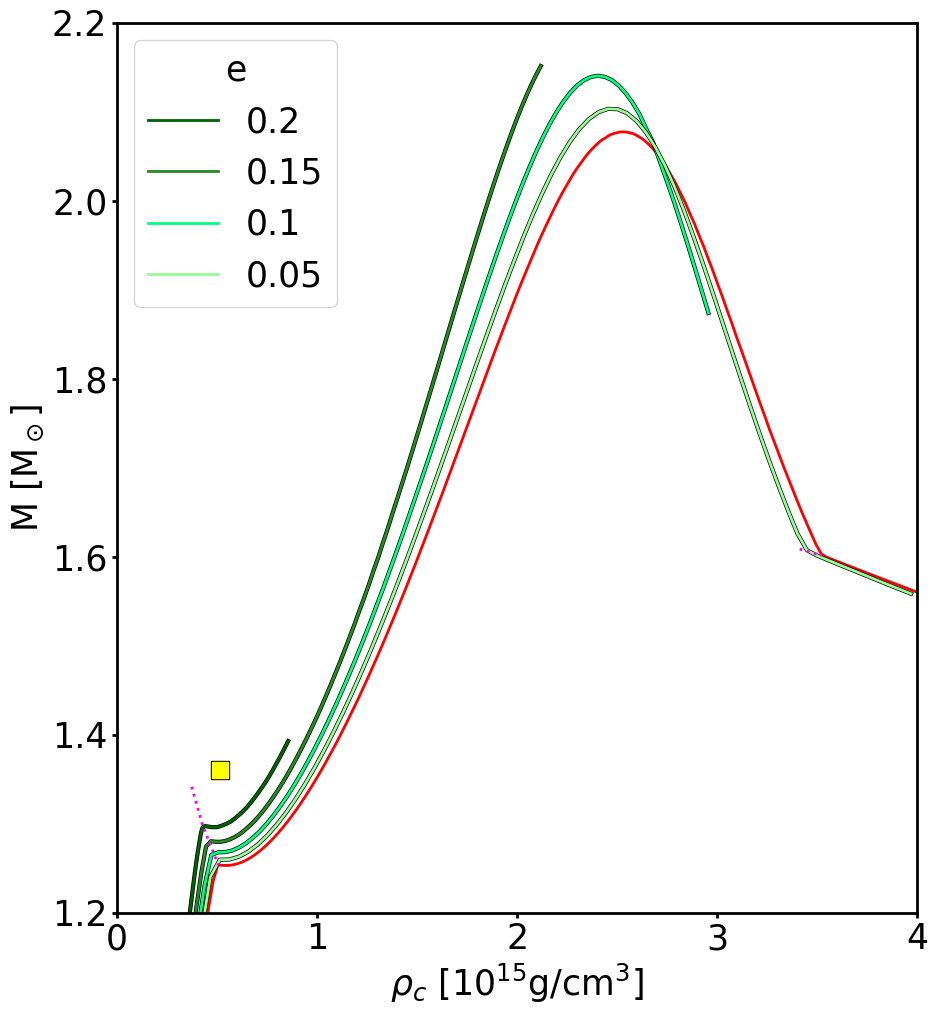}
  \includegraphics[width=0.31\textwidth]{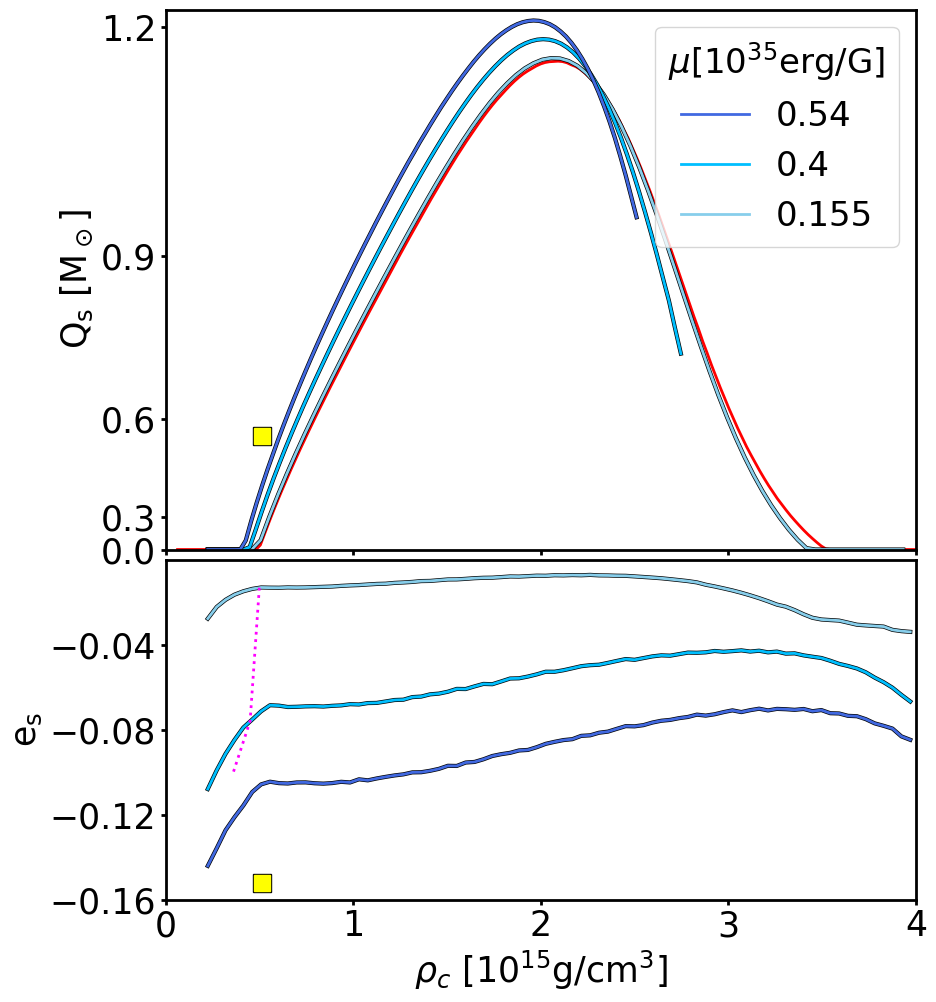}
  \caption{Sequences for models with purely poloidal magnetic field and $\beta _0=-6$. Left panel: Mass-density relation computed at fixed $B_\mathrm{max}$ (blue lines) compared with the un-magnetised sequence (red line). Middle panel: Mass-density relation computed at fixed $e$ (green lines) compared with the un-magnetised sequence (red line). Right panel: On top, scalar charge-density relation computed at fixed $\mu$ (blue lines) compared with the un-magnetised sequence (red line); on bottom, trace quadrupole deformation $e_\mathrm{s}$. In all panels, the dotted magenta lines represent the limit for spontaneous scalarisation and the yellow square represents model P of Fig.~\ref{fig:polo1}.}
  \label{fig:const_seq-p}
\end{figure*}

\subsection{Magnetised models with $\beta _0 =-5$}

In order to understand how our results depend on the specific choice of the STT parameter $\beta_0$, we have computed equilibrium configurations also for $\beta_0=-5$ and $\beta_0=-4.5$, closer to the limit for spontaneous scalarisation, both in the case of pure toroidal and purely poloidal magnetic fields.  For $\beta_0=-5$, the un-magnetised model with baryonic mass $M_o=1.680$M$_\odot$ is scalarised. It is then possible to compute deviations of various quantities with respect to their un-magnetised values, at fixed  baryonic mass $M_o=1.680$M$_\odot$, as was done for $\beta_0=-6$. In Fig.~\ref{fig:dev-b5}, we show how the quadrupole deformation $e$ changes with the maximum strength of the magnetic field $B_\mathrm{max}$. Again, we find that the scalarised part of the sequence shows a lower quadrupole deformation  than in GR, but now this difference is not as strong as for $\beta_0=-6$. In general $e$ is about $2/3$ of the value of the corresponding GR counterpart at the same $B_\mathrm{max}$, both in the toroidal and poloidal magnetic field case. For purely toroidal magnetic fields, there is some indication that the scalarised part reaches a maximum value  $B_\mathrm{max} \simeq 5.8\times 10^{17}$G, before it de-scalarises, and then reaches a new maximum corresponding to the GR value of $6.13\times 10^{17}$G. We can conclude that in STTs with  $\beta _0>-5$ the upper limit to  $B_\mathrm{max}$ is reached after the solution de-scalarises, while for  $\beta _0<-5$ it is reached for scalarised configurations. On the other hand in models with a purely poloidal magnetic field, we observe no evidence for de-scalarisation with increasing $k_\mathrm{pol}$. However, there seems to be an asymptote to a maximum value of $B_\mathrm{max}$ of $\simeq 7.5\times 10^{17}$G, slightly higher than in GR for $\beta_0=-5$. The same conclusions can be found looking at the deviations of other variables. What we see is that changes with respect to GR depend in a strongly non-linear way on the values of $\beta_0$.

In Fig.~\ref{fig:const_seq_b-5_t}, we repeat the same analysis of  Fig.~\ref{fig:const-toro_seq}, for purely toroidal fields. We show how the Komar mass and scalar charge change with central density holding fixed the magnetic flux $\Phi$, and the Komar mass for fixed values of the baryonic mass $M_0$. The region of de-scalarisation $\rho _{\mathrm c}=[\rho_{\mathrm{b}}, \rho_{\mathrm{t}}]$ is smaller, but the behaviour of the lower and upper bounds with magnetisation is the same.  The lower bound $\rho_\mathrm{b}$ moves to higher densities, from $\rho_\mathrm{b}=7.07\times 10^{14}$g cm$^{-3}$ for $\Phi =0$ to $\rho _{\mathrm c}=1.06\times 10^{15}$g cm$^{-3}$ for $\Phi =2\times 10^{30}$G cm$^2$, and the corresponding Komar (baryonic) mass  from $1.461$M$_\odot$ ($1.57$M$_\odot$) to 1.75M$_\odot$ (1.84M$_\odot$). Again we find no evidence suggesting the existence of an upper bound to the mass of the possible de-scalarised models. Analogously, the upper bound $\rho_\mathrm{t}$ for scalarised models increases, from $\rho_\mathrm{b}=2.65\times 10^{15}$g cm$^{-3}$ for $\Phi =0$ to $\rho _{\mathrm c}=3.05\times 10^{15}$g cm$^{-3}$ for $\Phi =2\times 10^{30}$G cm$^2$, and the corresponding Komar (baryonic) mass  from $1.67$M$_\odot$ ($1.83$M$_\odot$) to 1.77M$_\odot$ (1.85M$_\odot$).
Again, we find that for toroidal magnetic fields the density at which the maximum is reached increases, and the value of the maximum mass first remains almost constant at 1.81M$_\odot$, and then rises to 1.86M$_\odot$ for $\Phi=2\times 10^{30}$G~cm$^2$. In this case we also see that on sequences with $\Phi \geq 1.64\times 10^{30}$G cm$^2$ the mass of equilibrium models is always larger than the relative un-magnetised counterpart at the same central density. 

For poloidal magnetic fields, we observe in Fig.~\ref{fig:const_seq_b-5_p} a more regular trend, similar to the case with $\beta_0=-6$, where the maximum mass initially seems to remain unchanged to then rises at higher magnetisation. We find that, for poloidal fields, above a Komar mass of $1.7$M$_\odot$ there are no de-scalarised models. 

It is evident that now the magnetic field plays a more dominant role that the scalar field, and the general trends of the various sequences tend to approach what was found in GR. However, in the region where the scalar charge reaches its maximum, the trends are still in line with more scalarised configurations.

\begin{figure*}
  \centering
  \subfigure{\includegraphics[width=0.45\textwidth]{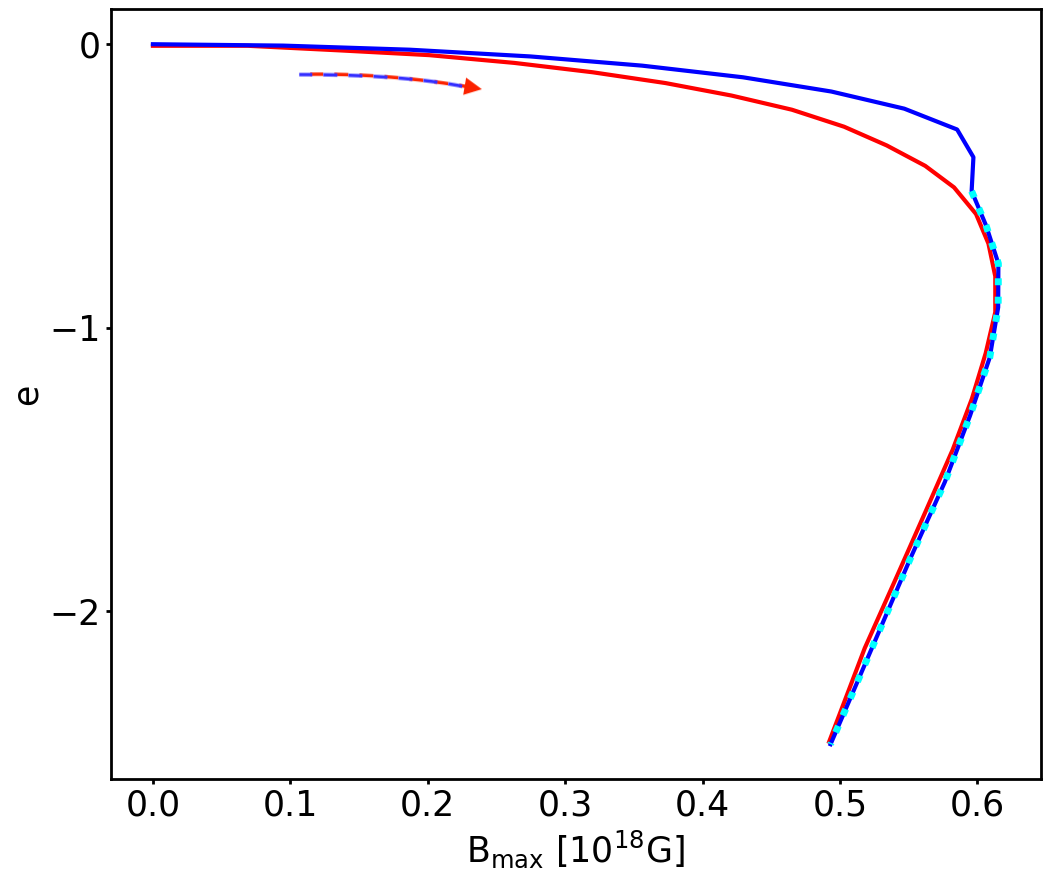}}
  \subfigure{\includegraphics[width=0.455\textwidth]{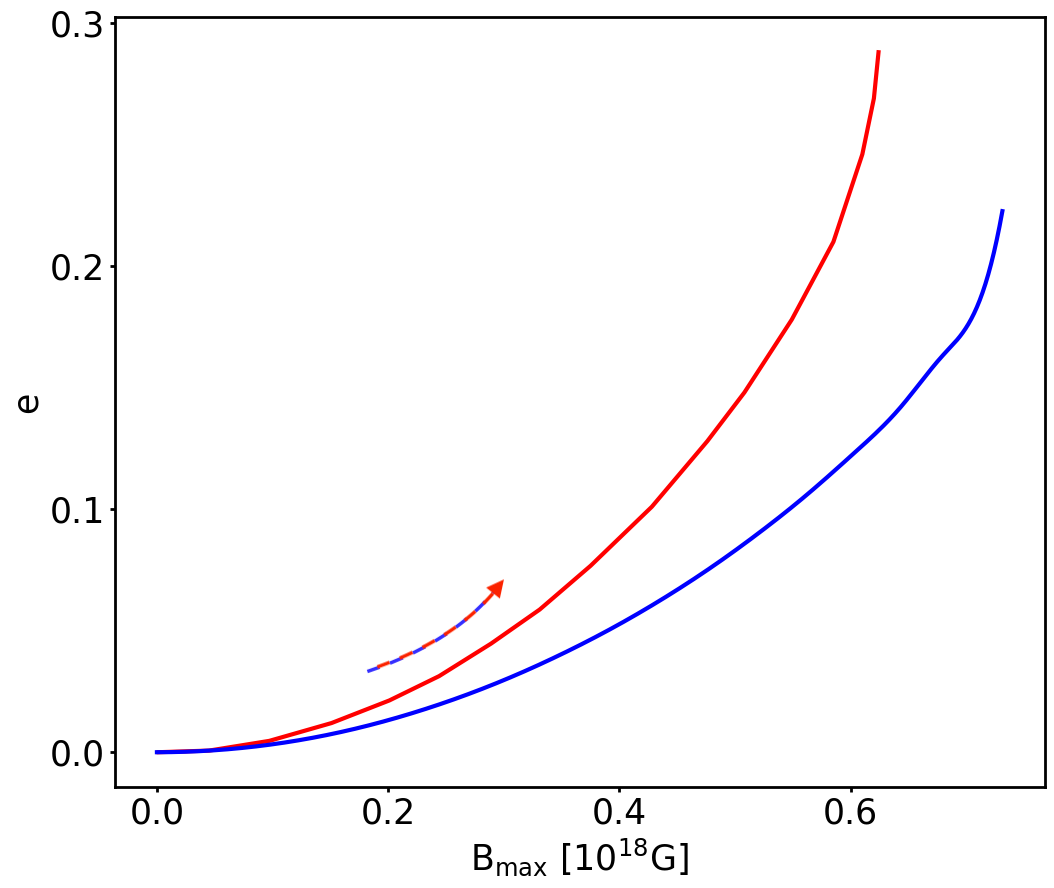}}
   \caption{Value of the quadrupole deformation $e$ along the equilibrium sequence with constant $M_0=1.68$M$_\odot$, as a function of $B_\mathrm{max}$,  for $\beta _0 =-5$ (blue lines) vs GR (red lines).  The cyan dotted line highlights the un-scalarised configurations. Left panel: Purely toroidal magnetic field; right panel: Purely poloidal magnetic field. The arrows show the direction of increasing magnetisation.}
   \label{fig:dev-b5}
\end{figure*}
\begin{figure*}
   	\centering
   	     \includegraphics[width=0.45\textwidth]{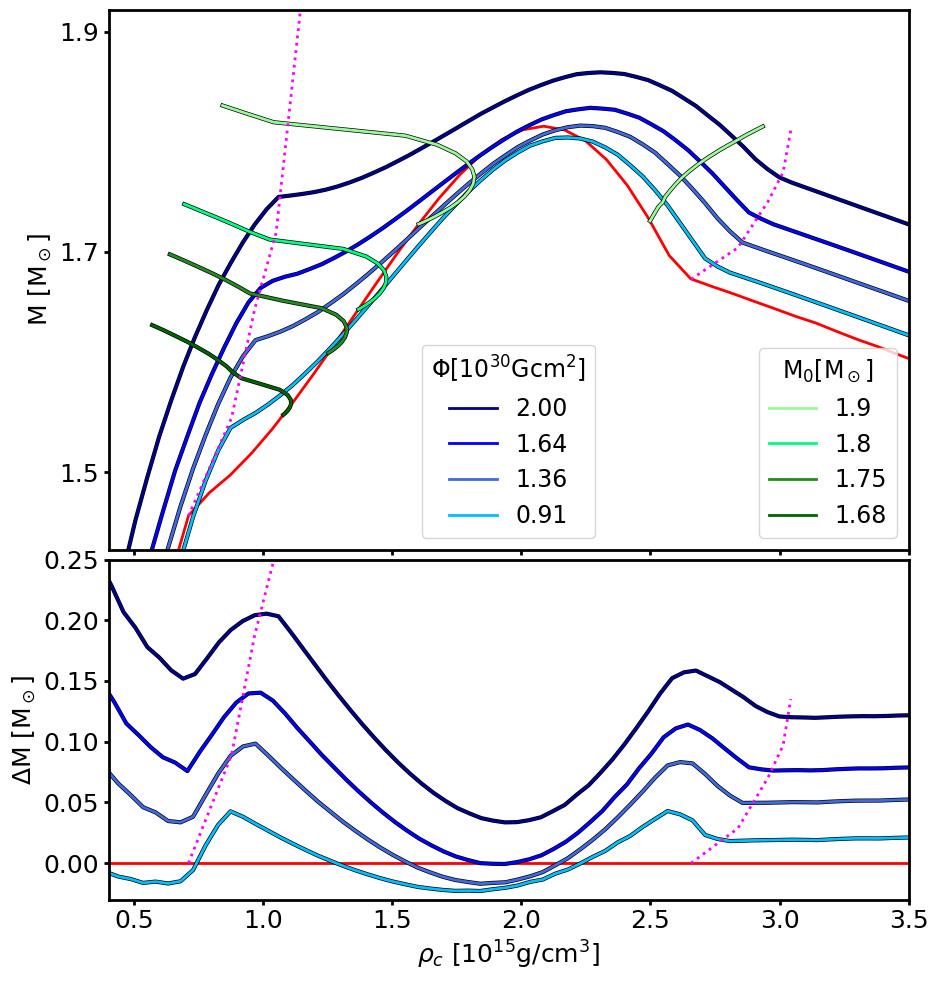}
   	     \includegraphics[width=0.47\textwidth]{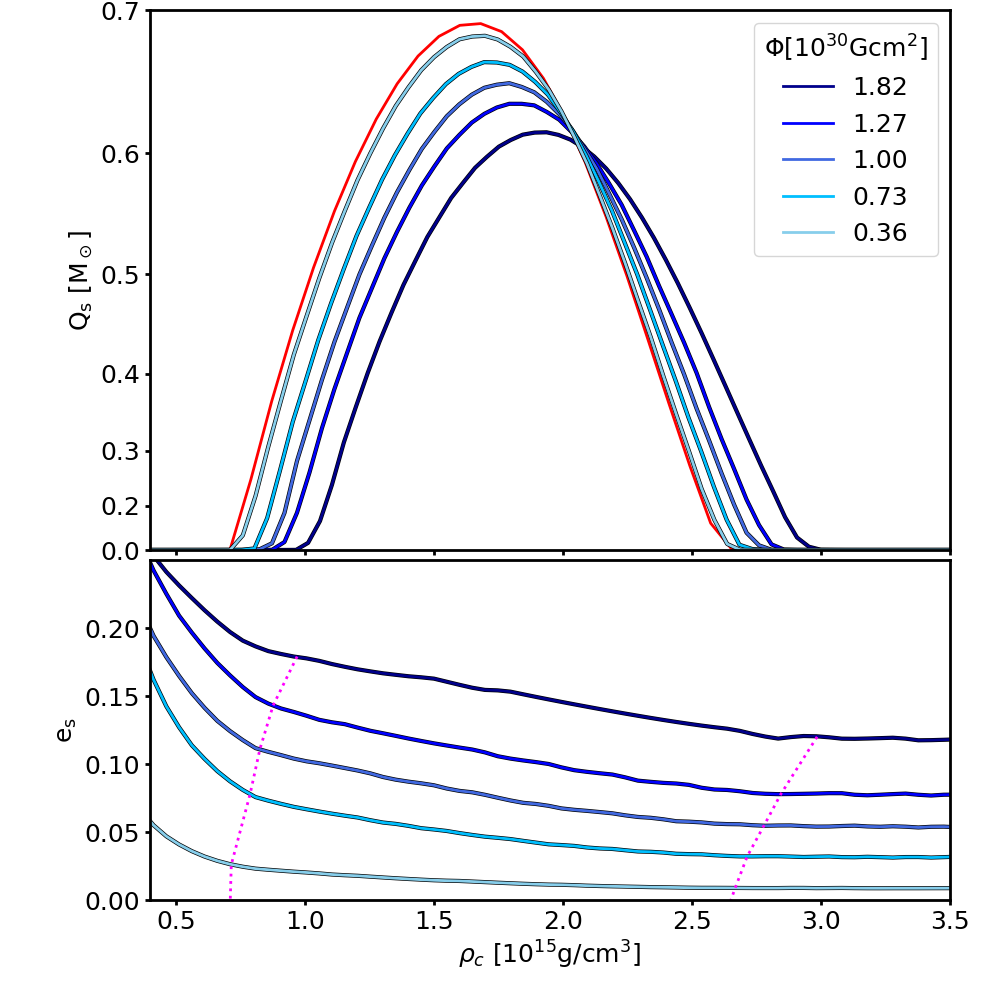}         
         \caption{Models with purely toroidal magnetic field and $\beta _0=-5$. Left panel: On top, sequences computed at fixed values of the magnetic flux $\Phi$ (blue lines) and at fixed baryonic mass (green lines), compared with the un-magnetised case (red line); on bottom, mass difference with respect to the un-magnetised case. Right panel: On top, scalar charge on sequences at fixed $\Phi$; on bottom, trace quadrupole $e_\mathrm{s}$ on the same sequences. The dotted magenta lines represent the limit for spontaneous scalarisation.}
         \label{fig:const_seq_b-5_t}
\end{figure*}
\begin{figure*}
   	\centering
   	     \includegraphics[width=0.47\textwidth]{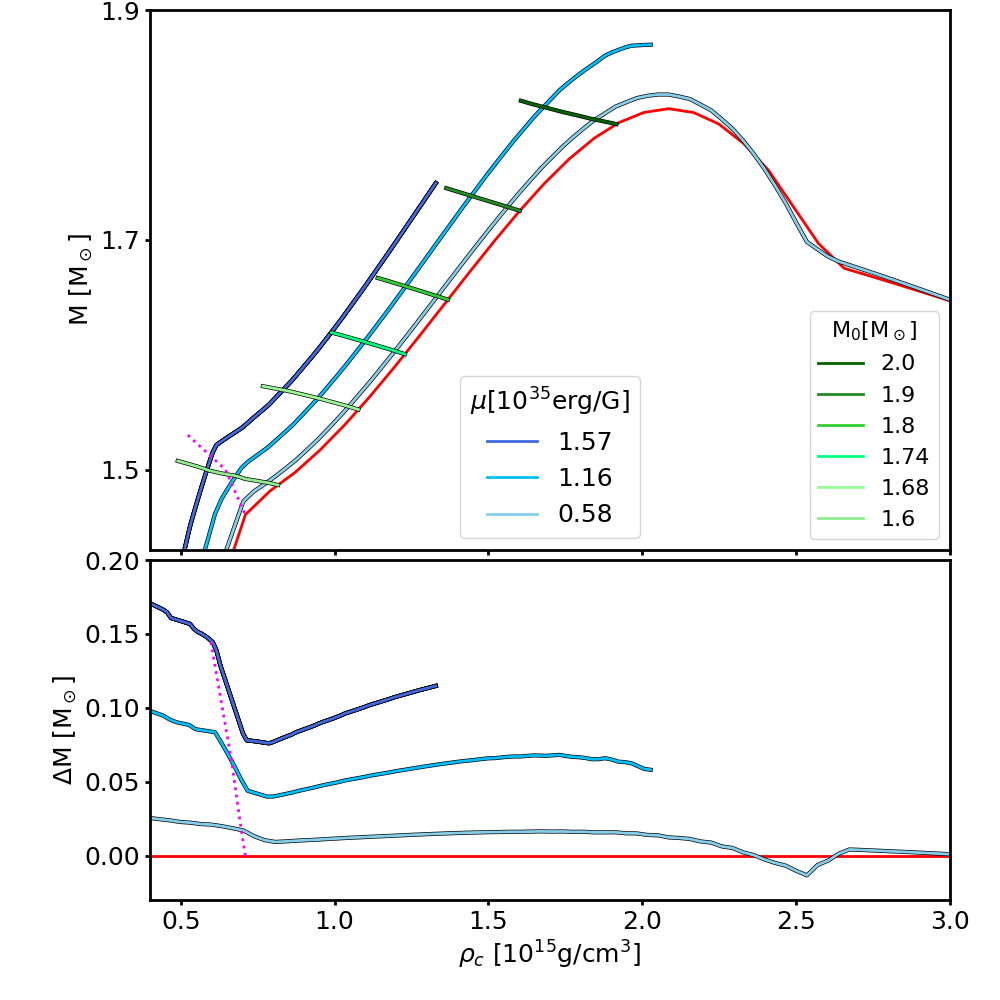}
         \includegraphics[width=0.47\textwidth]{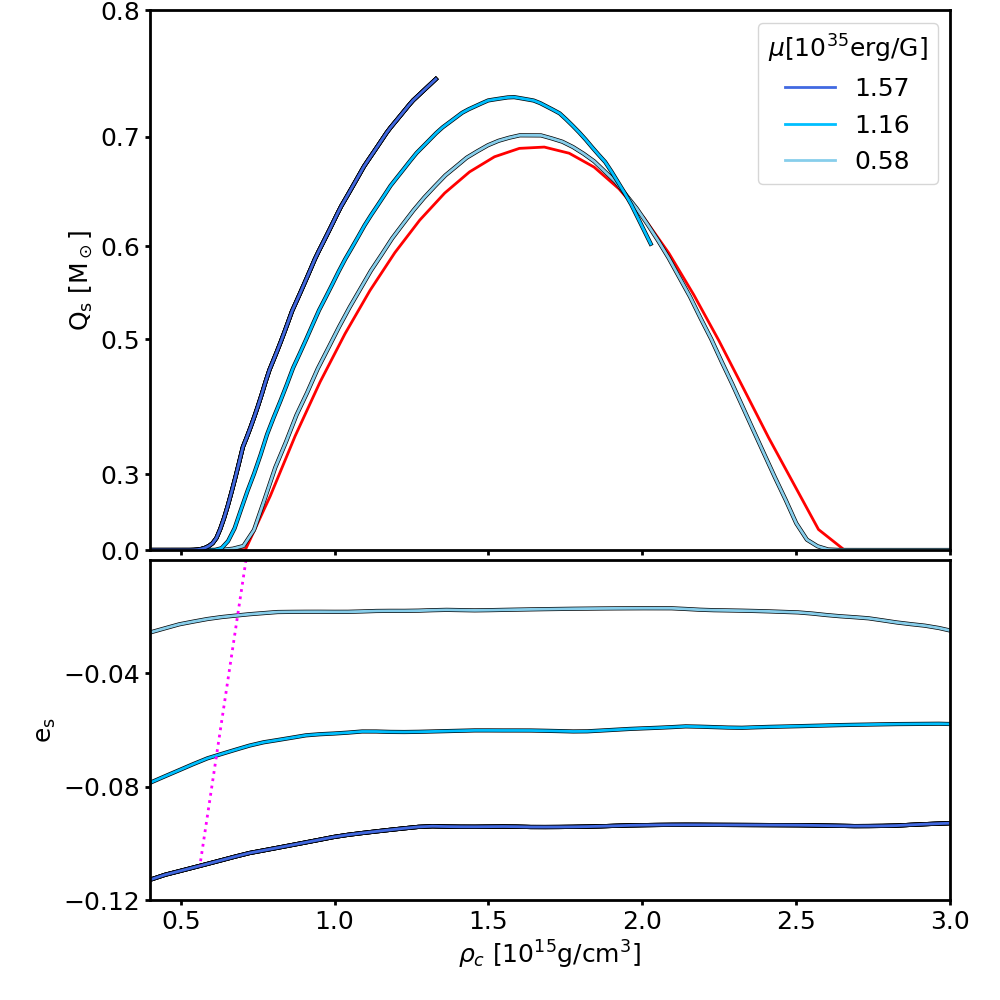}   
         \caption{Models with purely poloidal magnetic field and $\beta _0=-5$. Left panel: On top, sequences computed at fixed values of the magnetic dipole moment $\mu$ (blue lines) and at fixed baryonic mass (green lines), compared with the un-magnetised case (red line); on bottom, mass difference with respect to the un-magnetised case. Right panel: On top, scalar charge on sequences at fixed $\mu$; on bottom, trace quadrupole $e_\mathrm{s}$ on the same sequences. The dotted magenta lines represent the limit for spontaneous scalarisation.}
         \label{fig:const_seq_b-5_p}
\end{figure*}

\subsection{Magnetised models with $\beta _0 =-4.5$}
We consider here the case $\beta _0=-4.5$, which is close to the upper limit on massless STTs set by binary pulsar constraints \citep{freire_relativistic_2012,shao_constraining_2017,anderson_2019}. In Fig.~\ref{fig:const_seq_b-4.5}, we show how the Komar mass changes holding fixed the magnetic flux $\Phi$ for configurations with a purely toroidal magnetic field. The scalarised range is now strongly reduced. For the un-magnetised models, $\rho_\mathrm{b}=9.3\times 10^{14}$g cm$^{-3}$ and $\rho_\mathrm{t}=2.0\times 10^{15}$g cm$^{-3}$, with a Komar mass that changes from $1.58$M$_\odot$ to $1.71$M$_\odot$. As the magnetic flux increases, the typical scalarised trend in the mass-density relation becomes progressively less evident: already at $\Phi = 0.9 \times 10^{30}$G~cm$^2$ the seuqence is almost indistinguishable from GR. This is made even more evident looking at the scalar charges in Fig.~\ref{fig:const_seq_b-4.5}, where we observe simultaneously both a reduction of $Q_s$ and of the scalarisation range.

In case of a purely poloidal magnetic field, the trend is instead quite different, as can be seen in Fig.~\ref{fig:const_seq_b-4.5}. Increasing the magnetic flux $\Phi$, both the scalar charge and the scalarisation range increase, with $\rho_\mathrm{b}$ moving to lower values. The maximum mass rises, and there is no evidence for the de-scalarisation.

This difference, in part already present at lower $\beta_0$, can be understood if one recalls that spontaneous scalarisation can be seen, from a dynamical point of view, as an instability \citep{damour_tensor-scalar_1996}, which can be excited only if the minimum wavelength of unstable modes (a function of  $\beta_0$) is smaller or of the order of the typical highscale of the matter distribution (roughly the size of the compact star). Detailed calculations set this limit for NSs around $ \beta_0\approx -4.2,-4.0$. It is obvious, that close to this threshold limit, any process that modifies the distribution of matter in compact stars can have deep consequances on their spontaneous scalarisability. A strong toroidal magnetic field leads to a prolate distribution of density, that on average corresponds to a reduction of the typical  highscale of the matter distribution, potentially pushing the NS below the threshold for spontaneous scalarisation. On the other way a strong poloidal magnetic field leads to an oblate distribution of density, corresponding to an increase of the typical  highscale of the matter distribution, potentially pushing the NS above the threshold for spontaneous scalarisation.

\begin{figure*}
   	\centering
   	     \includegraphics[width=0.45\textwidth]{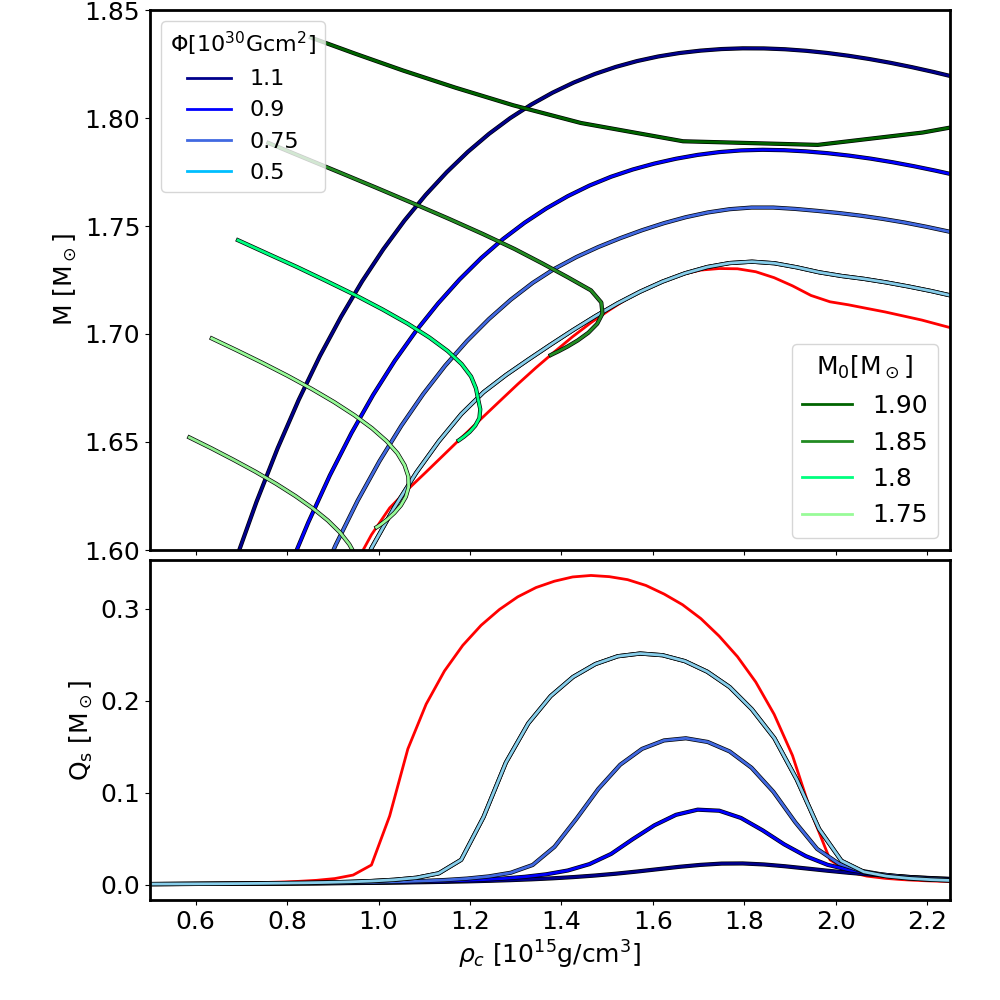}
         \includegraphics[width=0.45\textwidth]{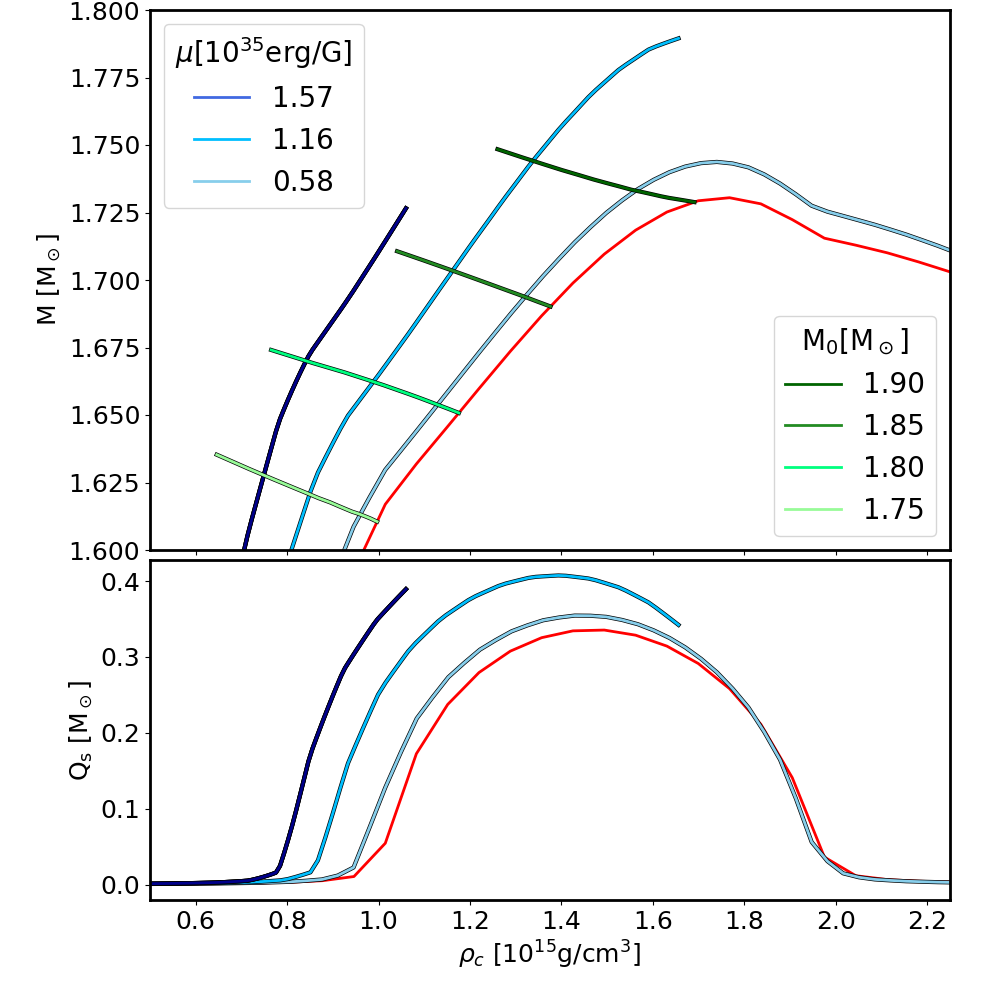}\\
         \caption{Left figure: Models with purely toroidal magnetic field and $\beta_0=-4.5$. Upper panel: Sequences computed at fixed values of the magnetic flux $\Phi$ (blue lines) and at fixed baryonic mass (green lines), compared with the un-magnetised case (red line). Bottom panel: Value of the scalar charge on the same sequences at fixed $\Phi$. Right figure: Models with purely poloidal magnetic field and $\beta_0=-4.5$. Upper panel: Sequences computed at fixed values of the magnetic dipole moment $\mu$ (blue lines) and at fixed baryonic mass (green lines), compared with the un-magnetised case (red line). Bottom panel: Value of the scalar charge on the same sequences at fixed $\mu$.}
         \label{fig:const_seq_b-4.5}
\end{figure*}

\subsection{On the stability of magnetised equilibrium models}

It is well known that NSs endowed with either a purely toroidal or a purely poloidal magnetic field are unstable against non-axisymmetric perturbations \citep{braithwaite_stable_2006,braithwaite_evolution_2006,braithwaite_axisymmetric_2009}. This is due to a magnetofluid instability that, on a typical Alfv\'{e}nic timescale,  leads to a reconfiguration of the magnetic field geometry toward a more tangled structure. Magnetic stability requires mixed configurations, with comparable amount of energy in the poloidal and toroidal components of the magnetic field.

With respect to axisymmetric perturbations, on the other hand, it is found that, purely poloidal magnetic fields are stable, while the stability of purely toroidal magnetic fields, against interchange modes, depends on their stratification. Toroidal configurations with $m=1$ are found to be stably stratified \citep{Schubert68a,Fricke69a}.

Independently of their magnetofluid stability, we are going to show that in STTs, NSs with purely toroidal magnetic fields, are also gravitationally unstable against spontaneous scalarisation.  The criterion for gravitational instability for non-rotating and un-magnetised NS is
\begin{equation}
	\frac{\partial M_0}{\partial \rho _{\mathrm c}} \leq 0 \quad ,
\end{equation}
where the equality defines the maximum mass. We note that in GR and STTs it is the baryonic mass that formally
enters the criterion, and not the Komar mass, given that the former is the dynamically conserved quantity. However in GR and STTs the Komar mass is always a monotonically increasing function of the baryonic mass and one can safely use it to evaluate stability.  This criterion can be generalised to magnetic configurations. Recalling that the flux-freezing condition of ideal MHD, ensures that the magnetic flux $\Phi$ is conserved in axisymmetry, one has that NSs with a purely toroidal magnetic field are unstable when
\begin{equation}
	\frac{\partial M_0}{\partial \rho _{\mathrm c}}\bigg{|}_{\Phi} \leq 0 \quad .
\end{equation}
In Fig.~\ref{fig:stab_toro}, we plot how the baryonic mass of various equilibrium configurations change with density at fixed values of the magnetic flux $\Phi$. It is immediately evident that each sequence shows four parts:
\begin{itemize}
\item a gravitationally stable de-scalarised GR part;
\item a gravitationally unstable scalarised part;
\item a gravitationally stable scalarised part (up to the density of the model of maximum mass for the entire sequence);
\item a gravitationally unstable scalarised part (beyond the density of the model of maximum mass for the entire sequence).
\end{itemize}
This is in sharp contrast to GR, where only two parts are found (stable and unstable), separated by the model with maximum mass.
In principle now we can have two maxima for the mass of NSs with purely toroidal magnetic fields: one corresponding to the de-scalarised part and one to the scalarised one. In the mass-density diagram there is a region where models are gravitationally unstable. Moreover, for any given value of $\Phi$, there is a range of masses where both de-scalarised and scalarised solutions are possible. On the other hand, there is a lower limit to the values of the magnetic flux that can support de-scalarised configurations of a given baryonic mass. Lowering the magnetic flux beyond this limit could lead to a gravitational instability where the star jumps from the de-scalarised branch to the scalarised one. This is a gravitational instability, unrelated to rearrangements of the magnetic field geometry, that will take place on a typical scalarisation timescale, of the order of the light crossing time of the NS. For example, with reference to Fig.~\ref{fig:stab_toro}, a de-scalarised configuration with M$_0=1.68$M$_\odot$ can only exist for $\Phi > 2.06 \times 10^{30}$G cm$^2$ and $\rho _{\mathrm c} < 7.12 \times 10^{14}$g cm$^{-3}$; below this limiting value of the magnetic flux, the NS will jump at the same baryonic mass but with a central density  $\rho _{\mathrm c} > 1.32 \times 10^{15}$g cm$^{-3}$, and a scalar charge  $Q_s = 0.8$M$_\odot$. Interestingly, these two limiting configurations have not just the same baryonic mass, and magnetic flux, but also the same Komar mass $M_\mathrm{k}=1.62$M$_\odot$. We have repeated this analysis also for higher values of $\beta_0$ and found that this effect already disappears at $\beta _0 = -5$. However, for $\beta _0 = -4.5$ we found that two configurations, one scalarised and the other un-scalarised, with the same baryonic mass still exist, but in this case they have the same central density.
\\\\
Independently of the specific choice of magnetic field distribution, that in our case is dictated by the request of an integrable form for the generalised Bernoulli equation, our results have shown that a strong toroidal magnetic field can support de-scalarised configurations, and that, in principle, if such magnetic field drops below a limiting value (for example because of non ideal processes or magnetic instabilities) such configuration can undergo a rapid `magnetically-induced spontaneous scalarisation'. 
In the case of purely poloidal configurations, the quantity that is dynamically conserved for axisymmetric perturbations is the net flux of the toroidal current $J^\phi$. This can be equivalently parametrised by the magnetic dipole moment. If we repeat the same analysis done in the toroidal case, considering sequences at fixed magnetic dipole moment, we see no evidence for the presence of an unstable part.

\begin{figure}
   	\centering
   	     \includegraphics[width=0.45\textwidth]{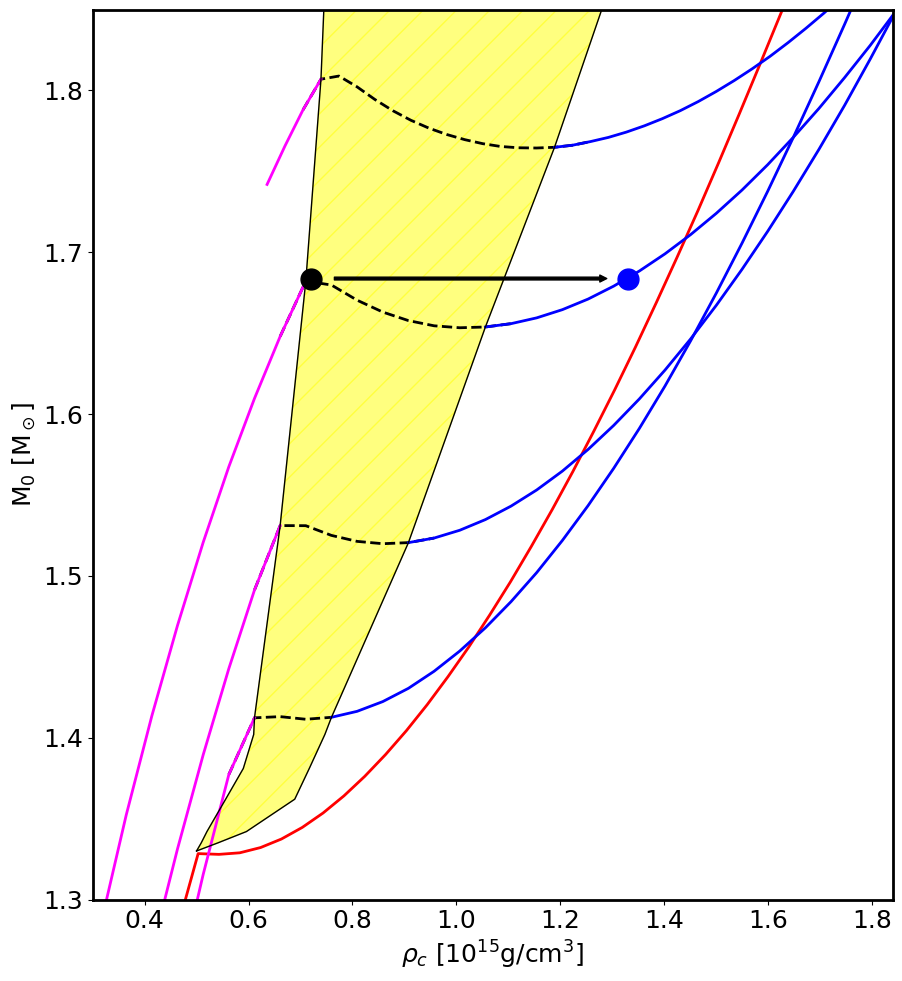}
             \caption{Sequences at fixed magnetic flux $\Phi$, computed in the case $\beta_0=-6$. The red curve is the un-magnetised solution. From bottom to top the other curves are computed at $\Phi=[2.55,2.06,1.46,0.91]\times 10^{30}$G cm$^2$. The various parts are:  gravitationally stable de-scalarised branch (solid magenta); gravitationally unstable scalarised branch (black dashed); gravitationally stable scalarised branch (solid blue). The yellow region corresponds to gravitationally unstable models. The black dot represents the de-scalarised configuration with M$_0=1.68$M$_\odot$ and $\Phi = 2.06 \times 10^{30}$G cm$^2$, while the arrow points to the blue dot where the configuration is expected to jump because of magnetically-induced spontaneous scalarisation.  }
         \label{fig:stab_toro}
\end{figure}

\section{Conclusions}
\label{sec:conclusions}
On the one hand, a proper understanding of the role of magnetic fields is fundamental  in the physics and phenomenology of NSs. Magnetic fields affect virtually all of their observational properties and can modify their structure to the point of affecting also their gravitational behaviour. On the other hand, several existing issues in our understanding of the physical Universe have led many theorists to postulate extensions of GR, some of which make interesting predictions on the structure of NSs. In particular, a class of theories known as scalar-tensor theories allow a phenomenon called spontaneous scalarisation, that in principle can lead to sizeable deviations in the structure of NSs from GR. Here, for the first time, we modelled and investigated the properties of magnetised NSs in STTs subject to spontaneous scalarisation, in the full non-linear regime, assuming either purely toroidal or purely poloidal magnetic fields. This is an extension and improvement of our previous work on magnetised NSs in GR.

We have shown how to develop a strategy, within the framework of the 3+1 formalism, to extend standard techniques developed for GRMHD to the case of STTs, by making simultaneous use of the Einstein-frame (where the metric equation have the same mathematical structure as in GR, and the same numerical schemes  can be applied) and the Jordan-frame (where the magnetofluid equations retain their conservative, quasi-hyperbolic form, and thus are amenable to be treated with standard finite volume or finite difference conservative schemes for fluid dynamics). In particular, for simplicity and ease of discussion, we have focused on the case of static configurations, illustrating how the equations that describe the density and magnetic field distribution change in the presence of a scalar field, and how the effects of a scalar field can be fully encapsulated in the conformal scaling factor $\mathcal{A}$.

Our formalism is based on the so called eXtended Conformally Flat Approximation (XCFC), which has proved to be very accurate in GR - even for strongly deformed NSs - and in the fully dynamical regime, also for systems undergoing collapse to black hole, as long as one is not interested in the GW emission.  The XCFC approach has several advantages in GR: the source terms of the metric equations are the same conserved variables evolved by the conservative algorithm for the fluid dynamics; the equations are decoupled and can be solved sequentially; local uniqueness is satisfied. We have verified that in STTs, the XCFC approach retains these properties. Even if, in principle, because of the sign of the scalar field term in the equation for $\alpha \psi$, local uniqueness could be violated, we have checked  that, practically, this is never the case, even for the most scalarised of our configurations.

We have shown that spontaneous scalarisation leads to multiple solutions for NSs, either weakly or strongly scalarised, and we have shown and characterised how the symmetry of the strongly scalarised branch is broken if one chooses a value $\alpha_0\neq 0$. In particular, we verified that our numerical algorithm always selects the $\mathcal{S}_{\mathrm{s}}^{-}$ branch. We also showed that the $\mathcal{S}_{\mathrm{s}}^{-}$ solutions are not always the ones with the largest deviation from GR in the mass-density diagram, but are always the ones with the largest scalar charge and smallest compactness.

In this paper, we carry out a detailed study of the properties of magnetised NSs in STT with spontaneous scalarisation, trying to characterise them as completely as possible, not just in term of their masses or radii, but also considering how the interplay of the magnetic and scalar fields affect their internal structure and deformation. We also tried to characterise the deformation of the scalar field, and introduced the parameter $e_\mathrm{s}$ related to the emission of quadrupolar scalar waves.  

In general, we find that the action of different configurations of the magnetic field on the overall structure of a NS leads to qualitatively similar results: a toroidal magnetic field produced prolate configurations, while a poloidal field leads to oblate one. However, significative changes are found when we proceed to a quantitative comparison.

When comparing STT to GR models, computed at the same central density $\rho_\mathrm{c}$ and maximum value of the magnetic field $B_\mathrm{max}$, we found that the distribution of density and magnetic field vary less than few percent. This suggests that GR models can be used as good proxy for the internal structure of magnetised NSs in STT. On the other hand, when for the same models we compare global integrated quantities like the mass, or the quadrupole deformation, we found deviations from GR up to 10-20\%. This difference can be easily understood recalling that while the distributions of density and magnetic field depend on the ratio $\alpha/\alpha_\mathrm{c}$ (i.e. on relative changes of the metric terms), the value of integrated quantities depends on the conformal factor $\psi^6$ through the volume element (i.e. on the absolute values of the metric terms). On top of this, the quadrupole deformation $e$, used to estimate the possible emission of GWs from deformed system, is properly computed in the E-frame, where the metric equations have the same mathematical structure of GR.

We have also investigated sequences at fixed baryonic mass, which is the conserved quantity from a dynamical and evolutionary perspective, and compared typical trends with those of GR for the same baryonic mass. We found that, in general, the presence of a scalar field reduces the deformability of NSs and tends to reduce the typical deviations from the spherically symmetric un-magnetised configuration. This also implies that with respect to GR, NSs at the same baryonic mass can host stronger magnetic fields. For configurations with purely toroidal magnetic fields we also showed that as the magnetisation rises the models de-scalarise. This effect was evaluated for various values of $\beta_0$ showing that there is a strong dependency.

We have then shown, using various parametrisations, how the mass-density relation changes with the magnetisation of the system, revealing both how this affects the region of spontaneous scalarisation and the location of the configuration with maximum mass, together with its value. In particular, we have shown that while for toroidal magnetic fields there is a de-scalarised region, for purely poloidal magnetic fields there is a limiting mass above which only scalarised solutions are possible. We have also shown that contrary to GR, where the maximum mass is  always an increasing function of magnetisation, in STTs, for purely toroidal magnetic fields, the maximum mass decreases with increasing magnetisation for systems with $B_\mathrm{max}$ lower than a threshold magnetic field, and then rises. We verified that the quadrupolar term arising from magnetic deformations in the source of the scalar field equation is of the same order of the one in Einstein's equations, suggesting comparable levels of gravitational losses in tensor and scalar waves.

In general, we found that for weakly magnetised models the presence of a scalar field dominates the properties of NSs, and its effect is to counter-balance the magnetic stresses, either by reducing the deformation, or leading to saturation of the values of the maximum mass.  We verified, by changing the value of $\beta_0$, that when scalarisation effects become smaller the typical trends of GR tend to be recovered, with the significative difference that while for purely toroidal fields a rise in magnetisation leads to de-scalarisation, for purely poloidal magnetic fields, on the contrary, it increases the total  scalar charge. Depending on its geometry, the magnetic field can either favour or suppress spontaneous scalarisation when $\beta_0$ is close to the threshold limit on the range of this effect.

Finally, we have also shown that the mutual interplay of a scalar and  toroidal magnetic field, in the presence of strong scalarisation effects, leads to unstable configurations and potentially to events of spontaneous scalarisation due to the loss of magnetic support - a `magnetically-induced spontaneous scalarisation'. 

This paper is mostly devoted to a global study of the properties of magnetised NSs in STT, with a particular focus on the comparison with their respective GR counterparts. For this reason, we adopted a simple polytropic EoS and considered only the two extreme cases of purely toroidal and purely poloidal magnetic fields, focusing the discussion on the case $\beta_0=-6$ to enhance and highlight the main differences. We plan to investigate in more detail, in a future work, how the deformability of NSs in STT depends on the choice of  $\beta_0$, and on the EoS \citep{pili_quark_2016}, and how it scales with the mass, radius, and compactness of NSs to see if it is possible to derive scaling laws that can parametrise the magnetic deformability, in a similar way to what has been previously done in GR \citep{pili_general_2017}.

We conclude by recalling that STTs are just a subset of a more extended class of alternative theories of gravity, TeVeS \citep{bekenstein_2004}, which predict also the possible existence of non-minimally coupled vector fields. As STTs, even theories with vector fields can present phenomena of `spontaneous vectorisation \citep{hellings_1973,heinsenberg_2014,kase_2018,kase_2020}. Interestingly the mathematics behind spontaneous vectorisation is not dissimilar to the one used to model non-linear current terms in magnetised NSs \citep{pili_axisymmetric_2014}, and spontaneous magnetic-vectorisation has already been treated and discussed within the framework of the standard techniques that we have illustrated here \citep{bucciantini_role_2015}. This shows that the algorithms and approaches we have introduced, even if developed in the context of the specific case of magnetic fields, have a far larger applicability to vector fields in general.

\begin{acknowledgements}
The authors acknowledge financial support from the ``Accordo Attuativo ASI-INAF n. 2017-14-H.0 Progetto: on the escape of cosmic rays and their impact on the background plasma'' and from the INFN Teongrav collaboration. We also thanks the referee for pointing us a few mistakes in the original text.
\end{acknowledgements}



\bibliographystyle{aa}
\bibliography{articolo.bib} 

\begin{thebibliography}{171}
\expandafter\ifx\csname natexlab\endcsname\relax\def\natexlab#1{#1}\fi

\bibitem[{Abbott {et~al.}(2016)Abbott, Abbott, Abbott, Abernathy, Acernese,
  Ackley, Adams, Adams, Addesso, Adhikari, Adya, Affeldt, Agathos, Agatsuma,
  Aggarwal, Aguiar, Aiello, Ain, Ajith, Allen, Allocca, Altin, Anderson,
  Anderson, Arai, Arain, Araya, Arceneaux, Areeda, Arnaud, Arun, Ascenzi,
  Ashton, Ast, Aston, Astone, Aufmuth, Aulbert, Babak, Bacon, Bader, Baker,
  Baldaccini, Ballardin, Ballmer, Barayoga, Barclay, Barish, Barker, Barone,
  Barr, Barsotti, Barsuglia, Barta, Bartlett, Barton, Bartos, Bassiri, Basti,
  Batch, Baune, Bavigadda, Bazzan, Behnke, Bejger, Belczynski, Bell, Bell,
  Berger, Bergman, Bergmann, Berry, Bersanetti, Bertolini, Betzwieser, Bhagwat,
  Bhandare, Bilenko, Billingsley, Birch, Birney, Birnholtz, Biscans, Bisht,
  Bitossi, Biwer, Bizouard, Blackburn, Blair, Blair, Blair, Bloemen, Bock,
  Bodiya, Boer, Bogaert, Bogan, Bohe, Bojtos, Bond, Bondu, Bonnand, Boom, Bork,
  Boschi, Bose, Bouffanais, Bozzi, Bradaschia, Brady, Braginsky, Branchesi,
  Brau, Briant, Brillet, Brinkmann, Brisson, Brockill, Brooks, Brown, Brown,
  Brown, Buchanan, Buikema, Bulik, Bulten, Buonanno, Buskulic, Buy, Byer,
  Cabero, Cadonati, Cagnoli, Cahillane, Bustillo, Callister, Calloni, Camp,
  Cannon, Cao, Capano, Capocasa, Carbognani, Caride, Diaz, Casentini, Caudill,
  Cavagli\`a, Cavalier, Cavalieri, Cella, Cepeda, Baiardi, Cerretani, Cesarini,
  Chakraborty, Chalermsongsak, Chamberlin, Chan, Chao, Charlton,
  Chassande-Mottin, Chen, Chen, Cheng, Chincarini, Chiummo, Cho, Cho, Chow,
  Christensen, Chu, Chua, Chung, Ciani, Clara, Clark, Cleva, Coccia, Cohadon,
  Colla, Collette, Cominsky, Constancio, Conte, Conti, Cook, Corbitt, Cornish,
  Corsi, Cortese, Costa, Coughlin, Coughlin, Coulon, Countryman, Couvares,
  Cowan, Coward, Cowart, Coyne, Coyne, Craig, Creighton, Creighton, Cripe,
  Crowder, Cruise, Cumming, Cunningham, Cuoco, Canton, Danilishin, D'Antonio,
  Danzmann, Darman, Da~Silva~Costa, Dattilo, Dave, Daveloza, Davier, Davies,
  Daw, Day, De, DeBra, Debreczeni, Degallaix, De~Laurentis, Del\'eglise,
  Del~Pozzo, Denker, Dent, Dereli, Dergachev, DeRosa, De~Rosa, DeSalvo,
  Dhurandhar, D\'{\i}az, Di~Fiore, Di~Giovanni, Di~Lieto, Di~Pace, Di~Palma,
  Di~Virgilio, Dojcinoski, Dolique, Donovan, Dooley, Doravari, Douglas, Downes,
  Drago, Drever, Driggers, Du, Ducrot, Dwyer, Edo, Edwards, Effler, Eggenstein,
  Ehrens, Eichholz, Eikenberry, Engels, Essick, Etzel, Evans, Evans, Everett,
  Factourovich, Fafone, Fair, Fairhurst, Fan, Fang, Farinon, Farr, Farr,
  Favata, Fays, Fehrmann, Fejer, Feldbaum, Ferrante, Ferreira, Ferrini,
  Fidecaro, Finn, Fiori, Fiorucci, Fisher, Flaminio, Fletcher, Fong, Fournier,
  Franco, Frasca, Frasconi, Frede, Frei, Freise, Frey, Frey, Fricke, Fritschel,
  Frolov, Fulda, Fyffe, Gabbard, Gair, Gammaitoni, Gaonkar, Garufi, Gatto,
  Gaur, Gehrels, Gemme, Gendre, Genin, Gennai, George, Gergely, Germain, Ghosh,
  Ghosh, Ghosh, Giaime, Giardina, Giazotto, Gill, Glaefke, Gleason, Goetz,
  Goetz, Gondan, Gonz\'alez, Castro, Gopakumar, Gordon, Gorodetsky, Gossan,
  Gosselin, Gouaty, Graef, Graff, Granata, Grant, Gras, Gray, Greco, Green,
  Greenhalgh, Groot, Grote, Grunewald, Guidi, Guo, Gupta, Gupta, Gushwa,
  Gustafson, Gustafson, Hacker, Hall, Hall, Hammond, Haney, Hanke, Hanks,
  Hanna, Hannam, Hanson, Hardwick, Harms, Harry, Harry, Hart, Hartman, Haster,
  Haughian, Healy, Heefner, Heidmann, Heintze, Heinzel, Heitmann, Hello,
  Hemming, Hendry, Heng, Hennig, Heptonstall, Heurs, Hild, Hoak, Hodge, Hofman,
  Hollitt, Holt, Holz, Hopkins, Hosken, Hough, Houston, Howell, Hu, Huang,
  Huerta, Huet, Hughey, Husa, Huttner, Huynh-Dinh, Idrisy, Indik, Ingram, Inta,
  Isa, Isac, Isi, Islas, Isogai, Iyer, Izumi, Jacobson, Jacqmin, Jang, Jani,
  Jaranowski, Jawahar, Jim\'enez-Forteza, Johnson, Johnson-McDaniel, Jones,
  Jones, Jonker, Ju, Haris, Kalaghatgi, Kalogera, Kandhasamy, Kang, Kanner,
  Karki, Kasprzack, Katsavounidis, Katzman, Kaufer, Kaur, Kawabe, Kawazoe,
  K\'ef\'elian, Kehl, Keitel, Kelley, Kells, Kennedy, Keppel, Key,
  Khalaidovski, Khalili, Khan, Khan, Khan, Khazanov, Kijbunchoo, Kim, Kim, Kim,
  Kim, Kim, Kim, King, King, Kinzel, Kissel, Kleybolte, Klimenko, Koehlenbeck,
  Kokeyama, Koley, Kondrashov, Kontos, Koranda, Korobko, Korth, Kowalska,
  Kozak, Kringel, Krishnan, Kr\'olak, Krueger, Kuehn, Kumar, Kumar, Kuo,
  Kutynia, Kwee, Lackey, Landry, Lange, Lantz, Lasky, Lazzarini, Lazzaro,
  Leaci, Leavey, Lebigot, Lee, Lee, Lee, Lee, Lenon, Leonardi, Leong, Leroy,
  Letendre, Levin, Levine, Li, Libson, Littenberg, Lockerbie, Logue, Lombardi,
  London, Lord, Lorenzini, Loriette, Lormand, Losurdo, Lough, Lousto, Lovelace,
  L\"uck, Lundgren, Luo, Lynch, Ma, MacDonald, Machenschalk, MacInnis, Macleod,
  Maga\~na Sandoval, Magee, Mageswaran, Majorana, Maksimovic, Malvezzi, Man,
  Mandel, Mandic, Mangano, Mansell, Manske, Mantovani, Marchesoni, Marion,
  M\'arka, M\'arka, Markosyan, Maros, Martelli, Martellini, Martin, Martin,
  Martynov, Marx, Mason, Masserot, Massinger, Masso-Reid, Matichard, Matone,
  Mavalvala, Mazumder, Mazzolo, McCarthy, McClelland, McCormick, McGuire,
  McIntyre, McIver, McManus, McWilliams, Meacher, Meadors, Meidam, Melatos,
  Mendell, Mendoza-Gandara, Mercer, Merilh, Merzougui, Meshkov, Messenger,
  Messick, Meyers, Mezzani, Miao, Michel, Middleton, Mikhailov, Milano, Miller,
  Millhouse, Minenkov, Ming, Mirshekari, Mishra, Mitra, Mitrofanov,
  Mitselmakher, Mittleman, Moggi, Mohan, Mohapatra, Montani, Moore, Moore,
  Moraru, Moreno, Morriss, Mossavi, Mours, Mow-Lowry, Mueller, Mueller, Muir,
  Mukherjee, Mukherjee, Mukherjee, Mukund, Mullavey, Munch, Murphy, Murray,
  Mytidis, Nardecchia, Naticchioni, Nayak, Necula, Nedkova, Nelemans, Neri,
  Neunzert, Newton, Nguyen, Nielsen, Nissanke, Nitz, Nocera, Nolting,
  Normandin, Nuttall, Oberling, Ochsner, O'Dell, Oelker, Ogin, Oh, Oh, Ohme,
  Oliver, Oppermann, Oram, O'Reilly, O'Shaughnessy, Ott, Ottaway, Ottens,
  Overmier, Owen, Pai, Pai, Palamos, Palashov, Palomba, Pal-Singh, Pan, Pan,
  Pankow, Pannarale, Pant, Paoletti, Paoli, Papa, Paris, Parker, Pascucci,
  Pasqualetti, Passaquieti, Passuello, Patricelli, Patrick, Pearlstone,
  Pedraza, Pedurand, Pekowsky, Pele, Penn, Perreca, Pfeiffer, Phelps, Piccinni,
  Pichot, Pickenpack, Piergiovanni, Pierro, Pillant, Pinard, Pinto, Pitkin,
  Poeld, Poggiani, Popolizio, Post, Powell, Prasad, Predoi, Premachandra,
  Prestegard, Price, Prijatelj, Principe, Privitera, Prix, Prodi, Prokhorov,
  Puncken, Punturo, Puppo, P\"urrer, Qi, Qin, Quetschke, Quintero,
  Quitzow-James, Raab, Rabeling, Radkins, Raffai, Raja, Rakhmanov, Ramet,
  Rapagnani, Raymond, Razzano, Re, Read, Reed, Regimbau, Rei, Reid, Reitze,
  Rew, Reyes, Ricci, Riles, Robertson, Robie, Robinet, Rocchi, Rolland,
  Rollins, Roma, Romano, Romano, Romanov, Romie, Rosi\ifmmode~\acute{n}\else
  \'{n}\fi{}ska, Rowan, R\"udiger, Ruggi, Ryan, Sachdev, Sadecki, Sadeghian,
  Salconi, Saleem, Salemi, Samajdar, Sammut, Sampson, Sanchez, Sandberg,
  Sandeen, Sanders, Sanders, Sassolas, Sathyaprakash, Saulson, Sauter, Savage,
  Sawadsky, Schale, Schilling, Schmidt, Schmidt, Schnabel, Schofield,
  Sch\"onbeck, Schreiber, Schuette, Schutz, Scott, Scott, Sellers, Sengupta,
  Sentenac, Sequino, Sergeev, Serna, Setyawati, Sevigny, Shaddock, Shaffer,
  Shah, Shahriar, Shaltev, Shao, Shapiro, Shawhan, Sheperd, Shoemaker,
  Shoemaker, Siellez, Siemens, Sigg, Silva, Simakov, Singer, Singer, Singh,
  Singh, Singhal, Sintes, Slagmolen, Smith, Smith, Smith, Smith, Son, Sorazu,
  Sorrentino, Souradeep, Srivastava, Staley, Steinke, Steinlechner,
  Steinlechner, Steinmeyer, Stephens, Stevenson, Stone, Strain, Straniero,
  Stratta, Strauss, Strigin, Sturani, Stuver, Summerscales, Sun, Sutton,
  Swinkels, Szczepa\ifmmode~\acute{n}\else \'{n}\fi{}czyk, Tacca, Talukder,
  Tanner, T\'apai, Tarabrin, Taracchini, Taylor, Theeg, Thirugnanasambandam,
  Thomas, Thomas, Thomas, Thorne, Thorne, Thrane, Tiwari, Tiwari, Tokmakov,
  Tomlinson, Tonelli, Torres, Torrie, T\"oyr\"a, Travasso, Traylor, Trifir\`o,
  Tringali, Trozzo, Tse, Turconi, Tuyenbayev, Ugolini, Unnikrishnan, Urban,
  Usman, Vahlbruch, Vajente, Valdes, Vallisneri, van Bakel, van Beuzekom,
  van~den Brand, Van Den~Broeck, Vander-Hyde, van~der Schaaf, van Heijningen,
  van Veggel, Vardaro, Vass, Vas\'uth, Vaulin, Vecchio, Vedovato, Veitch,
  Veitch, Venkateswara, Verkindt, Vetrano, Vicer\'e, Vinciguerra, Vine, Vinet,
  Vitale, Vo, Vocca, Vorvick, Voss, Vousden, Vyatchanin, Wade, Wade, Wade,
  Waldman, Walker, Wallace, Walsh, Wang, Wang, Wang, Wang, Wang, Ward, Ward,
  Warner, Was, Weaver, Wei, Weinert, Weinstein, Weiss, Welborn, Wen,
  We\ss{}els, Westphal, Wette, Whelan, Whitcomb, White, Whiting, Wiesner,
  Wilkinson, Willems, Williams, Williams, Williamson, Willis, Willke, Wimmer,
  Winkelmann, Winkler, Wipf, Wiseman, Wittel, Woan, Worden, Wright, Wu, Yablon,
  Yakushin, Yam, Yamamoto, Yancey, Yap, Yu, Yvert, Zadro\ifmmode~\dot{z}\else
  \.{z}\fi{}ny, Zangrando, Zanolin, Zendri, Zevin, Zhang, Zhang, Zhang, Zhang,
  Zhao, Zhou, Zhou, Zhu, Zucker, Zuraw, \& Zweizig}]{abbott_bbh_2016}
Abbott, B.~P., Abbott, R., Abbott, T.~D., {et~al.} 2016, \prl, 116, 061102

\bibitem[{Abbott {et~al.}(2017{\natexlab{a}})Abbott, Abbott, Abbott, Acernese,
  Ackley, Adams, Adams, Addesso, Adhikari, Adya, Affeldt, Afrough, Agarwal,
  Agathos, Agatsuma, Aggarwal, Aguiar, Aiello, Ain, Ajith, Allen, Allen,
  Allocca, Altin, Amato, Ananyeva, Anderson, Anderson, Angelova, Antier,
  Appert, Arai, Araya, Areeda, Arnaud, Arun, Ascenzi, Ashton, Ast, Aston,
  Astone, Atallah, Aufmuth, Aulbert, AultONeal, Austin, Avila-Alvarez, Babak,
  Bacon, Bader, Bae, Bailes, Baker, Baldaccini, Ballardin, Ballmer, Banagiri,
  Barayoga, Barclay, Barish, Barker, Barkett, Barone, Barr, Barsotti,
  Barsuglia, Barta, Barthelmy, Bartlett, Bartos, Bassiri, Basti, Batch, Bawaj,
  Bayley, Bazzan, Bécsy, Beer, Bejger, Belahcene, Bell, Berger, Bergmann,
  Bernuzzi, Bero, Berry, Bersanetti, Bertolini, Betzwieser, Bhagwat, Bhandare,
  Bilenko, Billingsley, Billman, Birch, Birney, Birnholtz, Biscans, Biscoveanu,
  Bisht, Bitossi, Biwer, Bizouard, Blackburn, Blackman, Blair, Blair, Blair,
  Bloemen, Bock, Bode, Boer, Bogaert, Bohe, Bondu, Bonilla, Bonnand, Boom,
  Bork, Boschi, Bose, Bossie, Bouffanais, Bozzi, Bradaschia, Brady, Branchesi,
  Brau, Briant, Brillet, Brinkmann, Brisson, Brockill, Broida, Brooks, Brown,
  Brown, Brunett, Buchanan, Buikema, Bulik, Bulten, Buonanno, Buskulic, Buy,
  Byer, Cabero, Cadonati, Cagnoli, Cahillane, Calderón~Bustillo, Callister,
  Calloni, Camp, Canepa, Canizares, Cannon, Cao, Cao, Capano, Capocasa,
  Carbognani, Caride, Carney, Carullo, Casanueva~Diaz, Casentini, Caudill,
  Cavaglià, Cavalier, Cavalieri, Cella, Cepeda, Cerdá-Durán, Cerretani,
  Cesarini, Chamberlin, Chan, Chao, Charlton, Chase, Chassande-Mottin,
  Chatterjee, Chatziioannou, Cheeseboro, Chen, Chen, Chen, Cheng, Chia,
  Chincarini, Chiummo, Chmiel, Cho, Cho, Chow, Christensen, Chu, Chua, Chua,
  Chung, Chung, Ciani, Ciolfi, Cirelli, Cirone, Clara, Clark, Clearwater,
  Cleva, Cocchieri, Coccia, Cohadon, Cohen, Colla, Collette, Cominsky,
  Constancio, Conti, Cooper, Corban, Corbitt, Cordero-Carrión, Corley,
  Cornish, Corsi, Cortese, Costa, Coughlin, Coughlin, Coulon, Countryman,
  Couvares, Covas, Cowan, Coward, Cowart, Coyne, Coyne, Creighton, Creighton,
  Cripe, Crowder, Cullen, Cumming, Cunningham, Cuoco, Dal~Canton, Dálya,
  Danilishin, D'Antonio, Danzmann, Dasgupta, Da~Silva~Costa, Dattilo, Dave,
  Davier, Davis, Daw, Day, De, DeBra, Degallaix, De~Laurentis, Deléglise,
  Del~Pozzo, Demos, Denker, Dent, De~Pietri, Dergachev, De~Rosa, DeRosa,
  De~Rossi, DeSalvo, de~Varona, Devenson, Dhurandhar, Díaz, Dietrich,
  Di~Fiore, Di~Giovanni, Di~Girolamo, Di~Lieto, Di~Pace, Di~Palma, Di~Renzo,
  Doctor, Dolique, Donovan, Dooley, Doravari, Dorrington, Douglas,
  Dovale~Álvarez, Downes, Drago, Dreissigacker, Driggers, Du, Ducrot, Dudi,
  Dupej, Dwyer, Edo, Edwards, Effler, Eggenstein, Ehrens, Eichholz, Eikenberry,
  Eisenstein, Essick, Estevez, Etienne, Etzel, Evans, Evans, Factourovich,
  Fafone, Fair, Fairhurst, Fan, Farinon, Farr, Farr, Fauchon-Jones, Favata,
  Fays, Fee, Fehrmann, Feicht, Fejer, Fernandez-Galiana, Ferrante, Ferreira,
  Ferrini, Fidecaro, Finstad, Fiori, Fiorucci, Fishbach, Fisher, Fitz-Axen,
  Flaminio, Fletcher, Fong, Font, Forsyth, Forsyth, Fournier, Frasca, Frasconi,
  Frei, Freise, Frey, Frey, Fries, Fritschel, Frolov, Fulda, Fyffe, Gabbard,
  Gadre, Gaebel, Gair, Gammaitoni, Ganija, Gaonkar, Garcia-Quiros, Garufi,
  Gateley, Gaudio, Gaur, Gayathri, Gehrels, Gemme, Genin, Gennai, George,
  George, Gergely, Germain, Ghonge, Ghosh, Ghosh, Ghosh, Giaime, Giardina,
  Giazotto, Gill, Glover, Goetz, Goetz, Gomes, Goncharov, González,
  Gonzalez~Castro, Gopakumar, Gorodetsky, Gossan, Gosselin, Gouaty, Grado,
  Graef, Granata, Grant, Gras, Gray, Greco, Green, Gretarsson, Groot, Grote,
  Grunewald, Gruning, Guidi, Guo, Gupta, Gupta, Gushwa, Gustafson, Gustafson,
  Halim, Hall, Hall, Hamilton, Hammond, Haney, Hanke, Hanks, Hanna, Hannam,
  Hannuksela, Hanson, Hardwick, Harms, Harry, Harry, Hart, Haster, Haughian,
  Healy, Heidmann, Heintze, Heitmann, Hello, Hemming, Hendry, Heng, Hennig,
  Heptonstall, Heurs, Hild, Hinderer, Ho, Hoak, Hofman, Holt, Holz, Hopkins,
  Horst, Hough, Houston, Howell, Hreibi, Hu, Huerta, Huet, Hughey, Husa,
  Huttner, Huynh-Dinh, Indik, Inta, Intini, Isa, Isac, Isi, Iyer, Izumi,
  Jacqmin, Jani, Jaranowski, Jawahar, Jiménez-Forteza, Johnson,
  Johnson-McDaniel, Jones, Jones, Jonker, Ju, Junker, Kalaghatgi, Kalogera,
  Kamai, Kandhasamy, Kang, Kanner, Kapadia, Karki, Karvinen, Kasprzack,
  Kastaun, Katolik, Katsavounidis, Katzman, Kaufer, Kawabe, Kéfélian, Keitel,
  Kemball, Kennedy, Kent, Key, Khalili, Khan, Khan, Khan, Khazanov, Kijbunchoo,
  Kim, Kim, Kim, Kim, Kim, Kim, Kimbrell, King, King, Kinley-Hanlon, Kirchhoff,
  Kissel, Kleybolte, Klimenko, Knowles, Koch, Koehlenbeck, Koley, Kondrashov,
  Kontos, Korobko, Korth, Kowalska, Kozak, Krämer, Kringel, Krishnan, Królak,
  Kuehn, Kumar, Kumar, Kumar, Kuo, Kutynia, Kwang, Lackey, Lai, Landry, Lang,
  Lange, Lantz, Lanza, Larson, Lartaux-Vollard, Lasky, Laxen, Lazzarini,
  Lazzaro, Leaci, Leavey, Lee, Lee, Lee, Lee, Lee, Lehmann, Lenon, Leon,
  Leonardi, Leroy, Letendre, Levin, Li, Linker, Littenberg, Liu, Liu, Lo,
  Lockerbie, London, Lord, Lorenzini, Loriette, Lormand, Losurdo, Lough,
  Lousto, Lovelace, Lück, Lumaca, Lundgren, Lynch, Ma, Macas, Macfoy,
  Machenschalk, MacInnis, Macleod, Magaña~Hernandez, Magaña-Sandoval,
  Magaña~Zertuche, Magee, Majorana, Maksimovic, Man, Mandic, Mangano, Mansell,
  Manske, Mantovani, Marchesoni, Marion, Márka, Márka, Markakis, Markosyan,
  Markowitz, Maros, Marquina, Marsh, Martelli, Martellini, Martin, Martin,
  Martynov, Marx, Mason, Massera, Masserot, Massinger, Masso-Reid,
  Mastrogiovanni, Matas, Matichard, Matone, Mavalvala, Mazumder, McCarthy,
  McClelland, McCormick, McCuller, McGuire, McIntyre, McIver, McManus, McNeill,
  McRae, McWilliams, Meacher, Meadors, Mehmet, Meidam, Mejuto-Villa, Melatos,
  Mendell, Mercer, Merilh, Merzougui, Meshkov, Messenger, Messick, Metzdorff,
  Meyers, Miao, Michel, Middleton, Mikhailov, Milano, Miller, Miller, Miller,
  Millhouse, Milovich-Goff, Minazzoli, Minenkov, Ming, Mishra, Mitra,
  Mitrofanov, Mitselmakher, Mittleman, Moffa, Moggi, Mogushi, Mohan, Mohapatra,
  Molina, Montani, Moore, Moraru, Moreno, Morisaki, Morriss, Mours, Mow-Lowry,
  Mueller, Muir, Mukherjee, Mukherjee, Mukherjee, Mukund, Mullavey, Munch,
  Muñiz, Muratore, Murray, Nagar, Napier, Nardecchia, Naticchioni, Nayak,
  Neilson, Nelemans, Nelson, Nery, Neunzert, Nevin, Newport, Newton, Ng,
  Nguyen, Nguyen, Nichols, Nielsen, Nissanke, Nitz, Noack, Nocera, Nolting,
  North, Nuttall, Oberling, O'Dea, Ogin, Oh, Oh, Ohme, Okada, Oliver,
  Oppermann, Oram, O'Reilly, Ormiston, Ortega, O'Shaughnessy, Ossokine,
  Ottaway, Overmier, Owen, Pace, Page, Page, Pai, Pai, Palamos, Palashov,
  Palomba, Pal-Singh, Pan, Pan, Pang, Pang, Pankow, Pannarale, Pant, Paoletti,
  Paoli, Papa, Parida, Parker, Pascucci, Pasqualetti, Passaquieti, Passuello,
  Patil, Patricelli, Pearlstone, Pedraza, Pedurand, Pekowsky, Pele, Penn,
  Perez, Perreca, Perri, Pfeiffer, Phelps, Piccinni, Pichot, Piergiovanni,
  Pierro, Pillant, Pinard, Pinto, Pirello, Pitkin, Poe, Poggiani, Popolizio,
  Porter, Post, Powell, Prasad, Pratt, Pratten, Predoi, Prestegard, Prijatelj,
  Principe, Privitera, Prix, Prodi, Prokhorov, Puncken, Punturo, Puppo,
  Pürrer, Qi, Quetschke, Quintero, Quitzow-James, Raab, Rabeling, Radkins,
  Raffai, Raja, Rajan, Rajbhandari, Rakhmanov, Ramirez, Ramos-Buades,
  Rapagnani, Raymond, Razzano, Read, Regimbau, Rei, Reid, Reitze, Ren, Reyes,
  Ricci, Ricker, Rieger, Riles, Rizzo, Robertson, Robie, Robinet, Rocchi,
  Rolland, Rollins, Roma, Romano, Romano, Romel, Romie, Rosińska, Ross, Rowan,
  Rüdiger, Ruggi, Rutins, Ryan, Sachdev, Sadecki, Sadeghian, Sakellariadou,
  Salconi, Saleem, Salemi, Samajdar, Sammut, Sampson, Sanchez, Sanchez,
  Sanchis-Gual, Sandberg, Sanders, Sassolas, Sathyaprakash, Saulson, Sauter,
  Savage, Sawadsky, Schale, Scheel, Scheuer, Schmidt, Schmidt, Schnabel,
  Schofield, Schönbeck, Schreiber, Schuette, Schulte, Schutz, Schwalbe, Scott,
  Scott, Seidel, Sellers, Sengupta, Sentenac, Sequino, Sergeev, Shaddock,
  Shaffer, Shah, Shahriar, Shaner, Shao, Shapiro, Shawhan, Sheperd, Shoemaker,
  Shoemaker, Siellez, Siemens, Sieniawska, Sigg, Silva, Singer, Singh, Singhal,
  Sintes, Slagmolen, Smith, Smith, Smith, Somala, Son, Sonnenberg, Sorazu,
  Sorrentino, Souradeep, Spencer, Srivastava, Staats, Staley, Steinke,
  Steinlechner, Steinlechner, Steinmeyer, Stevenson, Stone, Stops, Strain,
  Stratta, Strigin, Strunk, Sturani, Stuver, Summerscales, Sun, Sunil, Suresh,
  Sutton, Swinkels, Szczepańczyk, Tacca, Tait, Talbot, Talukder, Tanner,
  Tápai, Taracchini, Tasson, Taylor, Taylor, Tewari, Theeg, Thies, Thomas,
  Thomas, Thomas, Thorne, Thorne, Thrane, Tiwari, Tiwari, Tokmakov, Toland,
  Tonelli, Tornasi, Torres-Forné, Torrie, Töyrä, Travasso, Traylor,
  Trinastic, Tringali, Trozzo, Tsang, Tse, Tso, Tsukada, Tsuna, Tuyenbayev,
  Ueno, Ugolini, Unnikrishnan, Urban, Usman, Vahlbruch, Vajente, Valdes,
  Vallisneri, van Bakel, van Beuzekom, van~den Brand, Van Den~Broeck,
  Vander-Hyde, van~der Schaaf, van Heijningen, van Veggel, Vardaro, Varma,
  Vass, Vasúth, Vecchio, Vedovato, Veitch, Veitch, Venkateswara, Venugopalan,
  Verkindt, Vetrano, Viceré, Viets, Vinciguerra, Vine, Vinet, Vitale, Vo,
  Vocca, Vorvick, Vyatchanin, Wade, Wade, Wade, Walet, Walker, Wallace, Walsh,
  Wang, Wang, Wang, Wang, Wang, Ward, Warner, Was, Watchi, Weaver, Wei,
  Weinert, Weinstein, Weiss, Wen, Wessel, Weßels, Westerweck, Westphal, Wette,
  Whelan, Whitcomb, Whiting, Whittle, Wilken, Williams, Williams, Williamson,
  Willis, Willke, Wimmer, Winkler, Wipf, Wittel, Woan, Woehler, Wofford, Wong,
  Worden, Wright, Wu, Wysocki, Xiao, Yamamoto, Yancey, Yang, Yap, Yazback, Yu,
  Yu, Yvert, ZadroŻny, Zanolin, Zelenova, Zendri, Zevin, Zhang, Zhang, Zhang,
  Zhang, Zhao, Zhou, Zhou, Zhu, Zhu, Zimmerman, Zucker, Zweizig, Collaboration,
  \& Collaboration}]{abbott_gw170817:_2017}
Abbott, B.~P., Abbott, R., Abbott, T.~D., {et~al.} 2017{\natexlab{a}}, \prl,
  119, 161101

\bibitem[{Abbott {et~al.}(2017{\natexlab{b}})Abbott, Abbott, Abbott, Acernese,
  Ackley, Adams, Adams, Addesso, Adhikari, Adya, Affeldt, Afrough, Agarwal,
  Agathos, Agatsuma, Aggarwal, Aguiar, Aiello, Ain, Ajith, Allen, Allen,
  Allocca, Altin, Amato, Ananyeva, Anderson, Anderson, Angelova, Antier,
  Appert, Arai, Araya, Areeda, Arnaud, Arun, Ascenzi, Ashton, Ast, Aston,
  Astone, Atallah, Aufmuth, Aulbert, AultONeal, Austin, Avila-Alvarez, Babak,
  Bacon, Bader, Bae, Baker, Baldaccini, Ballardin, Ballmer, Banagiri, Barayoga,
  Barclay, Barish, Barker, Barkett, Barone, Barr, Barsotti, Barsuglia, Barta,
  Barthelmy, Bartlett, Bartos, Bassiri, Basti, Batch, Bawaj, Bayley, Bazzan,
  Bécsy, Beer, Bejger, Belahcene, Bell, Berger, Bergmann, Bero, Berry,
  Bersanetti, Bertolini, Betzwieser, Bhagwat, Bhandare, Bilenko, Billingsley,
  Billman, Birch, Birney, Birnholtz, Biscans, Biscoveanu, Bisht, Bitossi,
  Biwer, Bizouard, Blackburn, Blackman, Blair, Blair, Blair, Bloemen, Bock,
  Bode, Boer, Bogaert, Bohe, Bondu, Bonilla, Bonnand, Boom, Bork, Boschi, Bose,
  Bossie, Bouffanais, Bozzi, Bradaschia, Brady, Branchesi, Brau, Briant,
  Brillet, Brinkmann, Brisson, Brockill, Broida, Brooks, Brown, Brown, Brunett,
  Buchanan, Buikema, Bulik, Bulten, Buonanno, Buskulic, Buy, Byer, Cabero,
  Cadonati, Cagnoli, Cahillane, Calderón~Bustillo, Callister, Calloni, Camp,
  Canepa, Canizares, Cannon, Cao, Cao, Capano, Capocasa, Carbognani, Caride,
  Carney, Casanueva~Diaz, Casentini, Caudill, Cavaglià, Cavalier, Cavalieri,
  Cella, Cepeda, Cerdá-Durán, Cerretani, Cesarini, Chamberlin, Chan, Chao,
  Charlton, Chase, Chassande-Mottin, Chatterjee, Chatziioannou, Cheeseboro,
  Chen, Chen, Chen, Cheng, Chia, Chincarini, Chiummo, Chmiel, Cho, Cho, Chow,
  Christensen, Chu, Chua, Chua, Chung, Chung, Ciani, Ciolfi, Cirelli, Cirone,
  Clara, Clark, Clearwater, Cleva, Cocchieri, Coccia, Cohadon, Cohen, Colla,
  Collette, Cominsky, Constancio, Conti, Cooper, Corban, Corbitt,
  Cordero-Carrión, Corley, Cornish, Corsi, Cortese, Costa, Coughlin, Coughlin,
  Coulon, Countryman, Couvares, Covas, Cowan, Coward, Cowart, Coyne, Coyne,
  Creighton, Creighton, Cripe, Crowder, Cullen, Cumming, Cunningham, Cuoco,
  Dal~Canton, Dálya, Danilishin, D'Antonio, Danzmann, Dasgupta,
  Da~Silva~Costa, Dattilo, Dave, Davier, Davis, Daw, Day, De, DeBra, Degallaix,
  De~Laurentis, Deléglise, Del~Pozzo, Demos, Denker, Dent, De~Pietri,
  Dergachev, De~Rosa, DeRosa, De~Rossi, DeSalvo, de~Varona, Devenson,
  Dhurandhar, Díaz, Di~Fiore, Di~Giovanni, Di~Girolamo, Di~Lieto, Di~Pace,
  Di~Palma, Di~Renzo, Doctor, Dolique, Donovan, Dooley, Doravari, Dorrington,
  Douglas, Dovale~Álvarez, Downes, Drago, Dreissigacker, Driggers, Du, Ducrot,
  Dupej, Dwyer, Edo, Edwards, Effler, Ehrens, Eichholz, Eikenberry, Eisenstein,
  Essick, Estevez, Etienne, Etzel, Evans, Evans, Factourovich, Fafone, Fair,
  Fairhurst, Fan, Farinon, Farr, Farr, Fauchon-Jones, Favata, Fays, Fee,
  Fehrmann, Feicht, Fejer, Fernandez-Galiana, Ferrante, Ferreira, Ferrini,
  Fidecaro, Finstad, Fiori, Fiorucci, Fishbach, Fisher, Fitz-Axen, Flaminio,
  Fletcher, Fong, Font, Forsyth, Forsyth, Fournier, Frasca, Frasconi, Frei,
  Freise, Frey, Frey, Fries, Fritschel, Frolov, Fulda, Fyffe, Gabbard, Gadre,
  Gaebel, Gair, Gammaitoni, Ganija, Gaonkar, Garcia-Quiros, Garufi, Gateley,
  Gaudio, Gaur, Gayathri, Gehrels, Gemme, Genin, Gennai, George, George,
  Gergely, Germain, Ghonge, Ghosh, Ghosh, Ghosh, Giaime, Giardina, Giazotto,
  Gill, Glover, Goetz, Goetz, Gomes, Goncharov, González, Gonzalez~Castro,
  Gopakumar, Gorodetsky, Gossan, Gosselin, Gouaty, Grado, Graef, Granata,
  Grant, Gras, Gray, Greco, Green, Gretarsson, Griswold, Groot, Grote,
  Grunewald, Gruning, Guidi, Guo, Gupta, Gupta, Gushwa, Gustafson, Gustafson,
  Halim, Hall, Hall, Hamilton, Hammond, Haney, Hanke, Hanks, Hanna, Hannam,
  Hannuksela, Hanson, Hardwick, Harms, Harry, Harry, Hart, Haster, Haughian,
  Healy, Heidmann, Heintze, Heitmann, Hello, Hemming, Hendry, Heng, Hennig,
  Heptonstall, Heurs, Hild, Hinderer, Hoak, Hofman, Holt, Holz, Hopkins, Horst,
  Hough, Houston, Howell, Hreibi, Hu, Huerta, Huet, Hughey, Husa, Huttner,
  Huynh-Dinh, Indik, Inta, Intini, Isa, Isac, Isi, Iyer, Izumi, Jacqmin, Jani,
  Jaranowski, Jawahar, Jiménez-Forteza, Johnson, Jones, Jones, Jonker, Ju,
  Junker, Kalaghatgi, Kalogera, Kamai, Kandhasamy, Kang, Kanner, Kapadia,
  Karki, Karvinen, Kasprzack, Katolik, Katsavounidis, Katzman, Kaufer, Kawabe,
  Kéfélian, Keitel, Kemball, Kennedy, Kent, Key, Khalili, Khan, Khan, Khan,
  Khazanov, Kijbunchoo, Kim, Kim, Kim, Kim, Kim, Kim, Kimbrell, King, King,
  Kinley-Hanlon, Kirchhoff, Kissel, Kleybolte, Klimenko, Knowles, Koch,
  Koehlenbeck, Koley, Kondrashov, Kontos, Korobko, Korth, Kowalska, Kozak,
  Krämer, Kringel, Krishnan, Królak, Kuehn, Kumar, Kumar, Kumar, Kuo,
  Kutynia, Kwang, Lackey, Lai, Landry, Lang, Lange, Lantz, Lanza, Larson,
  Lartaux-Vollard, Lasky, Laxen, Lazzarini, Lazzaro, Leaci, Leavey, Lee, Lee,
  Lee, Lee, Lee, Lehmann, Lenon, Leonardi, Leroy, Letendre, Levin, Li, Linker,
  Littenberg, Liu, Lo, Lockerbie, London, Lord, Lorenzini, Loriette, Lormand,
  Losurdo, Lough, Lousto, Lovelace, Lück, Lumaca, Lundgren, Lynch, Ma, Macas,
  Macfoy, Machenschalk, MacInnis, Macleod, Magaña~Hernandez, Magaña-Sandoval,
  Magaña~Zertuche, Magee, Majorana, Maksimovic, Man, Mandic, Mangano, Mansell,
  Manske, Mantovani, Marchesoni, Marion, Márka, Márka, Markakis, Markosyan,
  Markowitz, Maros, Marquina, Marsh, Martelli, Martellini, Martin, Martin,
  Martynov, Mason, Massera, Masserot, Massinger, Masso-Reid, Mastrogiovanni,
  Matas, Matichard, Matone, Mavalvala, Mazumder, McCarthy, McClelland,
  McCormick, McCuller, McGuire, McIntyre, McIver, McManus, McNeill, McRae,
  McWilliams, Meacher, Meadors, Mehmet, Meidam, Mejuto-Villa, Melatos, Mendell,
  Mercer, Merilh, Merzougui, Meshkov, Messenger, Messick, Metzdorff, Meyers,
  Miao, Michel, Middleton, Mikhailov, Milano, Miller, Miller, Miller,
  Millhouse, Milovich-Goff, Minazzoli, Minenkov, Ming, Mishra, Mitra,
  Mitrofanov, Mitselmakher, Mittleman, Moffa, Moggi, Mogushi, Mohan, Mohapatra,
  Montani, Moore, Moraru, Moreno, Morriss, Mours, Mow-Lowry, Mueller, Muir,
  Mukherjee, Mukherjee, Mukherjee, Mukund, Mullavey, Munch, Muñiz, Muratore,
  Murray, Napier, Nardecchia, Naticchioni, Nayak, Neilson, Nelemans, Nelson,
  Nery, Neunzert, Nevin, Newport, Newton, Ng, Nguyen, Nguyen, Nichols, Nielsen,
  Nissanke, Nitz, Noack, Nocera, Nolting, North, Nuttall, Oberling, O'Dea,
  Ogin, Oh, Oh, Ohme, Okada, Oliver, Oppermann, Oram, O'Reilly, Ormiston,
  Ortega, O'Shaughnessy, Ossokine, Ottaway, Overmier, Owen, Pace, Page, Page,
  Pai, Pai, Palamos, Palashov, Palomba, Pal-Singh, Pan, Pan, Pang, Pang,
  Pankow, Pannarale, Pant, Paoletti, Paoli, Papa, Parida, Parker, Pascucci,
  Pasqualetti, Passaquieti, Passuello, Patil, Patricelli, Pearlstone, Pedraza,
  Pedurand, Pekowsky, Pele, Penn, Perez, Perreca, Perri, Pfeiffer, Phelps,
  Piccinni, Pichot, Piergiovanni, Pierro, Pillant, Pinard, Pinto, Pirello,
  Pitkin, Poe, Poggiani, Popolizio, Porter, Post, Powell, Prasad, Pratt,
  Pratten, Predoi, Prestegard, Price, Prijatelj, Principe, Privitera, Prodi,
  Prokhorov, Puncken, Punturo, Puppo, Pürrer, Qi, Quetschke, Quintero,
  Quitzow-James, Raab, Rabeling, Radkins, Raffai, Raja, Rajan, Rajbhandari,
  Rakhmanov, Ramirez, Ramos-Buades, Rapagnani, Raymond, Razzano, Read,
  Regimbau, Rei, Reid, Reitze, Ren, Reyes, Ricci, Ricker, Rieger, Riles, Rizzo,
  Robertson, Robie, Robinet, Rocchi, Rolland, Rollins, Roma, Romano, Romel,
  Romie, Rosińska, Ross, Rowan, Rüdiger, Ruggi, Rutins, Ryan, Sachdev,
  Sadecki, Sadeghian, Sakellariadou, Salconi, Saleem, Salemi, Samajdar, Sammut,
  Sampson, Sanchez, Sanchez, Sanchis-Gual, Sandberg, Sanders, Sassolas,
  Sathyaprakash, Saulson, Sauter, Savage, Sawadsky, Schale, Scheel, Scheuer,
  Schmidt, Schmidt, Schnabel, Schofield, Schönbeck, Schreiber, Schuette,
  Schulte, Schutz, Schwalbe, Scott, Scott, Seidel, Sellers, Sengupta, Sentenac,
  Sequino, Sergeev, Shaddock, Shaffer, Shah, Shahriar, Shaner, Shao, Shapiro,
  Shawhan, Sheperd, Shoemaker, Shoemaker, Siellez, Siemens, Sieniawska, Sigg,
  Silva, Singer, Singh, Singhal, Sintes, Slagmolen, Smith, Smith, Smith,
  Somala, Son, Sonnenberg, Sorazu, Sorrentino, Souradeep, Spencer, Srivastava,
  Staats, Staley, Steinke, Steinlechner, Steinlechner, Steinmeyer, Stevenson,
  Stone, Stops, Strain, Stratta, Strigin, Strunk, Sturani, Stuver,
  Summerscales, Sun, Sunil, Suresh, Sutton, Swinkels, Szczepańczyk, Tacca,
  Tait, Talbot, Talukder, Tanner, Tápai, Taracchini, Tasson, Taylor, Taylor,
  Tewari, Theeg, Thies, Thomas, Thomas, Thomas, Thorne, Thorne, Thrane, Tiwari,
  Tiwari, Tokmakov, Toland, Tonelli, Tornasi, Torres-Forné, Torrie, Töyrä,
  Travasso, Traylor, Trinastic, Tringali, Trozzo, Tsang, Tse, Tso, Tsukada,
  Tsuna, Tuyenbayev, Ueno, Ugolini, Unnikrishnan, Urban, Usman, Vahlbruch,
  Vajente, Valdes, van Bakel, van Beuzekom, van~den Brand, Van Den~Broeck,
  Vander-Hyde, van~der Schaaf, van Heijningen, van Veggel, Vardaro, Varma,
  Vass, Vasúth, Vecchio, Vedovato, Veitch, Veitch, Venkateswara, Venugopalan,
  Verkindt, Vetrano, Viceré, Viets, Vinciguerra, Vine, Vinet, Vitale, Vo,
  Vocca, Vorvick, Vyatchanin, Wade, Wade, Wade, Walet, Walker, Wallace, Walsh,
  Wang, Wang, Wang, Wang, Wang, Ward, Warner, Was, Watchi, Weaver, Wei,
  Weinert, Weinstein, Weiss, Wen, Wessel, Wessels, Westerweck, Westphal, Wette,
  Whelan, Whitcomb, Whiting, Whittle, Wilken, Williams, Williams, Williamson,
  Willis, Willke, Wimmer, Winkler, Wipf, Wittel, Woan, Woehler, Wofford, Wong,
  Worden, Wright, Wu, Wysocki, Xiao, Yamamoto, Yancey, Yang, Yap, Yazback, Yu,
  Yu, Yvert, Zadrożny, Zanolin, Zelenova, Zendri, Zevin, Zhang, Zhang, Zhang,
  Zhang, Zhao, Zhou, Zhou, Zhu, Zhu, Zimmerman, Zucker, Zweizig, Collaboration,
  Collaboration, Wilson-Hodge, Bissaldi, Blackburn, Briggs, Burns, Cleveland,
  Connaughton, Gibby, Giles, Goldstein, Hamburg, Jenke, Hui, Kippen, Kocevski,
  McBreen, Meegan, Paciesas, Poolakkil, Preece, Racusin, Roberts, Stanbro,
  Veres, von Kienlin, Gbm, Savchenko, Ferrigno, Kuulkers, Bazzano, Bozzo,
  Brandt, Chenevez, Courvoisier, Diehl, Domingo, Hanlon, Jourdain, Laurent,
  Lebrun, Lutovinov, Martin-Carrillo, Mereghetti, Natalucci, Rodi, Roques,
  Sunyaev, Ubertini, INTEGRAL, Aartsen, Ackermann, Adams, Aguilar, Ahlers,
  Ahrens, Samarai, Altmann, Andeen, Anderson, Ansseau, Anton, Argüelles,
  Auffenberg, Axani, Bagherpour, Bai, Barron, Barwick, Baum, Bay, Beatty,
  Becker~Tjus, Bernardini, Besson, Binder, Bindig, Blaufuss, Blot, Bohm,
  Börner, Bos, Bose, Böser, Botner, Bourbeau, Bourbeau, Bradascio, Braun,
  Brayeur, Brenzke, Bretz, Bron, Brostean-Kaiser, Burgman, Carver, Casey,
  Casier, Cheung, Chirkin, Christov, Clark, Classen, Coenders, Collin, Conrad,
  Cowen, Cross, Day, de~André, De~Clercq, DeLaunay, Dembinski, De~Ridder,
  Desiati, de~Vries, de~Wasseige, de~With, DeYoung, Díaz-Vélez, di~Lorenzo,
  Dujmovic, Dumm, Dunkman, Dvorak, Eberhardt, Ehrhardt, Eichmann, Eller,
  Evenson, Fahey, Fazely, Felde, Filimonov, Finley, Flis, Franckowiak,
  Friedman, Fuchs, Gaisser, Gallagher, Gerhardt, Ghorbani, Giang, Glauch,
  Glüsenkamp, Goldschmidt, Gonzalez, Grant, Griffith, Haack, Hallgren, Halzen,
  Hanson, Hebecker, Heereman, Helbing, Hellauer, Hickford, Hignight, Hill,
  Hoffman, Hoffmann, Hokanson-Fasig, Hoshina, Huang, Huber, Hultqvist,
  Hünnefeld, In, Ishihara, Jacobi, Japaridze, Jeong, Jero, Jones, Kalaczynski,
  Kang, Kappes, Karg, Karle, Kauer, Keivani, Kelley, Kheirandish, Kim, Kim,
  Kintscher, Kiryluk, Kittler, Klein, Kohnen, Koirala, Kolanoski, Köpke,
  Kopper, Kopper, Koschinsky, Koskinen, Kowalski, Krings, Kroll, Krückl,
  Kunnen, Kunwar, Kurahashi, Kuwabara, Kyriacou, Labare, Lanfranchi, Larson,
  Lauber, Lesiak-Bzdak, Leuermann, Liu, Lu, Lünemann, Luszczak, Madsen, Maggi,
  Mahn, Mancina, Maruyama, Mase, Maunu, McNally, Meagher, Medici, Meier, Menne,
  Merino, Meures, Miarecki, Micallef, Momenté, Montaruli, Moore, Moulai,
  Nahnhauer, Nakarmi, Naumann, Neer, Niederhausen, Nowicki, Nygren,
  Obertacke~Pollmann, Olivas, O'Murchadha, Palczewski, Pandya, Pankova,
  Peiffer, Pepper, Pérez de~los Heros, Pieloth, Pinat, Price, Przybylski,
  Raab, Rädel, Rameez, Rawlins, Rea, Reimann, Relethford, Relich, Resconi,
  Rhode, Richman, Robertson, Rongen, Rott, Ruhe, Ryckbosch, Rysewyk, Sälzer,
  Sanchez~Herrera, Sandrock, Sandroos, Santander, Sarkar, Sarkar, Satalecka,
  Schlunder, Schmidt, Schneider, Schoenen, Schöneberg, Schumacher, Seckel,
  Seunarine, Soedingrekso, Soldin, Song, Spiczak, Spiering, Stachurska,
  Stamatikos, Stanev, Stasik, Stettner, Steuer, Stezelberger, Stokstad,
  Stössl, Strotjohann, Stuttard, Sullivan, Sutherland, Taboada, Tatar,
  Tenholt, Ter-Antonyan, Terliuk, Tešić, Tilav, Toale, Tobin, Toscano, Tosi,
  Tselengidou, Tung, Turcati, Turley, Ty, Unger, Usner, Vandenbroucke,
  Van~Driessche, van Eijndhoven, Vanheule, van Santen, Vehring, Vogel, Vraeghe,
  Walck, Wallace, Wallraff, Wandler, Wandkowsky, Waza, Weaver, Weiss, Wendt,
  Werthebach, Whelan, Wiebe, Wiebusch, Wille, Williams, Wills, Wolf, Wood,
  Woolsey, Woschnagg, Xu, Xu, Xu, Yanez, Yodh, Yoshida, Yuan, Zoll,
  Collaboration, Balasubramanian, Mate, Bhalerao, Bhattacharya, Vibhute,
  Dewangan, Rao, Vadawale, Team, Svinkin, Hurley, Aptekar, Frederiks,
  Golenetskii, Kozlova, Lysenko, Oleynik, Tsvetkova, Ulanov, Cline,
  Collaboration, Li, Xiong, Zhang, Lu, Song, Cao, Chang, Chen, Chen, Chen,
  Chen, Chen, Chen, Cui, Cui, Deng, Dong, Du, Fu, Gao, Gao, Gao, Ge, Gu, Guan,
  Guo, Han, Hu, Huang, Huo, Jia, Jiang, Jiang, Jin, Jin, Li, Li, Li, Li, Li,
  Li, Li, Li, Li, Li, Li, Liang, Liao, Liu, Liu, Liu, Liu, Liu, Liu, Liu, Lu,
  Lu, Luo, Ma, Meng, Nang, Nie, Ou, Qu, Sai, Sun, Tan, Tao, Tao, Tuo, Wang,
  Wang, Wang, Wang, Wang, Wen, Wu, Wu, Xiao, Xu, Xu, Yan, Yang, Yang, Yang,
  Zhang, Zhang, Zhang, Zhang, Zhang, Zhang, Zhang, Zhang, Zhang, Zhang, Zhang,
  Zhang, Zhang, Zhang, Zhang, Zhang, Zhang, Zhang, Zhao, Zhao, Zhao, Zheng,
  Zhu, Zhu, Zou, Collaboration, Albert, André, Anghinolfi, Ardid, Aubert,
  Aublin, Avgitas, Baret, Barrios-Martí, Basa, Belhorma, Bertin, Biagi,
  Bormuth, Bourret, Bouwhuis, Brânzaș, Bruijn, Brunner, Busto, Capone,
  Caramete, Carr, Celli, Cherkaoui El~Moursli, Chiarusi, Circella, Coelho,
  Coleiro, Coniglione, Costantini, Coyle, Creusot, Díaz, Deschamps, De~Bonis,
  Distefano, Di~Palma, Domi, Donzaud, Dornic, Drouhin, Eberl, El~Bojaddaini,
  El~Khayati, Elsässer, Enzenhöfer, Ettahiri, Fassi, Felis, Fusco, Gay,
  Giordano, Glotin, Grégoire, Ruiz, Graf, Hallmann, van Haren, Heijboer,
  Hello, Hernández-Rey, Hössl, Hofestädt, Hugon, Illuminati, James, de~Jong,
  Jongen, Kadler, Kalekin, Katz, Kiessling, Kouchner, Kreter, Kreykenbohm,
  Kulikovskiy, Lachaud, Lahmann, Lefèvre, Leonora, Lotze, Loucatos, Marcelin,
  Margiotta, Marinelli, Martínez-Mora, Mele, Melis, Michael, Migliozzi,
  Moussa, Navas, Nezri, Organokov, Păvălaș, Pellegrino, Perrina, Piattelli,
  Popa, Pradier, Quinn, Racca, Riccobene, Sánchez-Losa, Saldaña, Salvadori,
  Samtleben, Sanguineti, Sapienza, Sieger, Spurio, Stolarczyk, Taiuti,
  Tayalati, Trovato, Turpin, Tönnis, Vallage, Van~Elewyck, Versari, Vivolo,
  Vizzoca, Wilms, Zornoza, Zúñiga, Collaboration, Beardmore, Breeveld,
  Burrows, Cenko, Cusumano, D'Aì, de~Pasquale, Emery, Evans, Giommi, Gronwall,
  Kennea, Krimm, Kuin, Lien, Marshall, Melandri, Nousek, Oates, Osborne,
  Pagani, Page, Palmer, Perri, Siegel, Sbarufatti, Tagliaferri, Tohuvavohu,
  Collaboration, Tavani, Verrecchia, Bulgarelli, Evangelista, Pacciani, Feroci,
  Pittori, Giuliani, Del~Monte, Donnarumma, Argan, Trois, Ursi, Cardillo,
  Piano, Longo, Lucarelli, Munar-Adrover, Fuschino, Labanti, Marisaldi,
  Minervini, Fioretti, Parmiggiani, Gianotti, Trifoglio, Di~Persio, Antonelli,
  Barbiellini, Caraveo, Cattaneo, Costa, Colafrancesco, D'Amico, Ferrari,
  Morselli, Paoletti, Picozza, Pilia, Rappoldi, Soffitta, Vercellone, Team,
  Foley, Coulter, Kilpatrick, Drout, Piro, Shappee, Siebert, Simon, Ulloa,
  Kasen, Madore, Murguia-Berthier, Pan, Prochaska, Ramirez-Ruiz, Rest,
  Rojas-Bravo, Team, Berger, Soares-Santos, Annis, Alexander, Allam, Balbinot,
  Blanchard, Brout, Butler, Chornock, Cook, Cowperthwaite, Diehl,
  Drlica-Wagner, Drout, Durret, Eftekhari, Finley, Fong, Frieman, Fryer,
  García-Bellido, Gruendl, Hartley, Herner, Kessler, Lin, Lopes, Lourenço,
  Margutti, Marshall, Matheson, Medina, Metzger, Muñoz, Muir, Nicholl, Nugent,
  Palmese, Paz-Chinchón, Quataert, Sako, Sauseda, Schlegel, Scolnic, Secco,
  Smith, Sobreira, Villar, Vivas, Wester, Williams, Yanny, Zenteno, Zhang,
  Abbott, Banerji, Bechtol, Benoit-Lévy, Bertin, Brooks, Buckley-Geer, Burke,
  Capozzi, Carnero~Rosell, Carrasco~Kind, Castander, Crocce, Cunha, D'Andrea,
  da~Costa, Davis, DePoy, Desai, Dietrich, Eifler, Fernandez, Flaugher,
  Fosalba, Gaztanaga, Gerdes, Giannantonio, Goldstein, Gruen, Gschwend,
  Gutierrez, Honscheid, James, Jeltema, Johnson, Johnson, Kent, Krause, Kron,
  Kuehn, Lahav, Lima, Maia, March, Martini, McMahon, Menanteau, Miller, Miquel,
  Mohr, Nichol, Ogando, Plazas, Romer, Roodman, Rykoff, Sanchez, Scarpine,
  Schindler, Schubnell, Sevilla-Noarbe, Sheldon, Smith, Smith, Stebbins,
  Suchyta, Swanson, Tarle, Thomas, Troxel, Tucker, Vikram, Walker, Wechsler,
  Weller, Carlin, Gill, Li, Marriner, Neilsen, Collaboration, Collaboration,
  Haislip, Kouprianov, Reichart, Sand, Tartaglia, Valenti, Yang, Collaboration,
  Benetti, Brocato, Campana, Cappellaro, Covino, D'Avanzo, D'Elia, Getman,
  Ghirlanda, Ghisellini, Limatola, Nicastro, Palazzi, Pian, Piranomonte,
  Possenti, Rossi, Salafia, Tomasella, Amati, Antonelli, Bernardini, Bufano,
  Capaccioli, Casella, Dadina, De~Cesare, Di~Paola, Giuffrida, Giunta, Israel,
  Lisi, Maiorano, Mapelli, Masetti, Pescalli, Pulone, Salvaterra, Schipani,
  Spera, Stamerra, Stella, Testa, Turatto, Vergani, Aresu, Bachetti, Buffa,
  Burgay, Buttu, Caria, Carretti, Casasola, Castangia, Carboni, Casu, Concu,
  Corongiu, Deiana, Egron, Fara, Gaudiomonte, Gusai, Ladu, Loru, Leurini,
  Marongiu, Melis, Melis, Migoni, Milia, Navarrini, Orlati, Ortu, Palmas,
  Pellizzoni, Perrodin, Pisanu, Poppi, Righini, Saba, Serra, Serrau, Stagni,
  Surcis, Vacca, Vargiu, Hunt, Jin, Klose, Kouveliotou, Mazzali, Møller, Nava,
  Piran, Selsing, Vergani, Wiersema, Toma, Higgins, Mundell,
  di~Serego~Alighieri, Gótz, Gao, Gomboc, Kaper, Kobayashi, Kopac, Mao,
  Starling, Steele, van~der Horst, TeAm, Acero, Atwood, Baldini, Barbiellini,
  Bastieri, Berenji, Bellazzini, Bissaldi, Blandford, Bloom, Bonino, Bottacini,
  Bregeon, Buehler, Buson, Cameron, Caputo, Caraveo, Cavazzuti, Chekhtman,
  Cheung, Chiang, Ciprini, Cohen-Tanugi, Cominsky, Costantin, Cuoco, D'Ammando,
  de~Palma, Digel, Di~Lalla, Di~Mauro, Di~Venere, Dubois, Fegan, Focke,
  Franckowiak, Fukazawa, Funk, Fusco, Gargano, Gasparrini, Giglietto, Giordano,
  Giroletti, Glanzman, Green, Grondin, Guillemot, Guiriec, Harding, Horan,
  Jóhannesson, Kamae, Kensei, Kuss, La~Mura, Latronico, Lemoine-Goumard,
  Longo, Loparco, Lovellette, Lubrano, Magill, Maldera, Manfreda, Mazziotta,
  McEnery, Meyer, Michelson, Mirabal, Monzani, Moretti, Morselli, Moskalenko,
  Negro, Nuss, Ojha, Omodei, Orienti, Orlando, Palatiello, Paliya, Paneque,
  Pesce-Rollins, Piron, Porter, Principe, Rainò, Rando, Razzano, Razzaque,
  Reimer, Reimer, Reposeur, Rochester, Saz~Parkinson, Sgrò, Siskind, Spada,
  Spandre, Suson, Takahashi, Tanaka, Thayer, Thayer, Thompson, Tibaldo, Torres,
  Torresi, Troja, Venters, Vianello, Zaharijas, Collaboration, Allison,
  Bannister, Dobie, Kaplan, Lenc, Lynch, Murphy, Sadler, Australia Telescope
  Compact~Array, Hotan, James, Oslowski, Raja, Shannon, Whiting, Australian
  SKA~Pathfinder, Arcavi, Howell, McCully, Hosseinzadeh, Hiramatsu, Poznanski,
  Barnes, Zaltzman, Vasylyev, Maoz, Group, Cooke, Bailes, Wolf, Deller, Lidman,
  Wang, Gendre, Andreoni, Ackley, Pritchard, Bessell, Chang, Möller, Onken,
  Scalzo, Ridden-Harper, Sharp, Tucker, Farrell, Elmer, Johnston,
  Venkatraman~Krishnan, Keane, Green, Jameson, Hu, Ma, Sun, Wu, Wang, Shang,
  Hu, Ashley, Yuan, Li, Tao, Zhu, Zhang, Suntzeff, Zhou, Yang, Orange, Morris,
  Cucchiara, Giblin, Klotz, Staff, Thierry, Schmidt, OzGrav, (Deeper, Wider,
  Program, AST3, Collaborations, Tanvir, Levan, Cano, de~Ugarte-Postigo,
  González-Fernández, Greiner, Hjorth, Irwin, Krühler, Mandel,
  Milvang-Jensen, O'Brien, Rol, Rosetti, Rosswog, Rowlinson, Steeghs, Thöne,
  Ulaczyk, Watson, Bruun, Cutter, Figuera~Jaimes, Fujii, Fruchter, Gompertz,
  Jakobsson, Hodosan, Jèrgensen, Kangas, Kann, Rabus, Schrøder, Stanway,
  Wijers, Collaboration, Lipunov, Gorbovskoy, Kornilov, Tyurina, Balanutsa,
  Kuznetsov, Vlasenko, Podesta, Lopez, Podesta, Levato, Saffe, Mallamaci,
  Budnev, Gress, Kuvshinov, Gorbunov, Vladimirov, Zimnukhov, Gabovich, Yurkov,
  Sergienko, Rebolo, Serra-Ricart, Tlatov, Ishmuhametova, Collaboration, Abe,
  Aoki, Aoki, Asakura, Baar, Barway, Bond, Doi, Finet, Fujiyoshi, Furusawa,
  Honda, Itoh, Kanda, Kawabata, Kawabata, Kim, Koshida, Kuroda, Lee, Liu,
  Matsubayashi, Miyazaki, Morihana, Morokuma, Motohara, Murata, Nagai,
  Nagashima, Nagayama, Nakaoka, Nakata, Ohsawa, Ohshima, Ohta, Okita, Saito,
  Saito, Sako, Sekiguchi, Sumi, Tajitsu, Takahashi, Takayama, Tamura, Tanaka,
  Tanaka, Terai, Tominaga, Tristram, Uemura, Utsumi, Yamaguchi, Yasuda,
  Yoshida, Zenko, J-GEM, Adams, Anupama, Bally, Barway, Bellm, Blagorodnova,
  Cannella, Chandra, Chatterjee, Clarke, Cobb, Cook, Copperwheat, De, Emery,
  Feindt, Foster, Fox, Frail, Fremling, Frohmaier, Garcia, Ghosh, Giacintucci,
  Goobar, Gottlieb, Grefenstette, Hallinan, Harrison, Heida, Helou, Ho, Horesh,
  Hotokezaka, Ip, Itoh, Jacobs, Jencson, Kasen, Kasliwal, Kassim, Kim, Kiran,
  Kuin, Kulkarni, Kupfer, Lau, Madsen, Mazzali, Miller, Miyasaka, Mooley,
  Myers, Nakar, Ngeow, Nugent, Ofek, Palliyaguru, Pavana, Perley, Peters, Pike,
  Piran, Qi, Quimby, Rana, Rosswog, Rusu, Sadler, Van~Sistine, Sollerman, Xu,
  Yan, Yatsu, Yu, Zhang, Zhao, GROWTH, JAGWAR, Caltech-NRAO, TTU-NRAO,
  Collaborations, Chambers, Huber, Schultz, Bulger, Flewelling, Magnier, Lowe,
  Wainscoat, Waters, Willman, Pan-STARRS, Ebisawa, Hanyu, Harita, Hashimoto,
  Hidaka, Hori, Ishikawa, Isobe, Iwakiri, Kawai, Kawai, Kawamuro, Kawase,
  Kitaoka, Makishima, Matsuoka, Mihara, Morita, Morita, Nakahira, Nakajima,
  Nakamura, Negoro, Oda, Sakamaki, Sasaki, Serino, Shidatsu, Shimomukai,
  Sugawara, Sugita, Sugizaki, Tachibana, Takao, Tanimoto, Tomida, Tsuboi,
  Tsunemi, Ueda, Ueno, Yamada, Yamaoka, Yamauchi, Yatabe, Yoneyama, Yoshii,
  Team, Coward, Crisp, Macpherson, Andreoni, Laugier, Noysena, Klotz, Gendre,
  Thierry, Turpin, Consortium, Im, Choi, Kim, Yoon, Lim, Lee, Lee, Kim, Ko,
  Joe, Kwon, Kim, Lim, Choi, Collaboration, Fynbo, Malesani, Xu,
  Optical~Telescope, Smartt, Jerkstrand, Kankare, Sim, Fraser, Inserra,
  Maguire, Leloudas, Magee, Shingles, Smith, Young, Kotak, Gal-Yam, Lyman,
  Homan, Agliozzo, Anderson, Angus, Ashall, Barbarino, Bauer, Berton,
  Botticella, Bulla, Cannizzaro, Cartier, Cikota, Clark, De~Cia, Della~Valle,
  Dennefeld, Dessart, Dimitriadis, Elias-Rosa, Firth, Flörs, Frohmaier,
  Galbany, González-Gaitán, Gromadzki, Gutiérrez, Hamanowicz, Harmanen,
  Heintz, Hernandez, Hodgkin, Hook, Izzo, James, Jonker, Kerzendorf,
  Kostrzewa-Rutkowska, Kromer, Kuncarayakti, Lawrence, Manulis, Mattila,
  McBrien, Müller, Nordin, O'Neill, Onori, Palmerio, Pastorello, Patat,
  Pignata, Podsiadlowski, Razza, Reynolds, Roy, Ruiter, Rybicki, Salmon, Pumo,
  Prentice, Seitenzahl, Smith, Sollerman, Sullivan, Szegedi, Taddia,
  Taubenberger, Terreran, Van~Soelen, Vos, Walton, Wright, Wyrzykowski, Yaron,
  pre="("{\textgreater}ePESSTO, Chen, Krühler, Schady, Wiseman, Greiner, Rau,
  Schweyer, Klose, Nicuesa~Guelbenzu, GROND, Palliyaguru, Tech~University,
  Shara, Williams, Vaisanen, Potter, Romero~Colmenero, Crawford, Buckley, Mao,
  Group, Díaz, Macri, García~Lambas, Mendes~de Oliveira, Nilo~Castellón,
  Ribeiro, Sánchez, Schoenell, Abramo, Akras, Alcaniz, Artola, Beroiz, Bonoli,
  Cabral, Camuccio, Chavushyan, Coelho, Colazo, Costa-Duarte, Cuevas~Larenas,
  Domínguez~Romero, Dultzin, Fernández, García, Girardini, Gonçalves,
  Gonçalves, Gurovich, Jiménez-Teja, Kanaan, Lares, Lopes~de Oliveira,
  López-Cruz, Melia, Molino, Padilla, Peñuela, Placco, Quiñones,
  Ramírez~Rivera, Renzi, Riguccini, Ríos-López, Rodriguez, Sampedro,
  Schneiter, Sodré, Starck, Torres-Flores, Tornatore, Zadrożny, Castillo,
  Collaboration, Castro-Tirado, Tello, Hu, Zhang, Cunniffe, Castellón,
  Hiriart, Caballero-García, Jelínek, Kubánek, Pérez~del Pulgar, Park,
  Jeong, Castro~Cerón, Pandey, Yock, Querel, Fan, Wang, Collaboration,
  Beardsley, Brown, Crosse, Emrich, Franzen, Gaensler, Horsley,
  Johnston-Hollitt, Kenney, Morales, Pallot, Sokolowski, Steele, Tingay, Trott,
  Walker, Wayth, Williams, Wu, Murchison Widefield~Array, Yoshida, Sakamoto,
  Kawakubo, Yamaoka, Takahashi, Asaoka, Ozawa, Torii, Shimizu, Tamura,
  Ishizaki, Cherry, Ricciarini, Penacchioni, Marrocchesi, Collaboration,
  Pozanenko, Volnova, Mazaeva, Minaev, Krugov, Kusakin, Reva, Moskvitin,
  Rumyantsev, Inasaridze, Klunko, Tungalag, Schmalz, Burhonov, Collaboration,
  Abdalla, Abramowski, Aharonian, Ait~Benkhali, Angüner, Arakawa, Arrieta,
  Aubert, Backes, Balzer, Barnard, Becherini, Becker~Tjus, Berge, Bernhard,
  Bernlöhr, Blackwell, Böttcher, Boisson, Bolmont, Bonnefoy, Bordas, Bregeon,
  Brun, Brun, Bryan, Büchele, Bulik, Capasso, Caroff, Carosi, Casanova,
  Cerruti, Chakraborty, Chaves, Chen, Chevalier, Colafrancesco, Condon, Conrad,
  Davids, Decock, Deil, Devin, deWilt, Dirson, Djannati-Ataï, Donath,
  O'C.~Drury, Dutson, Dyks, Edwards, Egberts, Emery, Ernenwein, Eschbach,
  Farnier, Fegan, Fernandes, Fiasson, Fontaine, Funk, Füssling, Gabici,
  Gallant, Garrigoux, Gaté, Giavitto, Giebels, Glawion, Glicenstein,
  Gottschall, Grondin, Hahn, Haupt, Hawkes, Heinzelmann, Henri, Hermann,
  Hinton, Hofmann, Hoischen, Holch, Holler, Horns, Ivascenko, Iwasaki,
  Jacholkowska, Jamrozy, Jankowsky, Jankowsky, Jingo, Jouvin, Jung-Richardt,
  Kastendieck, Katarzyński, Katsuragawa, Kerszberg, Khangulyan, Khélifi,
  King, Klepser, Klochkov, Kluźniak, Komin, Kosack, Krakau, Kraus, Krüger,
  Laffon, Lamanna, Lau, Lees, Lefaucheur, Lemière, Lemoine-Goumard, Lenain,
  Leser, Lohse, Lorentz, Liu, Lypova, Malyshev, Marandon, Marcowith, Mariaud,
  Marx, Maurin, Maxted, Mayer, Meintjes, Meyer, Mitchell, Moderski, Mohamed,
  Mohrmann, Morå, Moulin, Murach, Nakashima, de~Naurois, Ndiyavala,
  Niederwanger, Niemiec, Oakes, O'Brien, Odaka, Ohm, Ostrowski, Oya, Padovani,
  Panter, Parsons, Pekeur, Pelletier, Perennes, Petrucci, Peyaud, Piel, Pita,
  Poireau, Poon, Prokhorov, Prokoph, Pühlhofer, Punch, Quirrenbach, Raab,
  Rauth, Reimer, Reimer, Renaud, de~los Reyes, Rieger, Rinchiuso, Romoli,
  Rowell, Rudak, Rulten, Sahakian, Saito, Sanchez, Santangelo, Sasaki,
  Schlickeiser, Schüssler, Schulz, Schwanke, Schwemmer, Seglar-Arroyo,
  Settimo, Seyffert, Shafi, Shilon, Shiningayamwe, Simoni, Sol, Spanier,
  Spir-Jacob, Stawarz, Steenkamp, Stegmann, Steppa, Sushch, Takahashi,
  Tavernet, Tavernier, Taylor, Terrier, Tibaldo, Tiziani, Tluczykont, Trichard,
  Tsirou, Tsuji, Tuffs, Uchiyama, van~der Walt, van Eldik, van Rensburg, van
  Soelen, Vasileiadis, Veh, Venter, Viana, Vincent, Vink, Voisin, Völk,
  Vuillaume, Wadiasingh, Wagner, Wagner, Wagner, White, Wierzcholska, Willmann,
  Wörnlein, Wouters, Yang, Zaborov, Zacharias, Zanin, Zdziarski, Zech, Zefi,
  Ziegler, Zorn, Żywucka, Collaboration, Fender, Broderick, Rowlinson, Wijers,
  Stewart, ter Veen, Shulevski, Collaboration, Kavic, Simonetti, League, Tsai,
  Obenberger, Nathaniel, Taylor, Dowell, Liebling, Estes, Lippert, Sharma,
  Vincent, Farella, Wavelength~Array, Abeysekara, Albert, Alfaro, Alvarez,
  Arceo, Arteaga-Velázquez, Avila~Rojas, Ayala~Solares, Barber,
  Becerra~Gonzalez, Becerril, Belmont-Moreno, BenZvi, Berley, Bernal, Braun,
  Brisbois, Caballero-Mora, Capistrán, Carramiñana, Casanova, Castillo,
  Cotti, Cotzomi, Coutiño~de León, De~León, De~la Fuente, Diaz~Hernandez,
  Dichiara, Dingus, DuVernois, Díaz-Vélez, Ellsworth, Engel,
  Enríquez-Rivera, Fiorino, Fleischhack, Fraija, García-González, Garfias,
  Gerhardt, Gonzõlez~Muñoz, González, Goodman, Hampel-Arias, Harding,
  Hernandez, Hernandez-Almada, Hona, Hüntemeyer, Iriarte, Jardin-Blicq, Joshi,
  Kaufmann, Kieda, Lara, Lauer, Lennarz, León~Vargas, Linnemann, Longinotti,
  Raya, Luna-García, López-Coto, Malone, Marinelli, Martinez,
  Martinez-Castellanos, Martínez-Castro, Martínez-Huerta, Matthews,
  Miranda-Romagnoli, Moreno, Mostafá, Nellen, Newbold, Nisa, Noriega-Papaqui,
  Pelayo, Pretz, Pérez-Pérez, Ren, Rho, Rivière, Rosa-González, Rosenberg,
  Ruiz-Velasco, Salazar, Salesa~Greus, Sandoval, Schneider, Schoorlemmer,
  Sinnis, Smith, Springer, Surajbali, Tibolla, Tollefson, Torres, Ukwatta,
  Weisgarber, Westerhoff, Wisher, Wood, Yapici, Yodh, Younk, Zhou, Álvarez,
  Collaboration, Aab, Abreu, Aglietta, Albuquerque, Albury, Allekotte, Almela,
  Alvarez~Castillo, Alvarez-Muñiz, Anastasi, Anchordoqui, Andrada, Andringa,
  Aramo, Arsene, Asorey, Assis, Avila, Badescu, Balaceanu, Barbato,
  Barreira~Luz, Becker, Bellido, Berat, Bertaina, Bertou, Biermann, Biteau,
  Blaess, Blanco, Blazek, Bleve, Boháčová, Bonifazi, Borodai, Botti, Brack,
  Brancus, Bretz, Bridgeman, Briechle, Buchholz, Bueno, Buitink, Buscemi,
  Caballero-Mora, Caccianiga, Cancio, Canfora, Caruso, Castellina, Catalani,
  Cataldi, Cazon, Chavez, Chinellato, Chudoba, Clay, Cobos~Cerutti, Colalillo,
  Coleman, Collica, Coluccia, Conceição, Consolati, Contreras, Cooper, Coutu,
  Covault, Cronin, D'Amico, Daniel, Dasso, Daumiller, Dawson, Day, de~Almeida,
  de~Jong, De~Mauro, de~Mello~Neto, De~Mitri, de~Oliveira, de~Souza, Debatin,
  Deligny, Díaz~Castro, Diogo, Dobrigkeit, D'Olivo, Dorosti, Dos~Anjos, Dova,
  Dundovic, Ebr, Engel, Erdmann, Erfani, Escobar, Espadanal, Etchegoyen,
  Falcke, Farmer, Farrar, Fauth, Fazzini, Feldbusch, Fenu, Fick, Figueira,
  Filipčič, Freire, Fujii, Fuster, Gaïor, García, Gaté, Gemmeke,
  Gherghel-Lascu, Ghia, Giaccari, Giammarchi, Giller, Głas, Glaser, Golup,
  Gómez~Berisso, Gómez~Vitale, González, Gorgi, Gottowik, Grillo, Grubb,
  Guarino, Guedes, Halliday, Hampel, Hansen, Harari, Harrison, Harvey, Haungs,
  Hebbeker, Heck, Heimann, Herve, Hill, Hojvat, Holt, Homola, Hörandel,
  Horvath, Hrabovský, Huege, Hulsman, Insolia, Isar, Jandt, Johnsen,
  Josebachuili, Jurysek, Kääpä, Kampert, Keilhauer, Kemmerich, Kemp,
  Kieckhafer, Klages, Kleifges, Kleinfeller, Krause, Krohm, Kuempel,
  Kukec~Mezek, Kunka, Kuotb~Awad, Lago, LaHurd, Lang, Lauscher, Legumina,
  Leigui~de Oliveira, Letessier-Selvon, Lhenry-Yvon, Link, Lo~Presti, Lopes,
  López, López~Casado, Lorek, Luce, Lucero, Malacari, Mallamaci, Mandat,
  Mantsch, Mariazzi, Maris, Marsella, Martello, Martinez, Martínez~Bravo,
  Masías~Meza, Mathes, Mathys, Matthews, Matthiae, Mayotte, Mazur, Medina,
  Medina-Tanco, Melo, Menshikov, Merenda, Michal, Micheletti, Middendorf,
  Miramonti, Mitrica, Mockler, Mollerach, Montanet, Morello, Morlino, Müller,
  Müller, Muller, Müller, Mussa, Naranjo, Nguyen, Niculescu-Oglinzanu,
  Niechciol, Niemietz, Niggemann, Nitz, Nosek, Novotny, Nožka, Núñez,
  Oikonomou, Olinto, Palatka, Pallotta, Papenbreer, Parente, Parra, Paul, Pech,
  Pedreira, Pȩkala, Peña-Rodriguez, Pereira, Perlin, Perrone, Peters,
  Petrera, Phuntsok, Pierog, Pimenta, Pirronello, Platino, Plum, Poh, Porowski,
  Prado, Privitera, Prouza, Quel, Querchfeld, Quinn, Ramos-Pollan, Rautenberg,
  Ravignani, Ridky, Riehn, Risse, Ristori, Rizi, Rodrigues~de Carvalho,
  Rodriguez~Fernandez, Rodriguez~Rojo, Roncoroni, Roth, Roulet, Rovero, Ruehl,
  Saffi, Saftoiu, Salamida, Salazar, Saleh, Salina, Sánchez, Sanchez-Lucas,
  Santos, Santos, Sarazin, Sarmento, Sarmiento-Cano, Sato, Schauer, Scherini,
  Schieler, Schimp, Schmidt, Scholten, Schovánek, Schröder, Schröder,
  Schulz, Schumacher, Sciutto, Segreto, Shadkam, Shellard, Sigl, Silli,
  Šmída, Snow, Sommers, Sonntag, Soriano, Squartini, Stanca, Stanič,
  Stasielak, Stassi, Stolpovskiy, Strafella, Streich, Suarez, Suarez-Durán,
  Sudholz, Suomijärvi, Supanitsky, Šupík, Swain, Szadkowski, Taboada,
  Taborda, Timmermans, Todero~Peixoto, Tomankova, Tomé, Torralba~Elipe,
  Travnicek, Trini, Tueros, Ulrich, Unger, Urban, Valdés~Galicia, Valiño,
  Valore, van Aar, van Bodegom, van~den Berg, van Vliet, Varela,
  Vargas~Cárdenas, Vázquez, Veberič, Ventura, Vergara~Quispe, Verzi, Vicha,
  Villaseñor, Vorobiov, Wahlberg, Wainberg, Walz, Watson, Weber, Weindl,
  Wiedeński, Wiencke, Wilczyński, Wirtz, Wittkowski, Wundheiler, Yang,
  Yushkov, Zas, Zavrtanik, Zavrtanik, Zepeda, Zimmermann, Ziolkowski, Zong,
  Zuccarello, Collaboration, Kim, Schulze, Bauer, Corral-Santana,
  de~Gregorio-Monsalvo, González-López, Hartmann, Ishwara-Chandra, Martín,
  Mehner, Misra, Michałowski, Resmi, Collaboration, Paragi, Agudo, An,
  Beswick, Casadio, Frey, Jonker, Kettenis, Marcote, Moldon, Szomoru, van
  Langevelde, Yang, Team, Cwiek, Cwiok, Czyrkowski, Dabrowski, Kasprowicz,
  Mankiewicz, Nawrocki, Opiela, Piotrowski, Wrochna, Zaremba, Żarnecki,
  Collaboration, Haggard, Nynka, Ruan, University, Bland, Booler, Devillepoix,
  de~Gois, Hancock, Howie, Paxman, Sansom, Towner, Desert Fireball~Network,
  Tonry, Coughlin, Stubbs, Denneau, Heinze, Stalder, Weiland, ATLAS, Eatough,
  Kramer, Kraus, Time Resolution Universe~Survey, Troja, Piro,
  Becerra~González, Butler, Fox, Khandrika, Kutyrev, Lee, Ricci, Ryan,
  Sánchez-Ramírez, Veilleux, Watson, Wieringa, Burgess, van Eerten, Fontes,
  Fryer, Korobkin, Wollaeger, RIMAS, RATIR, Camilo, Foley, Goedhart,
  Makhathini, Oozeer, Smirnov, Fender, Woudt, \&
  South~Africa/MeerKAT}]{abbott_multi-messenger_2017}
Abbott, B.~P., Abbott, R., Abbott, T.~D., {et~al.} 2017{\natexlab{b}}, \apj,
  848, L12

\bibitem[{Abbott {et~al.}(2017{\natexlab{c}})Abbott, Abbott, Abbott, Acernese,
  Ackley, Adams, Adams, Addesso, Adhikari, Adya, Affeldt, Afrough, Agarwal,
  Agathos, Agatsuma, Aggarwal, Aguiar, Aiello, Ain, Ajith, Allen, Allen,
  Allocca, Aloy, Altin, Amato, Ananyeva, Anderson, Anderson, Angelova, Antier,
  Appert, Arai, Araya, Areeda, Arnaud, Arun, Ascenzi, Ashton, Ast, Aston,
  Astone, Atallah, Aufmuth, Aulbert, AultONeal, Austin, Avila-Alvarez, Babak,
  Bacon, Bader, Bae, Baker, Baldaccini, Ballardin, Ballmer, Banagiri, Barayoga,
  Barclay, Barish, Barker, Barkett, Barone, Barr, Barsotti, Barsuglia, Barta,
  Bartlett, Bartos, Bassiri, Basti, Batch, Bawaj, Bayley, Bazzan, Bécsy, Beer,
  Bejger, Belahcene, Bell, Berger, Bergmann, Bero, Berry, Bersanetti,
  Bertolini, Betzwieser, Bhagwat, Bhandare, Bilenko, Billingsley, Billman,
  Birch, Birney, Birnholtz, Biscans, Biscoveanu, Bisht, Bitossi, Biwer,
  Bizouard, Blackburn, Blackman, Blair, Blair, Blair, Bloemen, Bock, Bode,
  Boer, Bogaert, Bohe, Bondu, Bonilla, Bonnand, Boom, Bork, Boschi, Bose,
  Bossie, Bouffanais, Bozzi, Bradaschia, Brady, Branchesi, Brau, Briant,
  Brillet, Brinkmann, Brisson, Brockill, Broida, Brooks, Brown, Brown, Brunett,
  Buchanan, Buikema, Bulik, Bulten, Buonanno, Buskulic, Buy, Byer, Cabero,
  Cadonati, Cagnoli, Cahillane, Calderón~Bustillo, Callister, Calloni, Camp,
  Canepa, Canizares, Cannon, Cao, Cao, Capano, Capocasa, Carbognani, Caride,
  Carney, Casanueva~Diaz, Casentini, Caudill, Cavaglià, Cavalier, Cavalieri,
  Cella, Cepeda, Cerdá-Durán, Cerretani, Cesarini, Chamberlin, Chan, Chao,
  Charlton, Chase, Chassande-Mottin, Chatterjee, Chatziioannou, Cheeseboro,
  Chen, Chen, Chen, Cheng, Chia, Chincarini, Chiummo, Chmiel, Cho, Cho, Chow,
  Christensen, Chu, Chua, Chua, Chung, Chung, Ciani, Ciolfi, Cirelli, Cirone,
  Clara, Clark, Clearwater, Cleva, Cocchieri, Coccia, Cohadon, Cohen, Colla,
  Collette, Cominsky, Constancio, Conti, Cooper, Corban, Corbitt,
  Cordero-Carrión, Corley, Cornish, Corsi, Cortese, Costa, Coughlin, Coughlin,
  Coulon, Countryman, Couvares, Covas, Cowan, Coward, Cowart, Coyne, Coyne,
  Creighton, Creighton, Cripe, Crowder, Cullen, Cumming, Cunningham, Cuoco,
  Dal~Canton, Dálya, Danilishin, D'Antonio, Danzmann, Dasgupta,
  Da~Silva~Costa, Dattilo, Dave, Davier, Davis, Daw, Day, De, DeBra, Degallaix,
  De~Laurentis, Deléglise, Del~Pozzo, Demos, Denker, Dent, De~Pietri,
  Dergachev, De~Rosa, DeRosa, De~Rossi, DeSalvo, de~Varona, Devenson,
  Dhurandhar, Díaz, Di~Fiore, Di~Giovanni, Di~Girolamo, Di~Lieto, Di~Pace,
  Di~Palma, Di~Renzo, Doctor, Dolique, Donovan, Dooley, Doravari, Dorrington,
  Douglas, Dovale~Álvarez, Downes, Drago, Dreissigacker, Driggers, Du, Ducrot,
  Dupej, Dwyer, Edo, Edwards, Effler, Eggenstein, Ehrens, Eichholz, Eikenberry,
  Eisenstein, Essick, Estevez, Etienne, Etzel, Evans, Evans, Factourovich,
  Fafone, Fair, Fairhurst, Fan, Farinon, Farr, Farr, Fauchon-Jones, Favata,
  Fays, Fee, Fehrmann, Feicht, Fejer, Fernandez-Galiana, Ferrante, Ferreira,
  Ferrini, Fidecaro, Finstad, Fiori, Fiorucci, Fishbach, Fisher, Fitz-Axen,
  Flaminio, Fletcher, Fong, Font, Forsyth, Forsyth, Fournier, Frasca, Frasconi,
  Frei, Freise, Frey, Frey, Fries, Fritschel, Frolov, Fulda, Fyffe, Gabbard,
  Gadre, Gaebel, Gair, Gammaitoni, Ganija, Gaonkar, Garcia-Quiros, Garufi,
  Gateley, Gaudio, Gaur, Gayathri, Gehrels, Gemme, Genin, Gennai, George,
  George, Gergely, Germain, Ghonge, Ghosh, Ghosh, Ghosh, Giaime, Giardina,
  Giazotto, Gill, Glover, Goetz, Goetz, Gomes, Goncharov, González,
  Gonzalez~Castro, Gopakumar, Gorodetsky, Gossan, Gosselin, Gouaty, Grado,
  Graef, Granata, Grant, Gras, Gray, Greco, Green, Gretarsson, Groot, Grote,
  Grunewald, Gruning, Guidi, Guo, Gupta, Gupta, Gushwa, Gustafson, Gustafson,
  Halim, Hall, Hall, Hamilton, Hammond, Haney, Hanke, Hanks, Hanna, Hannam,
  Hannuksela, Hanson, Hardwick, Harms, Harry, Harry, Hart, Haster, Haughian,
  Healy, Heidmann, Heintze, Heitmann, Hello, Hemming, Hendry, Heng, Hennig,
  Heptonstall, Heurs, Hild, Hinderer, Hoak, Hofman, Holt, Holz, Hopkins, Horst,
  Hough, Houston, Howell, Hreibi, Hu, Huerta, Huet, Hughey, Husa, Huttner,
  Huynh-Dinh, Indik, Inta, Intini, Isa, Isac, Isi, Iyer, Izumi, Jacqmin, Jani,
  Jaranowski, Jawahar, Jiménez-Forteza, Johnson, Johnson-McDaniel, Jones,
  Jones, Jonker, Ju, Junker, Kalaghatgi, Kalogera, Kamai, Kandhasamy, Kang,
  Kanner, Kapadia, Karki, Karvinen, Kasprzack, Kastaun, Katolik, Katsavounidis,
  Katzman, Kaufer, Kawabe, Kéfélian, Keitel, Kemball, Kennedy, Kent, Key,
  Khalili, Khan, Khan, Khan, Khazanov, Kijbunchoo, Kim, Kim, Kim, Kim, Kim,
  Kim, Kimbrell, King, King, Kinley-Hanlon, Kirchhoff, Kissel, Kleybolte,
  Klimenko, Knowles, Koch, Koehlenbeck, Koley, Kondrashov, Kontos, Korobko,
  Korth, Kowalska, Kozak, Krämer, Kringel, Krishnan, Królak, Kuehn, Kumar,
  Kumar, Kumar, Kuo, Kutynia, Kwang, Lackey, Lai, Landry, Lang, Lange, Lantz,
  Lanza, Lartaux-Vollard, Lasky, Laxen, Lazzarini, Lazzaro, Leaci, Leavey, Lee,
  Lee, Lee, Lee, Lee, Lehmann, Lenon, Leonardi, Leroy, Letendre, Levin, Li,
  Linker, Littenberg, Liu, Lo, Lockerbie, London, Lord, Lorenzini, Loriette,
  Lormand, Losurdo, Lough, Lousto, Lovelace, Lück, Lumaca, Lundgren, Lynch,
  Ma, Macas, Macfoy, Machenschalk, MacInnis, Macleod, Magaña~Hernandez,
  Magaña-Sandoval, Magaña~Zertuche, Magee, Majorana, Maksimovic, Man, Mandic,
  Mangano, Mansell, Manske, Mantovani, Marchesoni, Marion, Márka, Márka,
  Markakis, Markosyan, Markowitz, Maros, Marquina, Martelli, Martellini,
  Martin, Martin, Martynov, Mason, Massera, Masserot, Massinger, Masso-Reid,
  Mastrogiovanni, Matas, Matichard, Matone, Mavalvala, Mazumder, McCarthy,
  McClelland, McCormick, McCuller, McGuire, McIntyre, McIver, McManus, McNeill,
  McRae, McWilliams, Meacher, Meadors, Mehmet, Meidam, Mejuto-Villa, Melatos,
  Mendell, Mercer, Merilh, Merzougui, Meshkov, Messenger, Messick, Metzdorff,
  Meyers, Miao, Michel, Middleton, Mikhailov, Milano, Miller, Miller, Miller,
  Millhouse, Milovich-Goff, Minazzoli, Minenkov, Ming, Mishra, Mitra,
  Mitrofanov, Mitselmakher, Mittleman, Moffa, Moggi, Mogushi, Mohan, Mohapatra,
  Montani, Moore, Moraru, Moreno, Morriss, Mours, Mow-Lowry, Mueller, Muir,
  Mukherjee, Mukherjee, Mukherjee, Mukund, Mullavey, Munch, Muñiz, Muratore,
  Murray, Napier, Nardecchia, Naticchioni, Nayak, Neilson, Nelemans, Nelson,
  Nery, Neunzert, Nevin, Newport, Newton, Ng, Nguyen, Nichols, Nielsen,
  Nissanke, Nitz, Noack, Nocera, Nolting, North, Nuttall, Oberling, O'Dea,
  Ogin, Oh, Oh, Ohme, Okada, Oliver, Oppermann, Oram, O'Reilly, Ormiston,
  Ortega, O'Shaughnessy, Ossokine, Ottaway, Overmier, Owen, Pace, Page, Page,
  Pai, Pai, Palamos, Palashov, Palomba, Pal-Singh, Pan, Pan, Pang, Pang,
  Pankow, Pannarale, Pant, Paoletti, Paoli, Papa, Parida, Parker, Pascucci,
  Pasqualetti, Passaquieti, Passuello, Patil, Patricelli, Pearlstone, Pedraza,
  Pedurand, Pekowsky, Pele, Penn, Perez, Perreca, Perri, Pfeiffer, Phelps,
  Piccinni, Pichot, Piergiovanni, Pierro, Pillant, Pinard, Pinto, Pirello,
  Pitkin, Poe, Poggiani, Popolizio, Porter, Post, Powell, Prasad, Pratt,
  Pratten, Predoi, Prestegard, Prijatelj, Principe, Privitera, Prodi,
  Prokhorov, Puncken, Punturo, Puppo, Pürrer, Qi, Quetschke, Quintero,
  Quitzow-James, Raab, Rabeling, Radkins, Raffai, Raja, Rajan, Rajbhandari,
  Rakhmanov, Ramirez, Ramos-Buades, Rapagnani, Raymond, Razzano, Read,
  Regimbau, Rei, Reid, Reitze, Ren, Reyes, Ricci, Ricker, Rieger, Riles, Rizzo,
  Robertson, Robie, Robinet, Rocchi, Rolland, Rollins, Roma, Romano, Romel,
  Romie, Rosińska, Ross, Rowan, Rüdiger, Ruggi, Rutins, Ryan, Sachdev,
  Sadecki, Sadeghian, Sakellariadou, Salconi, Saleem, Salemi, Samajdar, Sammut,
  Sampson, Sanchez, Sanchez, Sanchis-Gual, Sandberg, Sanders, Sassolas,
  Sathyaprakash, Saulson, Sauter, Savage, Sawadsky, Schale, Scheel, Scheuer,
  Schmidt, Schmidt, Schnabel, Schofield, Schönbeck, Schreiber, Schuette,
  Schulte, Schutz, Schwalbe, Scott, Scott, Seidel, Sellers, Sengupta, Sentenac,
  Sequino, Sergeev, Shaddock, Shaffer, Shah, Shahriar, Shaner, Shao, Shapiro,
  Shawhan, Sheperd, Shoemaker, Shoemaker, Siellez, Siemens, Sieniawska, Sigg,
  Silva, Singer, Singh, Singhal, Sintes, Slagmolen, Smith, Smith, Smith,
  Somala, Son, Sonnenberg, Sorazu, Sorrentino, Souradeep, Spencer, Srivastava,
  Staats, Staley, Steinke, Steinlechner, Steinlechner, Steinmeyer, Stevenson,
  Stone, Stops, Strain, Stratta, Strigin, Strunk, Sturani, Stuver,
  Summerscales, Sun, Sunil, Suresh, Sutton, Swinkels, Szczepańczyk, Tacca,
  Tait, Talbot, Talukder, Tanner, Tápai, Taracchini, Tasson, Taylor, Taylor,
  Tewari, Theeg, Thies, Thomas, Thomas, Thomas, Thorne, Thorne, Thrane, Tiwari,
  Tiwari, Tokmakov, Toland, Tonelli, Tornasi, Torres-Forné, Torrie, Töyrä,
  Travasso, Traylor, Trinastic, Tringali, Trozzo, Tsang, Tse, Tso, Tsukada,
  Tsuna, Tuyenbayev, Ueno, Ugolini, Unnikrishnan, Urban, Usman, Vahlbruch,
  Vajente, Valdes, van Bakel, van Beuzekom, van~den Brand, Van Den~Broeck,
  Vander-Hyde, van~der Schaaf, van Heijningen, van Veggel, Vardaro, Varma,
  Vass, Vasúth, Vecchio, Vedovato, Veitch, Veitch, Venkateswara, Venugopalan,
  Verkindt, Vetrano, Viceré, Viets, Vinciguerra, Vine, Vinet, Vitale, Vo,
  Vocca, Vorvick, Vyatchanin, Wade, Wade, Wade, Walet, Walker, Wallace, Walsh,
  Wang, Wang, Wang, Wang, Wang, Ward, Warner, Was, Watchi, Weaver, Wei,
  Weinert, Weinstein, Weiss, Wen, Wessel, Weßels, Westerweck, Westphal, Wette,
  Whelan, Whitcomb, Whiting, Whittle, Wilken, Williams, Williams, Williamson,
  Willis, Willke, Wimmer, Winkler, Wipf, Wittel, Woan, Woehler, Wofford, Wong,
  Worden, Wright, Wu, Wysocki, Xiao, Yamamoto, Yancey, Yang, Yap, Yazback, Yu,
  Yu, Yvert, Zadrożny, Zanolin, Zelenova, Zendri, Zevin, Zhang, Zhang, Zhang,
  Zhang, Zhao, Zhou, Zhou, Zhu, Zhu, Zimmerman, Zucker, Zweizig, Collaboration,
  Collaboration, Burns, Veres, Kocevski, Racusin, Goldstein, Connaughton,
  Briggs, Blackburn, Hamburg, Hui, von Kienlin, McEnery, Preece, Wilson-Hodge,
  Bissaldi, Cleveland, Gibby, Giles, Kippen, McBreen, Meegan, Paciesas,
  Poolakkil, Roberts, Stanbro, Gamma-ray Burst~Monitor, Savchenko, Ferrigno,
  Kuulkers, Bazzano, Bozzo, Brandt, Chenevez, Courvoisier, Diehl, Domingo,
  Hanlon, Jourdain, Laurent, Lebrun, Lutovinov, Mereghetti, Natalucci, Rodi,
  Roques, Sunyaev, Ubertini, \& (INTEGRAL}]{abbott_gravitational_2017}
Abbott, B.~P., Abbott, R., Abbott, T.~D., {et~al.} 2017{\natexlab{c}}, \apj,
  848, L13

\bibitem[{{Abbott} {et~al.}(2017){Abbott}, {Abbott}, {Abbott}, {Acernese},
  {Ackley}, {Adams}, {Adams}, {Addesso}, {Adhikari}, {Adya}, {Affeldt},
  {Afrough}, {(LIGO Scientific Collaboration}, \& {Virgo
  Collaboration}}]{abbott_2017post}
{Abbott}, B.~P., {Abbott}, R., {Abbott}, T.~D., {et~al.} 2017, \apjl, 851, L16

\bibitem[{Alcubierre(2008)}]{alcubierre_introduction_2008}
Alcubierre, M. 2008, Introduction to 3+1 Numerical Relativity, International
  Series of Monographs on Physics (OUP Oxford)

\bibitem[{{Anderson} {et~al.}(2019){Anderson}, {Freire}, \&
  {Yunes}}]{anderson_2019}
{Anderson}, D., {Freire}, P., \& {Yunes}, N. 2019, \cqg, 36, 225009

\bibitem[{{Anderson} \& {Yunes}(2019)}]{anderson_scalar_2019}
{Anderson}, D. \& {Yunes}, N. 2019, \cqg, 36, 165003

\bibitem[{Andreou {et~al.}(2019)Andreou, Franchini, Ventagli, \&
  Sotiriou}]{andreou_2019}
Andreou, N., Franchini, N., Ventagli, G., \& Sotiriou, T.~P. 2019, Phys. Rev.
  D, 99, 124022

\bibitem[{Barausse(2017)}]{barausse_testing_2017}
Barausse, E. 2017, in Proceedings of {The} 3rd {International} {Symposium} on
  ``{Quest} for the {Origin} of {Particles} and the {Universe}'', Vol. 294
  (SISSA Medialab), 029

\bibitem[{Barausse {et~al.}(2013)Barausse, Palenzuela, Ponce, \&
  Lehner}]{barausse_neutron-star_2013}
Barausse, E., Palenzuela, C., Ponce, M., \& Lehner, L. 2013, \prd, 87, 081506

\bibitem[{Bars \& Pope(1989)}]{bars_is_1989}
Bars, I. \& Pope, C.~N. 1989, \grg, 21, 545

\bibitem[{Bekenstein(2004)}]{bekenstein_2004}
Bekenstein, J.~D. 2004, \prd, 70, 083509

\bibitem[{Bekenstein \& Oron(2001)}]{Bekenstein_Oron00a}
Bekenstein, J.~D. \& Oron, A. 2001, \fph, 31, 895

\bibitem[{Bergmann(1968)}]{bergmann_comments_1968}
Bergmann, P.~G. 1968, \ijtp, 1, 25

\bibitem[{Berti {et~al.}(2015)Berti, Barausse, Cardoso, Gualtieri, Pani,
  Sperhake, Stein, Wex, Yagi, Baker, Burgess, Coelho, Doneva, Felice, Ferreira,
  Freire, Healy, Herdeiro, Horbatsch, Kleihaus, Klein, Kokkotas, Kunz, Laguna,
  Lang, Li, Littenberg, Matas, Mirshekari, Okawa, Radu, O'Shaughnessy,
  Sathyaprakash, Broeck, Winther, Witek, Aghili, Alsing, Bolen, Bombelli,
  Caudill, Chen, Degollado, Fujita, Gao, Gerosa, Kamali, Silva, Rosa,
  Sadeghian, Sampaio, Sotani, \& Zilhao}]{berti_testing_2015}
Berti, E., Barausse, E., Cardoso, V., {et~al.} 2015, \cqg, 32, 243001

\bibitem[{{Bertolami} \& {P{\'a}ramos}(2016)}]{bertolami_2016}
{Bertolami}, O. \& {P{\'a}ramos}, J. 2016, \grg, 48, 34

\bibitem[{Bocquet {et~al.}(1995)Bocquet, Bonazzola, Gourgoulhon, \&
  Novak}]{bocquet_rotating_1995}
Bocquet, M., Bonazzola, S., Gourgoulhon, E., \& Novak, J. 1995, \aap, 301, 757

\bibitem[{Bonanno {et~al.}(2003)Bonanno, Rezzolla, \&
  Urpin}]{bonanno_mean-field_2003}
Bonanno, A., Rezzolla, L., \& Urpin, V. 2003, \aap, 410, L33

\bibitem[{Braithwaite(2009)}]{braithwaite_axisymmetric_2009}
Braithwaite, J. 2009, \mnras, 397, 763

\bibitem[{Braithwaite \& Nordlund(2006)}]{braithwaite_stable_2006}
Braithwaite, J. \& Nordlund, A. 2006, \aap, 450, 1077

\bibitem[{Braithwaite \& Spruit(2006)}]{braithwaite_evolution_2006}
Braithwaite, J. \& Spruit, H.~C. 2006, \aap, 450, 1097

\bibitem[{Brans \& Dicke(1961)}]{brans_machs_1961}
Brans, C. \& Dicke, R.~H. 1961, \pr, 124, 925

\bibitem[{{Brax} {et~al.}(2017){Brax}, {Davis}, \& {Jha}}]{brax_2017}
{Brax}, P., {Davis}, A.-C., \& {Jha}, R. 2017, \prd, 95, 083514

\bibitem[{Brown(1993)}]{Brown93a}
Brown, J.~D. 1993, Class. Quant. Grav., 10, 1579

\bibitem[{Bucciantini \& Del~Zanna(2011)}]{bucciantini_general_2011}
Bucciantini, N. \& Del~Zanna, L. 2011, \aap, 528, A101

\bibitem[{Bucciantini \& Del~Zanna(2013)}]{bucciantini_fully_2013}
Bucciantini, N. \& Del~Zanna, L. 2013, \mnras, 428, 71

\bibitem[{Bucciantini {et~al.}(2015)Bucciantini, Pili, \&
  Zanna}]{bucciantini_role_2015}
Bucciantini, N., Pili, A.~G., \& Zanna, L.~D. 2015, \mnras, 447, 1

\bibitem[{Buchdahl(1970)}]{buchdahl_non-linear_1970}
Buchdahl, H.~A. 1970, \mnras, 150, 1

\bibitem[{Burrows {et~al.}(2007)Burrows, Dessart, Livne, Ott, \&
  Murphy}]{burrows_simulations_2007}
Burrows, A., Dessart, L., Livne, E., Ott, C.~D., \& Murphy, J. 2007, \apj, 664,
  416

\bibitem[{{Camelio} {et~al.}(2019){Camelio}, {Dietrich}, {Marques}, \&
  {Rosswog}}]{Camelio_Dietrich+19a}
{Camelio}, G., {Dietrich}, T., {Marques}, M., \& {Rosswog}, S. 2019, \prd, 100,
  123001

\bibitem[{Capozziello \& de~Laurentis(2011)}]{capozziello_extended_2011}
Capozziello, S. \& de~Laurentis, M. 2011, \prep, 509, 167

\bibitem[{Cartan(1986)}]{Cartan86a}
Cartan, {\'E}. 1986, On Manifolds with an Affine Connection and the Theory of
  General Relativity, Monographs and Textbooks in Physical Science
  (Bibliopolis)

\bibitem[{Carter(1969)}]{carter_killing_1969}
Carter, B. 1969, \jmp, 10, 70

\bibitem[{Carter(1970)}]{carter_commutation_1970}
Carter, B. 1970, \cmp, 17, 233

\bibitem[{{Carter}(2009)}]{carter_black_1973_a}
{Carter}, B. 2009, \grg, 41, 2873

\bibitem[{{Carter}(2010)}]{carter_black_1973_b}
{Carter}, B. 2010, \grg, 42, 653

\bibitem[{Chandrasekhar \& Fermi(1953)}]{chandrasekhar_problems_1953}
Chandrasekhar, S. \& Fermi, E. 1953, ApJ, 118, 116

\bibitem[{Chatterjee {et~al.}(2015)Chatterjee, Elghozi, Novak, \&
  Oertel}]{chatterjee_consistent_2015}
Chatterjee, D., Elghozi, T., Novak, J., \& Oertel, M. 2015, \mnras, 447, 3785

\bibitem[{Ciolfi {et~al.}(2019)Ciolfi, Kastaun, Kalinani, \&
  Giacomazzo}]{ciolfi_2019}
Ciolfi, R., Kastaun, W., Kalinani, J.~V., \& Giacomazzo, B. 2019, \prd, 100,
  023005

\bibitem[{{Ciolfi} \& {Rezzolla}(2013)}]{ciolfi_2013}
{Ciolfi}, R. \& {Rezzolla}, L. 2013, \mnras, 435, L43

\bibitem[{Cordero-Carrión {et~al.}(2009)Cordero-Carrión, Cerdá-Durán,
  Dimmelmeier, Jaramillo, Novak, \&
  Gourgoulhon}]{cordero-carrion_improved_2009}
Cordero-Carrión, I., Cerdá-Durán, P., Dimmelmeier, H., {et~al.} 2009, \prd,
  79, 024017

\bibitem[{{Dall'Osso} {et~al.}(2009){Dall'Osso}, {Shore}, \&
  {Stella}}]{Dall'Osso_Shore+09a}
{Dall'Osso}, S., {Shore}, S.~N., \& {Stella}, L. 2009, \mnras, 398, 1869

\bibitem[{Damour \& Esposito-Farèse(1993)}]{damour_nonperturbative_1993}
Damour, T. \& Esposito-Farèse, G. 1993, \prl, 70, 2220

\bibitem[{Damour \& Esposito-Farèse(1996)}]{damour_tensor-scalar_1996}
Damour, T. \& Esposito-Farèse, G. 1996, \prd, 54, 1474

\bibitem[{Damour {et~al.}(2002)Damour, Piazza, \&
  Veneziano}]{damour_runaway_2002}
Damour, T., Piazza, F., \& Veneziano, G. 2002, \prl, 89, 081601

\bibitem[{{Das} \& {Mukhopadhyay}(2015)}]{Das_Mukhopadhyay15a}
{Das}, U. \& {Mukhopadhyay}, B. 2015, \jcap, 2015, 016

\bibitem[{{De Felice} \& {Tanaka}(2010)}]{defelice_2010}
{De Felice}, A. \& {Tanaka}, T. 2010, Progress of Theoretical Physics, 124, 503

\bibitem[{De~Felice \& Tsujikawa(2010)}]{de_felice_f_2010}
De~Felice, A. \& Tsujikawa, S. 2010, \lrr, 13, 3

\bibitem[{{DeFelice} {et~al.}(2006){DeFelice}, {Hindmarsh}, \&
  {Trodden}}]{defelice_2006}
{DeFelice}, A., {Hindmarsh}, M., \& {Trodden}, M. 2006, \jcap, 2006, 005

\bibitem[{{Del Zanna} \& {Chiuderi}(1996)}]{del_zanna_exact_1996}
{Del Zanna}, L. \& {Chiuderi}, C. 1996, \aap, 310, 341

\bibitem[{Del~Zanna {et~al.}(2016)Del~Zanna, Papini, Landi, Bugli, \&
  Bucciantini}]{del_zanna_fast_2016}
Del~Zanna, L., Papini, E., Landi, S., Bugli, M., \& Bucciantini, N. 2016,
  \mnras, 460, 3753

\bibitem[{Del~Zanna {et~al.}(2007)Del~Zanna, Zanotti, Bucciantini, \&
  Londrillo}]{del_zanna_echo:_2007}
Del~Zanna, L., Zanotti, O., Bucciantini, N., \& Londrillo, P. 2007, \aap, 473,
  11

\bibitem[{Delphenich(2005)}]{Delphenich05a}
Delphenich, D. 2005, in {6th International Conference on Symmetry in Nonlinear
  Mathematical Physics (SNMP 05) Kiev, Ukraine, June 20-26, 2005}

\bibitem[{Del Zanna \& Bucciantini(2018)}]{del_zanna_chiral_2018}
Del Zanna, L. \& Bucciantini, N. 2018, Monthly Notices of the Royal
  Astronomical Society, 479, 657

\bibitem[{Deser(2000)}]{deser_infinities_2000}
Deser, S. 2000, \adp, 9, 299

\bibitem[{Dimmelmeier {et~al.}(2006)Dimmelmeier, Stergioulas, \&
  Font}]{dimmelmeier_non-linear_2006}
Dimmelmeier, H., Stergioulas, N., \& Font, J.~A. 2006, \mnras, 368, 1609

\bibitem[{Dohi {et~al.}(2020)Dohi, Kase, Kimura, Yamamoto, \&
  Hashimoto}]{dohi_2020}
Dohi, A., Kase, R., Kimura, R., Yamamoto, K., \& Hashimoto, M.-a. 2020
  [\eprint[arXiv]{2003.12571}]

\bibitem[{Doneva \& Yazadjiev(2016)}]{doneva_rapidly_2016}
Doneva, D.~D. \& Yazadjiev, S.~S. 2016, \jcap, 2016, 019

\bibitem[{{Doneva} \& {Yazadjiev}(2019)}]{doneva_topological_2019}
{Doneva}, D.~D. \& {Yazadjiev}, S.~S. 2019, arXiv e-prints, arXiv:1911.06908

\bibitem[{Doneva {et~al.}(2013)Doneva, Yazadjiev, Stergioulas, \&
  Kokkotas}]{doneva_rapidly_2014}
Doneva, D.~D., Yazadjiev, S.~S., Stergioulas, N., \& Kokkotas, K.~D. 2013,
  Phys. Rev. D, 88, 084060

\bibitem[{Doneva {et~al.}(2018)Doneva, Yazadjiev, Stergioulas, \&
  Kokkotas}]{doneva_differentially_2018}
Doneva, D.~D., Yazadjiev, S.~S., Stergioulas, N., \& Kokkotas, K.~D. 2018,
  Phys. Rev. D, 98, 104039

\bibitem[{Faraoni(2004)}]{faraoni_2004}
Faraoni, V. 2004, Cosmology in Scalar-Tensor Gravity, Fundamental Theories of
  Physics (Springer Netherlands)

\bibitem[{Ferrario {et~al.}(2015)Ferrario, Melatos, \&
  Zrake}]{ferrario_magnetic_2015}
Ferrario, L., Melatos, A., \& Zrake, J. 2015, \ssr, 191, 77

\bibitem[{Ferraro(1954)}]{ferraro_equilibrium_1954}
Ferraro, V. C.~A. 1954, ApJ, 119, 407

\bibitem[{Franzon {et~al.}(2016)Franzon, Dexheimer, \&
  Schramm}]{franzon_internal_2016}
Franzon, B., Dexheimer, V., \& Schramm, S. 2016, \prd, 94, 044018

\bibitem[{Freire {et~al.}(2012)Freire, Wex, Esposito-Farèse, Verbiest, Bailes,
  Jacoby, Kramer, Stairs, Antoniadis, \& Janssen}]{freire_relativistic_2012}
Freire, P. C.~C., Wex, N., Esposito-Farèse, G., {et~al.} 2012, \mnras, 423,
  3328

\bibitem[{{Fricke}(1969)}]{Fricke69a}
{Fricke}, K. 1969, \aap, 1, 388

\bibitem[{Frieben \& Rezzolla(2012)}]{frieben_equilibrium_2012}
Frieben, J. \& Rezzolla, L. 2012, \mnras, 427, 3406

\bibitem[{Friedman \& Stergioulas(2013)}]{friedman_2013}
Friedman, J.~L. \& Stergioulas, N. 2013, Rotating Relativistic Stars, Cambridge
  Monographs on Mathematical Physics (Cambridge University Press)

\bibitem[{Fujii \& Maeda(2003)}]{fujii_scalar-tensor_2003}
Fujii, Y. \& Maeda, K.-i. 2003, The {Scalar}-{Tensor} {Theory} of {Gravitation}

\bibitem[{Fujisawa \& Eriguchi(2015)}]{fujisawa_appearance_2015}
Fujisawa, K. \& Eriguchi, Y. 2015, PASJ, 67, 53

\bibitem[{{Gao} {et~al.}(2016){Gao}, {Zhang}, \& {L{\"u}}}]{gao_2015}
{Gao}, H., {Zhang}, B., \& {L{\"u}}, H.-J. 2016, \prd, 93, 044065

\bibitem[{Gerosa {et~al.}(2016)Gerosa, Sperhake, \&
  Ott}]{gerosa_numerical_2016}
Gerosa, D., Sperhake, U., \& Ott, C.~D. 2016, \cqg, 33, 135002

\bibitem[{{Gong} {et~al.}(2018){Gong}, {Papantonopoulos}, \& {Yi}}]{gong_2018}
{Gong}, Y., {Papantonopoulos}, E., \& {Yi}, Z. 2018, \epjc, 78, 738

\bibitem[{Gourgoulhon(2012)}]{gourgoulhon_3+1_2012}
Gourgoulhon, {\'E}. 2012, 3+1 Formalism in General Relativity: Bases of
  Numerical Relativity, Lecture Notes in Physics (Springer Berlin Heidelberg)

\bibitem[{Green {et~al.}(1988)Green, Schwartz, \&
  Witten}]{green_superstring_1988}
Green, M.~B., Schwartz, J.~H., \& Witten, E. 1988, Astronomische Nachrichten,
  309, 297

\bibitem[{{Hagihara} {et~al.}(2019){Hagihara}, {Era}, {Iikawa}, {Takeda}, \&
  {Asada}}]{hagihara_2019}
{Hagihara}, Y., {Era}, N., {Iikawa}, D., {Takeda}, N., \& {Asada}, H. 2019,
  arXiv:1912.06340

\bibitem[{Harada(1998)}]{harada_neutron_1998}
Harada, T. 1998, \prd, 57, 4802

\bibitem[{Hawking(1972)}]{hawking_black_1972}
Hawking, S.~W. 1972, \cmp, 25, 167

\bibitem[{Hawking \& Ellis(1973)}]{hawking_ellis_1973}
Hawking, S.~W. \& Ellis, G. F.~R. 1973, The Large Scale Structure of
  Space-Time, Cambridge Monographs on Mathematical Physics (Cambridge
  University Press)

\bibitem[{{Heisenberg}(2014)}]{heinsenberg_2014}
{Heisenberg}, L. 2014, \jcap, 2014, 015

\bibitem[{{Hellings} \& {Nordtvedt}(1973)}]{hellings_1973}
{Hellings}, R.~W. \& {Nordtvedt}, K. 1973, \prd, 7, 3593

\bibitem[{{Iosif} \& {Stergioulas}(2014)}]{iosif_2014}
{Iosif}, P. \& {Stergioulas}, N. 2014, \grg, 46, 1800

\bibitem[{Isenberg(2008)}]{isenberg_waveless_2008}
Isenberg, J.~A. 2008, \ijmpd, 17, 265

\bibitem[{{Just}(1959)}]{Just59a}
{Just}, K. 1959, Z. Naturforsch. A, 14, 751

\bibitem[{{Kase} {et~al.}(2018){Kase}, {Minamitsuji}, \&
  {Tsujikawa}}]{kase_2018}
{Kase}, R., {Minamitsuji}, M., \& {Tsujikawa}, S. 2018, \prd, 97, 084009

\bibitem[{{Kase} {et~al.}(2020){Kase}, {Minamitsuji}, \&
  {Tsujikawa}}]{kase_2020}
{Kase}, R., {Minamitsuji}, M., \& {Tsujikawa}, S. 2020, arXiv:2001.10701

\bibitem[{Kawamura {et~al.}(2016)Kawamura, Giacomazzo, Kastaun, Ciolfi,
  Endrizzi, Baiotti, \& Perna}]{kawamura_2016}
Kawamura, T., Giacomazzo, B., Kastaun, W., {et~al.} 2016, \prd, 94, 064012

\bibitem[{Kiuchi {et~al.}(2009)Kiuchi, Kotake, \&
  Yoshida}]{kiuchi_equilibrium_2009}
Kiuchi, K., Kotake, K., \& Yoshida, S. 2009, \apj, 698, 541

\bibitem[{Kiuchi \& Yoshida(2008)}]{kiuchi_relativistic_2008}
Kiuchi, K. \& Yoshida, S. 2008, \prd, 78, 044045

\bibitem[{Konno(2001)}]{konno_moments_2001}
Konno, K. 2001, \aap, 372, 594

\bibitem[{Kundt \& Trümper(1966)}]{kundt_orthogonal_1966}
Kundt, W. \& Trümper, M. 1966, Z. Phys.~A, 192, 419

\bibitem[{Langlois {et~al.}(2018)Langlois, Saito, Yamauchi, \&
  Noui}]{langlois_2018}
Langlois, D., Saito, R., Yamauchi, D., \& Noui, K. 2018, \prd, 97, 061501

\bibitem[{{Lasky} {et~al.}(2011){Lasky}, {Zink}, {Kokkotas}, \&
  {Glampedakis}}]{lasky_2011}
{Lasky}, P.~D., {Zink}, B., {Kokkotas}, K.~D., \& {Glampedakis}, K. 2011,
  \apjl, 735, L20

\bibitem[{Lovelock(1971)}]{lovelock_einstein_1971}
Lovelock, D. 1971, \jmp, 12, 498

\bibitem[{{Margalit} \& {Metzger}(2017)}]{margalit_2017}
{Margalit}, B. \& {Metzger}, B.~D. 2017, \apjl, 850, L19

\bibitem[{Matsuda \& Nariai(1973)}]{matsuda_hydrodynamic_1973}
Matsuda, T. \& Nariai, H. 1973, \ptp, 49, 1195

\bibitem[{Mendes \& Ortiz(2016)}]{mendes_highly_2016}
Mendes, R.~F. \& Ortiz, N. 2016, \prd, 93, 124035

\bibitem[{{Metzger} {et~al.}(2011){Metzger}, {Giannios}, {Thompson},
  {Bucciantini}, \& {Quataert}}]{Metzger_Giannios+11a}
{Metzger}, B.~D., {Giannios}, D., {Thompson}, T.~A., {Bucciantini}, N., \&
  {Quataert}, E. 2011, \mnras, 413, 2031

\bibitem[{Miketinac(1975)}]{miketinac_structure_1975}
Miketinac, M.~J. 1975, \apss, 35, 349

\bibitem[{{Miller} {et~al.}(2019){Miller}, {Lamb}, {Dittmann}, {Bogdanov},
  {Arzoumanian}, {Gendreau}, {Guillot}, {Harding}, {Ho}, {Lattimer}, {Ludlam},
  {Mahmoodifar}, {Morsink}, {Ray}, {Strohmayer}, {Wood}, {Enoto}, {Foster},
  {Okajima}, {Prigozhin}, \& {Soong}}]{miller_2019}
{Miller}, M.~C., {Lamb}, F.~K., {Dittmann}, A.~J., {et~al.} 2019, \apjl, 887,
  L24

\bibitem[{Monaghan(1965)}]{monaghan_magnetic_1965}
Monaghan, J.~J. 1965, MNRAS, 131, 105

\bibitem[{Monaghan(1966)}]{monaghan_magnetic_1966}
Monaghan, J.~J. 1966, \mnras, 134, 275

\bibitem[{{Nordtvedt}(1970)}]{nord_1970}
{Nordtvedt}, Kenneth, J. 1970, \apj, 161, 1059

\bibitem[{Novak(1998{\natexlab{a}})}]{novak_neutron_1998}
Novak, J. 1998{\natexlab{a}}, \prd, 58, 064019

\bibitem[{Novak(1998{\natexlab{b}})}]{novak_spherical_1998}
Novak, J. 1998{\natexlab{b}}, \prd, 57, 4789

\bibitem[{Oppenheimer \& Volkoff(1939)}]{Oppenheimer_Volkoff39a}
Oppenheimer, J.~R. \& Volkoff, G.~M. 1939, Phys. Rev., 55, 374

\bibitem[{Oron(2002)}]{oron_relativistic_2002}
Oron, A. 2002, \prd, 66, 023006

\bibitem[{Ostriker \& Hartwick(1968)}]{ostriker_rapidly_1968}
Ostriker, J.~P. \& Hartwick, F. D.~A. 1968, \apj, 153, 797

\bibitem[{Ott {et~al.}(2007)Ott, Dimmelmeier, Marek, Janka, Zink, Hawke, \&
  Schnetter}]{ott_rotating_2007}
Ott, C.~D., Dimmelmeier, H., Marek, A., {et~al.} 2007, \cqg, 24, S139

\bibitem[{{{\"O}zel} \& {Freire}(2016)}]{ozel_2016}
{{\"O}zel}, F. \& {Freire}, P. 2016, \araa, 54, 401

\bibitem[{Pani \& Berti(2014)}]{pani_i-love-q_2014}
Pani, P. \& Berti, E. 2014, \prd, 90, 024025, arXiv: 1405.4547

\bibitem[{Papantonopoulos(2015)}]{papantonopoulos_modifications_2015}
Papantonopoulos, E. 2015, Modifications of {Einstein}'s {Theory} of {Gravity}
  at {Large} {Distances}, Lecture {Notes} in {Physics} (Springer International
  Publishing)

\bibitem[{{Papitto} {et~al.}(2014){Papitto}, {Torres}, {Rea}, \&
  {Tauris}}]{papitto_2014}
{Papitto}, A., {Torres}, D.~F., {Rea}, N., \& {Tauris}, T.~M. 2014, \aap, 566,
  A64

\bibitem[{{Pappas} {et~al.}(2019){Pappas}, {Doneva}, {Sotiriou}, {Yazadjiev},
  \& {Kokkotas}}]{pappas_multipole_2019}
{Pappas}, G., {Doneva}, D.~D., {Sotiriou}, T.~P., {Yazadjiev}, S.~S., \&
  {Kokkotas}, K.~D. 2019, \prd, 99, 104014

\bibitem[{Peebles \& Ratra(2003)}]{peebles_cosmological_2003}
Peebles, P. J.~E. \& Ratra, B. 2003, \rmp, 75, 559

\bibitem[{Pili {et~al.}(2014)Pili, Bucciantini, \&
  Del~Zanna}]{pili_axisymmetric_2014}
Pili, A.~G., Bucciantini, N., \& Del~Zanna, L. 2014, \mnras, 439, 3541

\bibitem[{Pili {et~al.}(2015)Pili, Bucciantini, \&
  Del~Zanna}]{pili_general_2015}
Pili, A.~G., Bucciantini, N., \& Del~Zanna, L. 2015, \mnras, 447, 2821

\bibitem[{Pili {et~al.}(2017)Pili, Bucciantini, \&
  Del~Zanna}]{pili_general_2017}
Pili, A.~G., Bucciantini, N., \& Del~Zanna, L. 2017, \mnras, 470, 2469

\bibitem[{{Pili} {et~al.}(2016){Pili}, {Bucciantini}, {Drago}, {Pagliara}, \&
  {Del Zanna}}]{pili_quark_2016}
{Pili}, A.~G., {Bucciantini}, N., {Drago}, A., {Pagliara}, G., \& {Del Zanna},
  L. 2016, \mnras, 462, L26

\bibitem[{{Popov}(2016)}]{popov_origins_2016}
{Popov}, S.~B. 2016, \aat, 29, 183

\bibitem[{Prendergast(1956)}]{prendergast_equilibrium_1956}
Prendergast, K.~H. 1956, \apj, 123, 498

\bibitem[{Price \& Rosswog(2006)}]{price_2006}
Price, D.~J. \& Rosswog, S. 2006, \sci, 312, 719

\bibitem[{{Quiros}(2019)}]{quiros_2019}
{Quiros}, I. 2019, \ijmpd, 28, 1930012

\bibitem[{{Raithel} {et~al.}(2018){Raithel}, {\"O}zel, \&
  {Psaltis}}]{raithel_2018}
{Raithel}, C.~A., {\"O}zel, F., \& {Psaltis}, D. 2018, \apjl, 857, L23

\bibitem[{{Ramazano{\v{g}}lu}(2017)}]{rama_2017}
{Ramazano{\v{g}}lu}, F.~M. 2017, \prd, 96, 064009

\bibitem[{Ramazanoğlu \& Pretorius(2016)}]{ramazanoglu_spontaneous_2016}
Ramazanoğlu, F.~M. \& Pretorius, F. 2016, \prd, 93, 064005, arXiv: 1601.07475

\bibitem[{Rheinhardt \& Geppert(2005)}]{rheinhardt_proto-neutron-star_2005}
Rheinhardt, M. \& Geppert, U. 2005, \aap, 435, 201

\bibitem[{Roberts(1955)}]{roberts_equilibrium_1955}
Roberts, P.~H. 1955, \apj, 122, 508

\bibitem[{{Rowlinson} {et~al.}(2013){Rowlinson}, {O'Brien}, {Metzger},
  {Tanvir}, \& {Levan}}]{rowlinson_2013}
{Rowlinson}, A., {O'Brien}, P.~T., {Metzger}, B.~D., {Tanvir}, N.~R., \&
  {Levan}, A.~J. 2013, \mnras, 430, 1061

\bibitem[{Roxburgh(1966)}]{roxburgh_magnetostatic_1966}
Roxburgh, I.~W. 1966, \mnras, 132, 347

\bibitem[{{Salgado}(2006)}]{Salgado06a}
{Salgado}, M. 2006, Classical and Quantum Gravity, 23, 4719

\bibitem[{{Salgado} {et~al.}(2008){Salgado}, {Mart{\'\i}nez del R{\'\i}o},
  {Alcubierre}, \& {N{\'u}{\~n}ez}}]{Salgado_Martinez-del-Rio+08a}
{Salgado}, M., {Mart{\'\i}nez del R{\'\i}o}, D., {Alcubierre}, M., \&
  {N{\'u}{\~n}ez}, D. 2008, \prd, 77, 104010

\bibitem[{{Salgado} {et~al.}(1998){Salgado}, {Sudarsky}, \&
  {Nucamendi}}]{salgado_1998}
{Salgado}, M., {Sudarsky}, D., \& {Nucamendi}, U. 1998, \prd, 58, 124003

\bibitem[{{Santiago} \& {Silbergleit}(2000)}]{Santiago_Silbergleit00a}
{Santiago}, D.~I. \& {Silbergleit}, A.~S. 2000, General Relativity and
  Gravitation, 32, 565

\bibitem[{{Schubert}(1968)}]{Schubert68a}
{Schubert}, G. 1968, \apj, 151, 1099

\bibitem[{Schärer {et~al.}(2014)Schärer, Angélil, Bondarescu, Jetzer, \&
  Lundgren}]{scharer_testing_2014}
Schärer, A., Angélil, R., Bondarescu, R., Jetzer, P., \& Lundgren, A. 2014,
  \prd, 90, 123005

\bibitem[{Shao {et~al.}(2017)Shao, Sennett, Buonanno, Kramer, \&
  Wex}]{shao_constraining_2017}
Shao, L., Sennett, N., Buonanno, A., Kramer, M., \& Wex, N. 2017, \prx, 7,
  041025

\bibitem[{Shibata \& Sekiguchi(2005)}]{shibata_magnetohydrodynamics_2005}
Shibata, M. \& Sekiguchi, Y.-i. 2005, \prd, 72, 044014

\bibitem[{Shibata {et~al.}(2014)Shibata, Taniguchi, Okawa, \&
  Buonanno}]{shibata_coalescence_2014}
Shibata, M., Taniguchi, K., Okawa, H., \& Buonanno, A. 2014, \prd, 89, 084005

\bibitem[{Silva {et~al.}(2015)Silva, Macedo, Berti, \&
  Crispino}]{silva_slowly_2015}
Silva, H.~O., Macedo, C. F.~B., Berti, E., \& Crispino, L. C.~B. 2015, Class.
  Quant. Grav., 32, 145008

\bibitem[{{Silva} {et~al.}(2018){Silva}, {Sakstein}, {Gualtieri}, {Sotiriou},
  \& {Berti}}]{silva_2018}
{Silva}, H.~O., {Sakstein}, J., {Gualtieri}, L., {Sotiriou}, T.~P., \& {Berti},
  E. 2018, \prl, 120, 131104

\bibitem[{Sotani(2012)}]{sotani_slowly_2012}
Sotani, H. 2012, \prd, 86, 124036, arXiv: 1211.6986

\bibitem[{{Sotani} \& {Kokkotas}(2005)}]{sotani_2005}
{Sotani}, H. \& {Kokkotas}, K.~D. 2005, \prd, 71, 124038

\bibitem[{Sotiriou(2006)}]{sotiriou_fr_2006}
Sotiriou, T.~P. 2006, \cqg, 23, 5117

\bibitem[{{Spruit}(2009)}]{spruit_source_2009}
{Spruit}, H.~C. 2009, in IAU Symposium, Vol. 259, Cosmic Magnetic Fields: From
  Planets, to Stars and Galaxies, ed. K.~G. {Strassmeier}, A.~G. {Kosovichev},
  \& J.~E. {Beckman}, 61--74

\bibitem[{{Staykov} {et~al.}(2019){Staykov}, {Doneva}, {Popchev}, \&
  {Yazadjiev}}]{staykov_2019}
{Staykov}, K.~V., {Doneva}, D.~D., {Popchev}, D., \& {Yazadjiev}, S.~S. 2019,
  in American Institute of Physics Conference Series, Vol. 2075, American
  Institute of Physics Conference Series, 040006

\bibitem[{{Staykov} {et~al.}(2018){Staykov}, {Popchev}, {Doneva}, \&
  {Yazadjiev}}]{staykov_2018}
{Staykov}, K.~V., {Popchev}, D., {Doneva}, D.~D., \& {Yazadjiev}, S.~S. 2018,
  \epjc, 78, 586

\bibitem[{Suvorov(2018)}]{suvorov_monopolar_2018}
Suvorov, A.~G. 2018, \prd, 98, 084026

\bibitem[{Taniguchi {et~al.}(2015)Taniguchi, Shibata, \&
  Buonanno}]{taniguchi_quasiequilibrium_2015}
Taniguchi, K., Shibata, M., \& Buonanno, A. 2015, \prd, 91, 024033

\bibitem[{Tayler(1973)}]{tayler_adiabatic_1973}
Tayler, R.~J. 1973, \mnras, 161, 365

\bibitem[{Tolman(1939)}]{Tolman39a}
Tolman, R.~C. 1939, Phys. Rev., 55, 364

\bibitem[{{Tomei} {et~al.}(2020){Tomei}, {Del Zanna}, {Bugli}, \&
  {Bucciantini}}]{tomei2020}
{Tomei}, N., {Del Zanna}, L., {Bugli}, M., \& {Bucciantini}, N. 2020, \mnras,
  491, 2346

\bibitem[{Tomimura \& Eriguchi(2005)}]{tomimura_new_2005}
Tomimura, Y. \& Eriguchi, Y. 2005, \mnras, 359, 1117

\bibitem[{Touboul {et~al.}(2017)Touboul, Métris, Rodrigues, André, Baghi,
  Bergé, Boulanger, Bremer, Carle, Chhun, Christophe, Cipolla, Damour, Danto,
  Dittus, Fayet, Foulon, Gageant, Guidotti, Hagedorn, Hardy, Huynh, Inchauspe,
  Kayser, Lala, Lämmerzahl, Lebat, Leseur, Liorzou, List, Löffler, Panet,
  Pouilloux, Prieur, Rebray, Reynaud, Rievers, Robert, Selig, Serron, Sumner,
  Tanguy, \& Visser}]{touboul_microscope_2017}
Touboul, P., Métris, G., Rodrigues, M., {et~al.} 2017, \prl, 119, 231101

\bibitem[{Trimble(1987)}]{trimble_existence_1987}
Trimble, V. 1987, \araa, 25, 425

\bibitem[{Ury\ifmmode~\bar{u}\else \={u}\fi{}
  {et~al.}(2019)Ury\ifmmode~\bar{u}\else \={u}\fi{}, Yoshida, Gourgoulhon,
  Markakis, Fujisawa, Tsokaros, Taniguchi, \& Eriguchi}]{uryu_new_2019}
Ury\ifmmode~\bar{u}\else \={u}\fi{}, K. b.~o., Yoshida, S., Gourgoulhon, E.,
  {et~al.} 2019, Phys. Rev. D, 100, 123019

\bibitem[{Uryū {et~al.}(2014)Uryū, Gourgoulhon, Markakis, Fujisawa, Tsokaros,
  \& Eriguchi}]{uryu_equilibrium_2014}
Uryū, K., Gourgoulhon, E., Markakis, C.~M., {et~al.} 2014, \prd, 90, 101501

\bibitem[{{van~Dantzig} \& {Dirac}(1934)}]{van-Dantzig_Dirac34a}
{van~Dantzig}, D. \& {Dirac}, P.~A.~M. 1934, Math. Proc. Camb. Philos. Soc.,
  30, 421

\bibitem[{Wagoner(1970)}]{wagoner_scalar-tensor_1970}
Wagoner, R.~V. 1970, \prd, 1, 3209

\bibitem[{Will(2014)}]{will_confrontation_2014}
Will, C.~M. 2014, \lrr, 17

\bibitem[{Wilson \& Mathews(2003)}]{wilson_mathews_2003}
Wilson, J.~R. \& Mathews, G.~J. 2003, Relativistic Numerical Hydrodynamics,
  Cambridge Monographs on Mathematical Physics (Cambridge University Press)

\bibitem[{Wilson {et~al.}(1996)Wilson, Mathews, \&
  Marronetti}]{wilson_relativistic_1996}
Wilson, J.~R., Mathews, G.~J., \& Marronetti, P. 1996, \prd, 54, 1317

\bibitem[{Woltjer(1960)}]{woltjer_magnetostatic_1960}
Woltjer, L. 1960, \apj, 131, 227

\bibitem[{Wright(1973)}]{wright_pinch_1973}
Wright, G. a.~E. 1973, \mnras, 162, 339

\bibitem[{{Yakovlev} {et~al.}(2005){Yakovlev}, {Gnedin}, {Gusakov}, {Kaminker},
  {Levenfish}, \& {Potekhin}}]{yako_2005}
{Yakovlev}, D.~G., {Gnedin}, O.~Y., {Gusakov}, M.~E., {et~al.} 2005, \nphysa,
  752, 590

\bibitem[{Yazadjiev(2012)}]{yazadjiev_relativistic_2012}
Yazadjiev, S.~S. 2012, \prd, 85, 044030

\bibitem[{{Yazadjiev} {et~al.}(2016){Yazadjiev}, {Doneva}, \&
  {Popchev}}]{yaza_2016}
{Yazadjiev}, S.~S., {Doneva}, D.~D., \& {Popchev}, D. 2016, \prd, 93, 084038

\bibitem[{Yoshida {et~al.}(2006)Yoshida, Yoshida, \&
  Eriguchi}]{yoshida_twisted-torus_2006}
Yoshida, S., Yoshida, S., \& Eriguchi, Y. 2006, \apj, 651, 462

\bibitem[{Zhang {et~al.}(2019)Zhang, Niu, \& Zhao}]{zhang_2019}
Zhang, X., Niu, R., \& Zhao, W. 2019, \prd, 100, 024038

\end{thebibliography}



\appendix
\section{XCFC for a rotating NS}\label{app:rotating}
We show here how the standard techniques of \texttt{XNS}, based on the XCFC approach to the solution of the metric functions, can be adapted to take into account the presence of a scalar field. For simplicity we are going to consider here only un-magnetised rotators. The generalisation to magnetised ones is trivial and strictly follows what was done in \citet{pili_general_2017}.
In the E-frame the standard set of XCFC equations is:
\begin{align}
  &\Delta_LW^i =8\pi f^{ij}\hat{S}_j \quad , \\
  &\Delta \psi  = -2\pi\hat{E} \psi^{-1}-\tfrac{1}{8}f_{ik}f^{jl}\hat{A}^{ij}\hat{A}^{kl}\psi^{-7} \quad , \\
  &\Delta (\alpha \psi) = [2\pi(\hat{E}+2\hat{S}) \psi^{-2}+\tfrac{7}{8}f_{ik}f^{jl}\hat{A}^{ij}\hat{A}^{kl}\psi^{-8}]\alpha\psi , \\
  &\Delta_L\beta^i =16\pi \alpha \psi^{-6} f^{ij}\hat{S}_j+2\hat{A}^{ij}\hat{\nabla} _j(\alpha \psi^{-6}) \quad , 
\end{align}  
where $f_{ij}$is the flat 3-metric, $\hat{\nabla}_i $ is the flat covariant derivative ($\hat{\nabla}_k f_{ij}=0 $), and $\Delta = \hat{\nabla}_i \hat{\nabla}^i$ is the usual Laplacian operator in flat 3-space.  $\Delta_L$ is defined as:
\begin{align}
  \Delta_LX^i = \Delta X^i +\tfrac{1}{3}\hat{\nabla}^i(\hat{\nabla}_j X^j)
\end{align}
and 
\begin{align}
 \hat{A}^{ij}=  \hat{\nabla}^i W^j +  \hat{\nabla}^j W^i -\tfrac{2}{3}f^{ij} (\hat{\nabla}_k W^k) \quad .
\end{align}
The source terms come from the 3+1 decomposition of the energy-momentum tensor in the E-frame:
\begin{align}
 & \hat{E}= \psi^{6} \bar{n}_\mu\bar{n}_\nu (\bar{T}^{\mu\nu}_{\mathrm p} +\bar{T}^{\mu\nu}_{\mathrm s}) \quad , \\
 & \hat{S}_j= \psi^{6} \bar{n}_\mu\bar{\gamma}_{j\nu} (\bar{T}^{\mu\nu}_{\mathrm p} +\bar{T}^{\mu\nu}_{\mathrm s}) \quad , \\
 & \hat{S}= \psi^{6} \bar{\gamma}^{j}_{\mu}\bar{\gamma}_{j\nu} (\bar{T}^{\mu\nu}_{\mathrm p} +\bar{T}^{\mu\nu}_{\mathrm s}) \quad .
\end{align}
For stationary ($\partial_t =0$) and axisymmetric  ($\partial_\phi =0$) configurations, for the metric given by Eq.~\ref{eq:cfcmetric} (where the only non vanishing component of the shift vector is $\beta^\phi$), assuming that the only non vanishing component of the velocity is $v^\phi$, it can be shown that
\begin{align}
  & \hat{E} = \psi^{6} \left\{ \mathcal{A}^4\left[ \Gamma^2(e+p)-p \right]+\tfrac{1}{8\pi} Q^2\right\} \quad , \\
  & \hat{S}_r = \hat{S}_\theta=0 \quad , \\ 
 & \hat{S}_\phi= \psi^{6} \mathcal{A}^3(e+p)\Gamma v_\phi \quad , \\
 & \hat{S}= \psi^{6}  \left\{ \mathcal{A}^4\left[ \Gamma^2(e+p)v^2+3p \right]-\tfrac{1}{8\pi} Q^2\right\} \quad , 
\end{align}
where $e$, $p$, and $v^2=\tilde{\gamma}_{ij} v^i v^j$ are all in the J-frame and $Q^2= \bar{\gamma}_{ij} Q^i Q^j$ is instead in the E-frame.

If on a time slice the values of the physical quantities are provided as well as the scalar field, then the XCFC set of equations can be solved for the metric component in the E-frame. It is evident that the XCFC scheme retains its main interesting property of decoupling the various equations, allowing to solve them separately, one after the other.
This holds also in the more general time dependent case. In fact 3+1 schemes for GRHD and MHD evolve the conserved quantities in the J-frame $\psi^{6}  \mathcal{A}^3 \tilde{E}_{\mathrm p}$ and $\psi^{6}  \mathcal{A}^3 \tilde{S}^{\mathrm p} _i$. Combined with a scheme that evolves also the 3+1 components of the scalar field $P$ and $Q^i$, the XCFC equations can then be used to solve for the metric.

\section{The S-TOV system}\label{app:stov}

The S-TOV system of equations can be derived setting $B^i=0$ in Eqs.~\ref{eq:psi},\ref{eq:alpsi},\ref{eq:scal1},\ref{eq:euler}:
\begin{align}
	&\frac{4}{\psi} \frac{d\psi}{dr} =\xi \quad ,\label{eq:psitov} \\
	&\frac{d \chi}{dr} = Q_r \quad ,\label{eq:chitov} \\
	&\frac{d \xi}{dr}=-\frac{\xi ^2}{4}-\frac{2}{r}\xi - 8\pi \psi ^4 \mathcal{A}^4 \left( \rho h -p \right) - Q_r^2 \quad ,\label{eq:xitov} \\
	&\frac{d\alpha}{dr}=\frac{\alpha}{4+2r\xi}\left( -\frac{r}{2}\xi^2-2\xi+16\pi r \mathcal{A}^4 p\psi^4-2rQ_r^2 \right) ,\label{eq:alphatov} \\
	&\frac{d}{dr}\left(\mathcal{A}^4 p \right)=-\frac{\mathcal{A}^4 \rho h}{\alpha}\frac{d \alpha }{dr}+\alpha _\mathrm{s}(\chi) \mathcal{A}^4 \left( 4p-\rho h \right)Q_r \quad ,\label{eq:ptov} \\
	&\frac{d Q_r}{dr}=-Q_r \left[ \frac{1}{\alpha} \frac{d\alpha}{dr}+\frac{\xi}{2}   +\frac{2}{r} \right]-4\pi \psi ^4 \alpha _\mathrm{s}(\chi) \mathcal{A}^4 \left( 4p - \rho h \right) .\label{eq:Xitov}
\end{align}
These must be supplemented by a barotropic EoS $p=p(\rho)$, $\varepsilon=\varepsilon(\rho)$.
This system can be solved, given the value at $r=0$ of the density $\rho_{\mathrm c}$, the conformal factor $\psi_{\mathrm c}$ and the scalar field $\chi_{\mathrm c}$ (recalling that all radial derivatives of scalar quantities vanish in $r=0$). The value of the lapse function at the center,  $\alpha_{\mathrm c}$, is irrelevant to the solution per se, since only its derivative appears in Eqs.~\ref{eq:psitov}-\ref{eq:Xitov}. This means that the lapse function is derived minus an arbitrary constant, which is then chosen in order to satisfy the correct asymptotic behaviour at $r\rightarrow \infty$.

The correct STT solution satisfies the following requirements:
\begin{itemize}
\item the ratio $C=\alpha Q_r/2\partial_r \alpha$ must be constant outside the NS, because it can be shown that it is equal to the ratio $Q_{\mathrm s}/2M$ between the net scalar charge and twice the Komar mass in the E-frame;
\item in vacuum $\alpha$ and $\psi$ must behave like the Just metric \citep{Just59a} in isotropic coordinates.   
\end{itemize}
Given that the Just metric in isotropic coordinates has no analytical form, we provide here an approximation that proves to be accurate with a precision $\sim 10^{-4}$, already at a couple of NS radii. If one writes the metric terms $\psi$ and $\alpha$, outside of the NS surface, as
\begin{align}\label{eq:isodef}
  \psi ^4(r) &= \left[ 1+\frac{1}{2r}\sum ^\infty _{i=0} \frac{m_i}{r^i} \right]^4, \\ \alpha ^2(r) &= \left[ 1-\frac{1}{2r}\sum ^\infty _{i=0} \frac{n_i}{r^i} \right]^2\left[ 1+\frac{1}{2r}\sum ^\infty _{i=0} \frac{m_i}{r^i} \right]^{-2} \quad ,
\end{align}
one finds that the first values of $m_i$ for $i>0$ are:
\begin{align}
  m_1 &= -C^2 m_0^2 \quad ,\\
  m_2 &= -C^2 m_0^3/6 \quad ,\\
  m_3 &=  -C^2(1+3C^2) m_0^4/12 \quad ,\\
  m_4 &=  -C^2(3+11C^2) m_0^5/120 \quad ,\\
  m_5 &=  -C^2(9+58C^2 + 90C^4) m_0^4/720 \quad ,\\
  m_6 &=  -C^2(45+334C^2+618C^4) m_0^5/10080 \quad ,
\end{align}
and $n_i =(-1)^i m_i$. When $Q_{\mathrm s}=0$ one finds $m_0 = n_0$ and $m_i=n_i =0$ for $i>0$, recovering the GR solution.

\section{Global quantities}\label{app:glob}
In this appendix we list the main global quantities used in the paper. We give their general form, valid also in the case of a non-static, but stationary, spacetime.

The Komar mass in the E-frame is
\begin{equation}
\begin{split}
	\bar{M}_\mathrm{k}&\coloneqq 2\int _{\Sigma _t}\left( \bar{T}_{\mu \nu}-\frac{1}{2}\bar{T}g_{\mu \nu} \right) n^\mu \xi ^\nu \sqrt{\gamma}d^3 x = \\
	&=2\pi \int  \mathcal{A}^4 \bigg[ 2p+ (\varepsilon+\rho +p) \Gamma ^2 \left( 1 + v^i v_i - 2\alpha ^{-1} \mathcal{A}v_i \beta ^i \right) +\\
	&+E^i E_i + B^i B_i + \epsilon _{ijk} \alpha ^{-1}\beta ^i \mathcal{A}^2 E^j B^k \bigg]\sqrt{-g} dr d\theta \quad ,
\end{split}
\end{equation}
where $\Sigma _t$ is a spacelike hypersurface of constant coordinate time  $t$ and $\xi ^\nu$ is the timelike Killing vector associated to the stationarity of the spacetime. In our static case it reduces to
\begin{equation}\label{eq:mk}
	\bar{M}_\mathrm{k}=2\pi \int  \mathcal{A}^4 \left[ \varepsilon+\rho +3p + B^i B_i \right]\alpha \psi ^6 r^2 \sin \theta dr d\theta  .
\end{equation}
The baryonic mass, which is the same in the E-frame and in the J-frame, is
\begin{equation}
	M_\mathrm{0} = \int _{\Sigma _t} \mathcal{A}^3 \rho \Gamma \sqrt{\gamma}d^3 x \quad ,
\end{equation}
which in our case is
\begin{equation}
	M_\mathrm{0} = 2\pi \int \mathcal{A}^3 \rho \psi ^6 r^2 \sin \theta dr d\theta \quad .
\end{equation}
The proper mass in the E-frame is
\begin{equation}
	\bar{M}_\mathrm{p} = 2\pi \int \mathcal{A}^4 \left(\varepsilon+\rho \right) \psi ^6 r^2 \sin \theta dr d\theta \quad .
\end{equation}
The scalar charge of the star in the E-frame, $\bar{Q}_\mathrm{s}$, is defined as the monopole component of the scalar field at asymptotically large radii:
\begin{equation}
	\lim_{r\to\infty} \chi (r) \coloneqq \frac{\bar{Q}_\mathrm{s}}{r} \quad .
\end{equation}
By integrating Eq.~\ref{eq:scal} over a spherical volume of asymptotically large radius, using Stokes' theorem and using the fact that $\bar{T}_\mathrm{p}=0$ outside the star's surface, we obtain 
\begin{equation}\label{eq:qs}
	\bar{Q}_\mathrm{s} = 2\pi \int \alpha \alpha _\mathrm{s}(\chi) \mathcal{A}^4 T_\mathrm{p} \psi ^6 r^2 \sin \theta dr d\theta \quad ,
\end{equation}
where $T_\mathrm{p}=3p-\varepsilon-\rho$.
The circumferential radius in the J-frame is
\begin{equation}\label{eq:rc}
	\tilde{R}_\mathrm{c} \coloneqq \left[ \mathcal{A}\psi ^2 r \right]_{\theta = \pi /2}
\end{equation}
The magnetic energy in the J-frame is
\begin{equation}
	\tilde{\mathcal{H}}\coloneqq \int _{\Sigma _t} \frac{1}{2} \left(B_i B^i + E_i E^i \right) \mathcal{A}^{3}  \sqrt{\gamma}d^3x \quad ,
\end{equation}
which reduces to
\begin{equation}
	\tilde{\mathcal{H}}=\pi \int B_i B^i \mathcal{A}^{3} \psi ^6 r^2 \sin \theta dr d\theta \quad .
\end{equation}
The binding energy of the star in the E-frame is defined as
\begin{equation}\label{eq:w}
	\bar{W}\coloneqq M_\mathrm{p}-M_\mathrm{k}+\bar{\mathcal{H}}\quad .
\end{equation}
The flux of the toroidal magnetic field, which is the same in the E-frame and in the J-frame, is
\begin{equation}\label{eq:magflux}
	\Phi = \int r \mathcal{A}^2 \sqrt{B^i B_i} \psi ^4 dr d\theta \quad .
\end{equation}
The magnetic dipole moment in the J-frame is
\begin{equation}\label{eq:magdip}
	\tilde{\mu} = \check{A}_\phi \frac{4r^3}{4r+M_\mathrm{k}} \bigg{|}_{r \gg R_\mathrm{c}} \quad .
\end{equation}
The value converges already for $r\simeq 5-10 R_\mathrm{c}$.

The quadrupole deformation of the star in the E-frame is defined as
\begin{equation}\label{eq:e}
	\bar{e} \coloneqq \frac{\bar{I}^\mathrm{p}_{zz}+\bar{I}^\mathrm{s}_{zz}-\bar{I}^\mathrm{p}_{xx}-\bar{I}^\mathrm{s}_{xx}}{\bar{I}^\mathrm{p}_{zz}} \quad ,
\end{equation}
where the physical and scalar field moments of inertia around the polar axis $z$ and the $x$ axis are, respectively,
\begin{align}
	&\bar{I}^\mathrm{p}_{zz} = 2\pi \int \mathcal{A}^4 \left(\varepsilon+\rho \right) r^4 \sin ^3 \theta dr d\theta \quad ,\\
	&\bar{I}^\mathrm{s}_{zz} = -\frac{1}{4} \int  \psi ^4 Q^2 r^4 \sin ^3 \theta dr d\theta \quad ,\\
	&\bar{I}^\mathrm{p}_{xx} = \pi \int \mathcal{A}^4 \left(\varepsilon+\rho \right) r^4 \sin \theta \left( 1+\cos ^2 \theta \right) dr d\theta \quad ,\\
	&\bar{I}^\mathrm{s}_{xx} = -\frac{1}{8} \int \psi ^4 Q^2 r^4 \sin \theta \left( 1+\cos ^2 \theta \right) dr d\theta \quad .
\end{align}
We note that we defined the moments of inertia of the scalar field in the same way as the usual physical ones: as integrals of the energy density $\bar{T}_\mathrm{s} ^{00}$.

The quadrupolar deformation of the trace $\bar{e}_\mathrm{s}$ is related to the quadrupolar and monopolar distributions of the scalar field at asimptotically large radii, and is defined as
\begin{equation}\label{eq:es}
	\bar{e}_\mathrm{s} \coloneqq   \frac{\int \alpha _\mathrm{s}(\chi) \mathcal{A}^4 T_\mathrm{p} \left( 2-3\sin ^2 \theta \right) r^4 \sin \theta dr d\theta}{r^2_\mathrm{e} \int \alpha _\mathrm{s}(\chi) \mathcal{A}^4 T_\mathrm{p} r^2 \sin \theta dr d\theta} \quad .
\end{equation}
We note that the denominator is the Newtonian equivalent of $\bar{Q}_{\mathrm s}\bar{R}^2_\mathrm{c}$.


\label{lastpage}
\end{document}